\documentclass[10pt,twocolumn,letterpaper]{article}

\usepackage[pagenumbers]{cvpr} % To force page numbers, e.g. for an arXiv version
\definecolor{cvprblue}{rgb}{0.21,0.49,0.74}
\usepackage[pagebackref,breaklinks,colorlinks,allcolors=cvprblue]{hyperref}

\usepackage{url}            % simple URL typesetting
\usepackage{booktabs}       % professional-quality tables
\usepackage{amsfonts}       % blackboard math symbols
\usepackage{nicefrac}       % compact symbols for 1/2, etc.
\usepackage{microtype}      % microtypography
\usepackage{xcolor}         % colors
\usepackage{graphicx}
\usepackage{subcaption}
\usepackage{bm}
\usepackage{upgreek}
\usepackage{amsmath}
\usepackage{bbm}
\usepackage{xcolor}
\usepackage{booktabs}
\usepackage{multirow}
\usepackage{makecell}
\usepackage{soul}
\usepackage{algorithmicx}
\usepackage{algpseudocode}
\usepackage[ruled,linesnumbered]{algorithm2e}
\usepackage{fontawesome}
\usepackage{wrapfig}
\usepackage{tikz}
\usepackage{bbding}
\usepackage{pifont}

\def\paperID{xxxx} % *** Enter the Paper ID here
\def\confName{CVPR}
\def\confYear{2025}

\title{Neural Compression for 3D Geometry Sets}

\author{Siyu Ren\\
City University of Hong Kong\\
{\tt\small siyuren2-c@my.cityu.edu.hk}
\and
Junhui Hou$^*$\\
City University of Hong Kong\\
{\tt\small jh.hou@cityu.edu.hk}
\and
Weiyao Lin\\
Shanghai Jiao Tong University\\
{\tt\small wylin@sjtu.edu.cn}
\and
Wenping Wang\\
Texas A\&M University\\
{\tt\small wenping@cs.hku.hk}
}
\begin{document}
\maketitle
\begin{abstract}
We present NeCGS, the first neural compression paradigm, which can compress a geometry set encompassing thousands of detailed and diverse 3D mesh models by up to 900 times 
with high accuracy and preservation of detailed geometric structures. Specifically, we first propose TSDF-Def, a new implicit representation that is capable of \textbf{accurately} representing irregular 3D mesh models with various structures into regular 4D tensors of \textbf{uniform} and \textbf{compact} size, where 3D surfaces can be extracted through the deformable marching cubes. 
Then we construct a quantization-aware auto-decoder network architecture to regress these 4D tensors to explore the local geometric similarity within each shape and across different shapes for redundancy removal, resulting in more compact representations, including an embedded feature of a smaller size associated with each 3D model and a network parameter shared by all models.  
We finally encode the resulting features and network parameters into bitstreams through entropy coding. Besides, our NeCGS can handle the dynamic scenario well, where new 3D models are constantly added to a compressed set. Extensive experiments and ablation studies demonstrate the significant advantages of our NeCGS over state-of-the-art methods both quantitatively and qualitatively. 
%\textcolor{magenta}{We have included the source code in the \textit{Supplemental Material}}.
\textcolor{magenta}{The source code is available at \href{https://github.com/rsy6318/NeCGS}{https://github.com/rsy6318/NeCGS}.}

\end{abstract}    
\begin{figure*}[h] %[t]
    \centering
    \subfloat[Original ]{\includegraphics[width=0.45\linewidth]{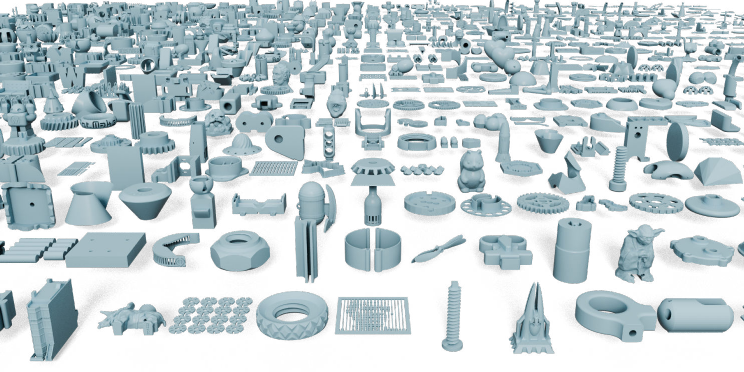}\vspace{-0.3cm}}\quad
    \subfloat[Decompressed]{\includegraphics[width=0.45\linewidth]{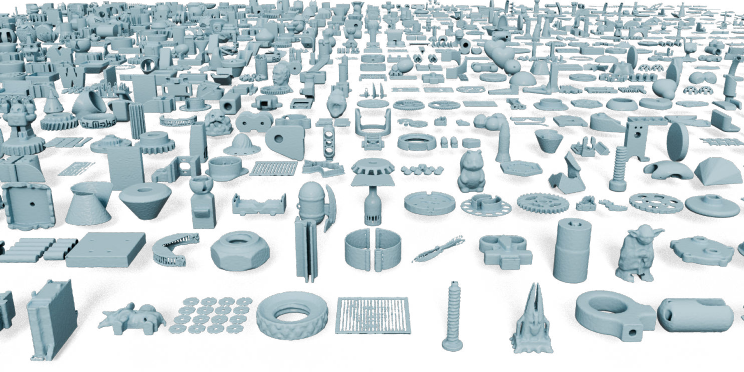}\vspace{-0.3cm}} %\\
    \vspace{-0.3cm}
    \caption{
  \small (\textbf{a}) A geometry set of 1000 3D shapes with various structures from the Thingi10K dataset \cite{THINGI10K}, and the data size is \textbf{335.92} MB. (\textbf{b}) The decompressed 3D shapes after being compressed into \textbf{2.064} MB by our NeCGS.  
   \color{blue}{\faSearch~} Zoom in for details.} \vspace{-0.4cm}
\end{figure*}
\section{Introduction} 
3D mesh models find widespread applications across computer graphics, virtual reality, robotics, and autonomous driving. As geometric data becomes increasingly complex and voluminous, effective compression techniques have become critical for efficient storage and transmission.  

In contrast to 2D images and videos typically structured as  \textit{regular} 2D or 3D tensors, 3D mesh models are commonly represented as \textit{irregular} polygons, posing challenges for compression. Thus, a natural idea is to re-structure mesh models and then leverage image or video compression techniques to compress them. A common practice involves converting mesh models into voxelized point clouds, with the ability to reconstruct the original mesh models using surface reconstruction methods \cite{SPSR, IMLS}. 
Based on this, MPEG has recently developed two types of 3D point cloud compression (PCC) standards \cite{PCCSTANDARD, GPCCVPCC2}: geometry-based PCC (GPCC) for static models and video-based PCC (VPCC) for sequential models. Additionally, the field has seen the emergence of various learning-based PCC techniques \cite{LEARNINGPCC1, LEARNINGPCC2, LEARNINGPCC3, LEARNINGPCC4, PCGCV2} driven by advances in deep learning, improving overall compression efficiency. However, voxelized point clouds demand high resolutions (typically $2^{10}$ or more) to accurately represent geometric shapes, introducing redundancy and limiting compression efficiency. Moreover, these methods primarily focus on individual 3D models or temporally linked 3D model sequences, and they face challenges when dealing with diverse, unrelated 3D shape datasets.

An alternative regular representation method involves employing implicit fields of 3D mesh models, such as signed distance fields (SDFs) and truncated SDFs (TSDFs). This kind of approach entails computing the implicit field value at each uniformly distributed grid point, thereby forming a regular tensor. Mesh models can be reconstructed from these tensors using techniques like Marching Cubes \cite{MARCHINGCUBE} or its variants \cite{MESHUDF, GEOUDF}. In comparison to point clouds, implicit tensorss offer a way to represent mesh models at a relatively lower resolution. However, to capture intricate and detailed geometric structures, larger SDF/TSDF tensors are often necessary when dealing with more complex 3D models. Consequently, when working with geometry sets that encompass a variety of categories, the need for SDF/TSDF tensors of varying sizes arises, sometimes requiring very large tensors to ensure accuracy. This diversity poses significant challenges during the subsequent compression phase.  Recently, neural implicit representation methods, such as DeepSDF \cite{DEEPSDF}, utilize multilayer perceptrons (MLPs) to regress the SDFs of any given query points. While this representation achieves high accuracy for single or similar models (e.g., chairs, tables), the limited receptive field of MLPs makes it challenging to represent large numbers of models in different categories, which is a more common scenario in practice.

We propose NeCGS, a novel framework for compressing large sets of 3D mesh models with diverse categories. Our NeCGS framework consists of two stages: regular geometry representation and compact neural compression. In the first stage, all 3D mesh models are precisely represented into 4D tensors of uniform and compact size through the proposed efficient and effective implicit representation, called TSDF-Def, from which the 3D models can be recovered using the associated deformable marching cubes. In the second stage, we develop a quantization-aware auto-decoder to regress these 4D tensors to explore the local geometric similarity within each shape and across different shapes. The embedded features and decoder parameters of smaller data sizes represent these models, and compressing these components allows us to compress the entire geometry set.
We conducted extensive experiments on various datasets, demonstrating that our NeCGS framework achieves higher compression efficiency compared to existing geometry compression methods when handling large numbers of models. Our NeCGS can achieve a compression ratio of nearly 900 on some datasets, compressing hundreds or even thousands of different models into 1$\sim$2 MB while preserving detailed structures. In addition, our NeCGS can adapt to the dynamic scenario well, where new 3D models are constantly added to the compressed set.

In summary, the main contributions of this paper are: 
\begin{itemize}
    \item an efficient, accurate, and compact implicit representation for 3D mesh models; and 
    \item a novel neural compression framework for 3D geometry sets; and 
    \item a state-of-the-art benchmark for 3D geometry set compression. 
\end{itemize}

\section{Related Work}
\subsection{Geometry Representation}

\noindent\textbf{Explicit Geometry Representation.} 
Among the explicit representations, voxelization \cite{IFNET} is the most intuitive. In this method, geometry models are represented by regularly distributed grids, effectively converting them into 3D `images'. While this approach simplifies the processing of geometry models using image processing techniques, it requires a high resolution to accurately represent the models, which demands substantial memory and limits its application.
Another widely used geometry representation method is the point cloud, which consists of discrete points sampled from the surfaces of models. This method has become a predominant approach for surface representation \cite{ICP,POINTNET,POINTNET2}. However, the discrete nature of the points imposes constraints on its use in downstream tasks such as rendering and editing.
Triangle meshes offer a more precise and efficient geometry representation. By approximating surfaces with numerous triangles, they achieve higher accuracy and efficiency for certain downstream tasks. 

\vspace{0.5em}
\noindent\textbf{Implicit Geometry Representation.} 
Implicit representations use the isosurface of a function or field to represent surfaces. The most widely used implicit representations include Binary Occupancy Field (BOF) \cite{SPSR, OCCNET}, Signed Distance Field (SDF) \cite{DEEPSDF, DEEPIMLS}, and Truncated Signed Distance Field (TSDF) \cite{DYNAMICFUSION}, from which the model's surface can be easily extracted. However, these methods are limited to representing watertight models. The Unsigned Distance Field (UDF) \cite{NDF}, which is the absolute value of the SDF, can represent more general models, not just watertight ones. Despite this advantage, extracting surfaces from UDF is challenging, which limits its application. 

\vspace{0.5em}
\noindent\textbf{Conversion between Geometry Representations.} 
Geometry representations can be converted between explicit and implicit forms. Various methods \cite{PSR, SPSR, IMLS, IMLS2, OCCNET, DEEPIMLS, GEOUDF} are available for calculating the implicit field from given models. Conversely, when converting from implicit to explicit forms, Marching Cubes \cite{MARCHINGCUBE} and its derivatives \cite{DMTET, FLEXICUBE, MESHUDF, GEOUDF} can reconstruct continuous surfaces from various implicit fields.

\begin{figure*}[h]%[t]
    \centering
    \includegraphics[width=0.9\linewidth]{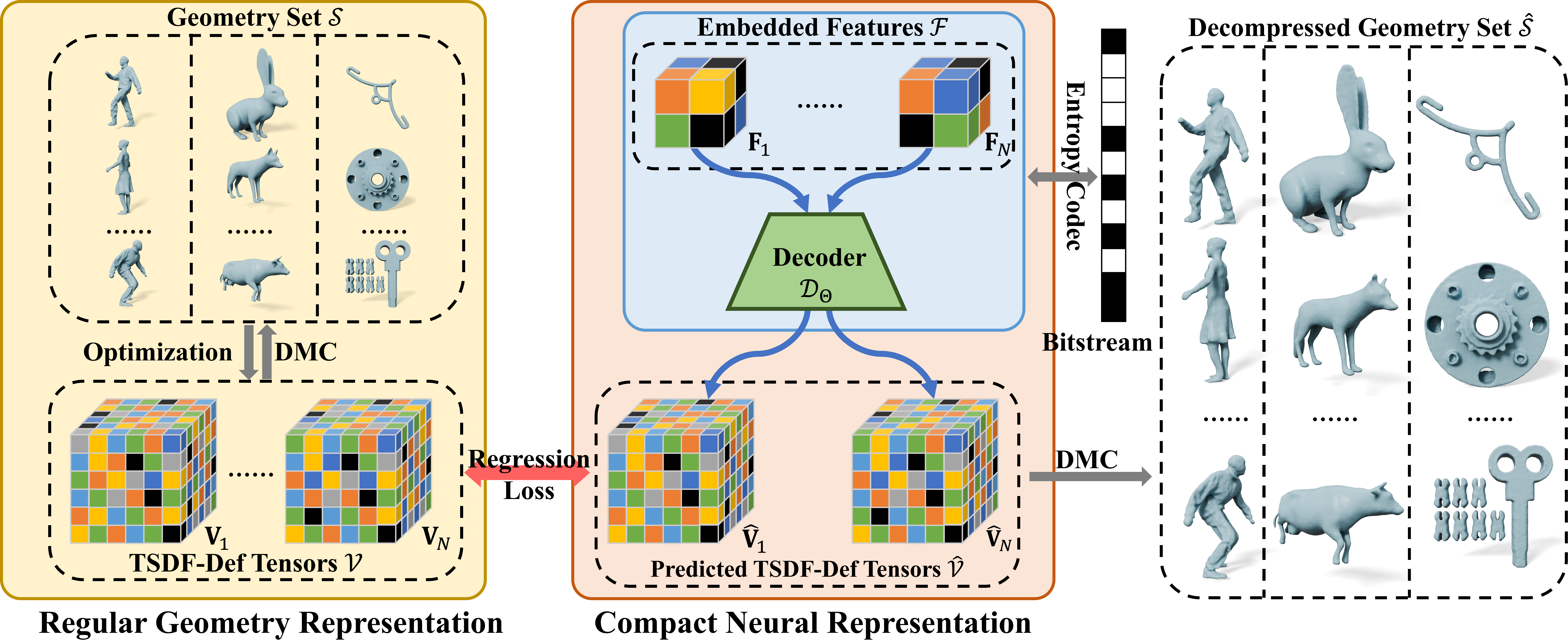}
    \vspace{-0.2cm}
    \caption{\small The pipeline of NeCGS for compressing a geometry set. It first represents irregular 3D mesh models into regular 4D tensors of uniform and compact size through the proposed TSDF-Def, a new implicit representation, and an auto-decoder network is utilized to regress these tensors. Then the embedded features and decoder parameters are compressed into bitstreams through entropy coding. When decompressing the models, the decompressed embedded features are fed into the decoder with the decompressed parameters from the bitstreams, reconstructing the TSDF-Def tensors, and the models can be extracted from them.
    }
    \label{PIPELINE} \vspace{-0.3cm}
\end{figure*}

\subsection{3D Geometry Data Compression}
\noindent\textbf{Single 3D Geometric Model Compression.}
In recent decades, compression techniques for images and videos have rapidly advanced \cite{IMGCOMPRESS1, IMGCOMPRESS2, IMGCOMPRESS3, NERV, HNERV}. However, the irregular nature of geometry data makes it more challenging to compress compared to images and video, which are represented as volumetric data. A natural approach is to convert geometry data into voxelized point clouds, treating them as 3D `images', and then applying image and video compression techniques to them. Following this intuition, MPEG developed the GPCC standards \cite{GPCCVPCC, GPCCVPCC2, GPCCVPCC3}, where triangle meshes or triangle soup approximates the surfaces of 3D models, enabling the compression of models with more complex structures.
Subsequently, several improved methods \cite{IMPROVEPCC2, IMPROVEPCC3, IMPROVEPCC4, IMPROVEPCC7} and learning-based methods \cite{LEARNPCC, LEARNPCC2, LEARNPCC3, LEARNPCC4, LEARNPCC5, LEARNPCC6, PCGCV2} have been proposed to further enhance compression performance. However, these methods rely on voxelized point clouds to represent geometry models, which is inefficient and memory-intensive, limiting their compression efficiency.
In contrast to the previously mentioned methods, Draco \cite{DRACO} uses a kd-tree-based coding method to compress vertices and employs the EdgeBreaker algorithm to encode the topological relationships of the geometry data. Draco utilizes uniform quantization to control the compression ratio, but its performance decreases at higher compression ratios. 

\vspace{0.5em}
\noindent\textbf{Multiple Model Compression.}
Compared to compressing single 3D geometric models, compressing multiple objects is significantly more challenging. SLRMA \cite{SLRMA} addresses this by using a low-rank matrix to approximate vertex matrices, thus compressing sequential models. Mekuria et al. \cite{PCCCODEC} proposed the first codec for compressing 
sequential point clouds, where each frame is coded using Octree subdivision through an 8-bit occupancy code. Building on this concept, MPEG developed the VPCC standards \cite{GPCCVPCC, GPCCVPCC2, GPCCVPCC3}, which utilize 3D-to-2D projection and encode time-varying projected planes, depth maps, and other data using video codecs.
Several improved methods \cite{IMPROVEVPCC, IMPROVEVPCC2, IMPROVEVPCC3, IMPROVEVPCC4} have been proposed to enhance the compression of sequential models. Recently, shape priors like SMPL \cite{SMPL} and SMAL \cite{SMAL} have been introduced, allowing the pose and shape of a template frame to be altered using only a few parameters. Pose-driven geometry compression methods \cite{POSEDRIVE0, POSEDRIVE, POSEDRIVE2} leverage this approach to achieve high compression efficiency. However, these methods are limited to sequences of corresponding geometry data and cannot handle sets of unrelated geometry data, which is more common in practice.
\section{Proposed Method}
 \textbf{Overview.} Given a set of  $N$ 3D \textit{mesh} models containing diverse categories and structures, denoted as $\mathcal{S}=\{\mathbf{S}_i\}_{i=1}^{N}$, 
we aim to compress them into a bitstream while maintaining the quality of the decompressed models as much as possible. To this end, we propose a neural compression paradigm called NeCGS. As shown in Fig. \ref{PIPELINE}, NeCGS consists of two main modules, i.e., Regular Geometry Representation (RGR) and Compact Neural Representation (CNR). Specifically, RGR first represents each \textit{irregular} mesh model within $\mathcal{S}$ into a \textit{regular} 4D tensor, namely TSDF-Def that \textit{mplicitly} describes the geometry structure of the model, via a rendering-based optimization, thus leading to a set of 4D tensors $\mathcal{V}:=\{\mathbf{V}_i\}_{i=1}^{N}$ with $\mathbf{V}_i$ corresponding to $\mathbf{S}_i$. Then CNR further obtains a more compact neural representation of $\mathcal{V}$, where a \textit{quantization-aware} auto-decoder-based network regresses these 4D tensors to explore the local geometric similarity both within each shape and across different shapes for redundancy removal, producing embedded features for them. Finally, the embedded features, along with the network parameters, are encoded into a bitstream through a typical entropy coding method to achieve compression. 
Note that our NeCGS can adapt to the dynamic scenario well, where new 3D models are constantly added to the compressed set.

\subsection{Regular Geometry Representation} \label{SGR}
Unlike 2D images and videos, where pixels are uniformly distributed on 2D \textit{regular} girds, the \textit{irregular} characteristic of 3D mesh models %\JHDEL{the geometry models in $\mathcal{S}$} 
makes it challenging %\JHDEL{difficult} 
to compress them efficiently and effectively. 
We propose to convert each 3D mesh model to a 4D regular TSDF-Def tensor, which implicitly represents the geometry structure of the model. Such a regular representation can describe the model precisely, and its regular nature proves beneficial for compression in the subsequent stage.
%is helpfuf for the compression in the next stage.
\begin{figure}[t]
    \centering  %\vspace{-0.5cm}
    \includegraphics[width=0.32\linewidth]{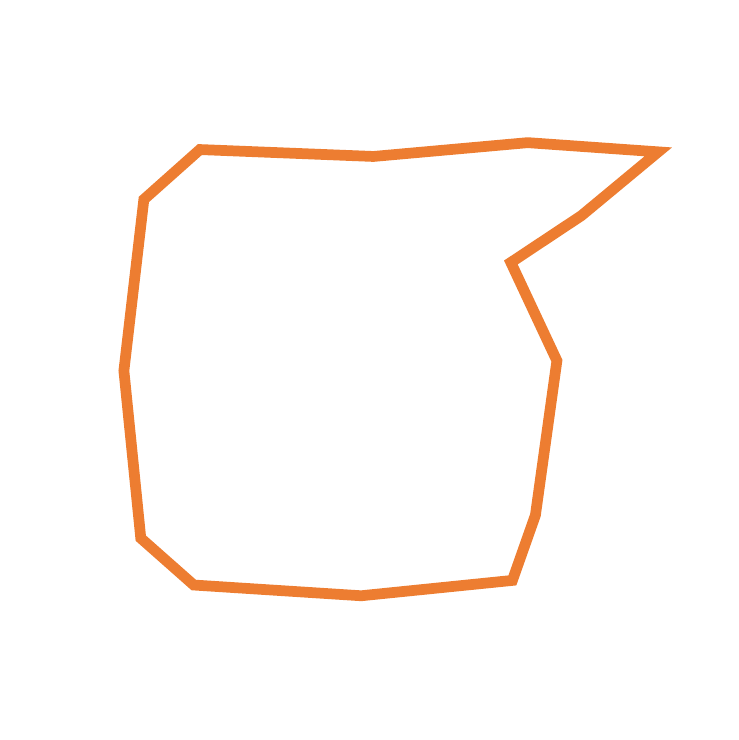} 
    \includegraphics[width=0.32\linewidth]{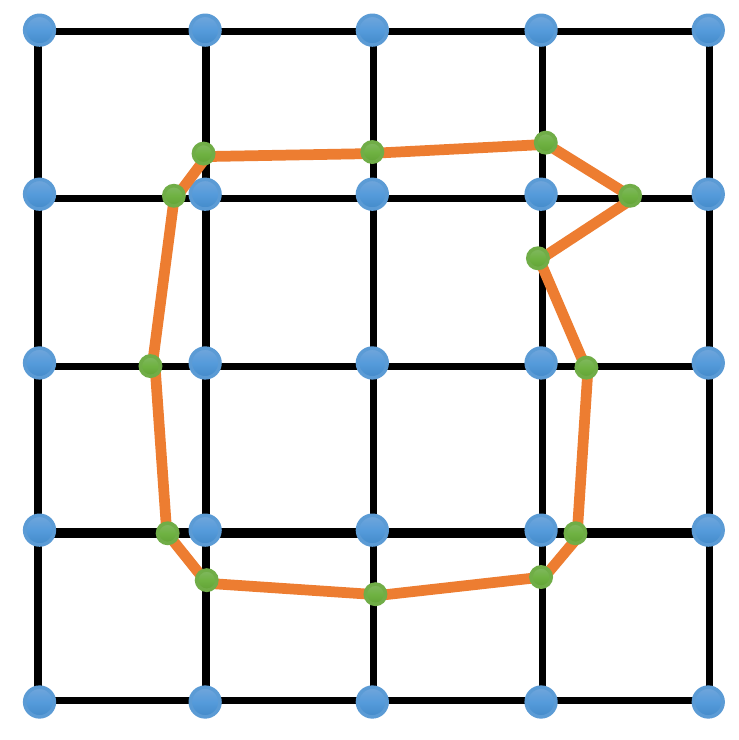} 
    \includegraphics[width=0.32\linewidth]{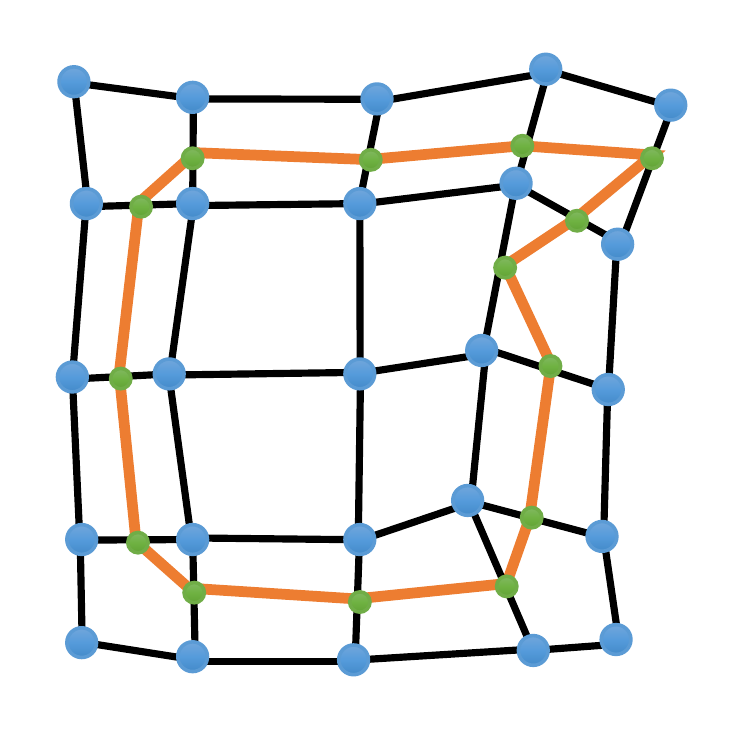}
    \vspace{-0.2cm}
    \caption{\small  \textbf{Left:} The original surface. \textbf{Middle:} Surface recovered from traditional TSDF through  MC. \textbf{Right:} Surface recovered from our TSDF-Def tensor through DMC. The \textcolor[rgb]{0.33,0.607,0.854}{blue} points refer to the deformable grid points, the \textcolor[rgb]{0.443,0.701,0.270}{green} points refer to the vertices of the extracted surfaces, and the \textcolor[rgb]{0.929,0.490,0.192}{orange} lines refer to the faces of the extracted surfaces.}
    \label{DMC} \vspace{-0.4cm}
\end{figure}

\vspace{0.5em}
\noindent\textbf{Definition of TSDF-Def.} 
Inspired by recent shape extracting methods \cite{DMTET, FLEXICUBE}, we propose TSDF-Def, which extends the traditional TSDF by introducing an additional deformation for each grid point to adjust the detailed structure during the extraction of models, as shown in Fig. \ref{DMC}.  
Specifically, each irregular 3D mesh model $\mathbf{S}$ is represented as a regular 4D tensor, denoted as $\mathbf{V}\in\mathbb{R}^{\scriptscriptstyle K\times K\times K\times 4}$, where $K$ is the resolution. The value of the grid point located at $(u,v,w)$ is corresponding to $\mathbf{V}(u,v,w): =[\mathtt{TSDF}(u,v,w), \Delta u, \Delta v, \Delta w]$, where $(\Delta u, \Delta v, \Delta w)$ are the deformation for the grid point and $1\leq u,v,w\leq K$.
Accordingly, we design differentiable \textit{Deformable Marching Cubes} (DMC), the variant of the Marching Cubes method \cite{MARCHINGCUBE}, for surface extraction from a TSDF-Def tensor.

\vspace{0.5em}
\noindent \textbf{Remark}. 
Traditional SDFs/TSDFs need to increase the tensor size/resolution to capture intricate and detailed geometric structures of 3D models. Consequently, when working with a geometry set encompassing a variety of categories, the need for SDF/TSDF tensors of varying sizes arises, sometimes requiring \textit{very large} tensors to ensure accuracy. This diversity is clearly unfavorable for the subsequent compression process. In contrast, our proposed TSDF-Def tensor introduces deformation for each grid, allowing it to represent detailed structures of models at a much \textbf{lower} and \textbf{uniform} resolution, which benefits the following compression process. See the experimental comparisons in Sec. \ref{SEC:ABLATION:STUDY}.

\vspace{0.5em}
\noindent\textbf{Optimization of TSDF-Def Tensors.}
In the context of representing a 3D mesh model as a 4D tensor through TSDF-Def, the primary goal is to minimize geometry distortion in the reconstructed surface from it. Usually, the majority of an object's regions could be precisely captured using TSDF alone, with only a few areas featuring intricate geometric details that necessitate deformation. Hence, superfluous deformations throughout the optimization process should be reduced.
Consequently, we formulate the optimization of the TSDF-Def tensor as \vspace{-0.2cm}  
\begin{equation}
     \min_{\mathbf{V}}\mathcal{E}_{\rm Rec}(\texttt{DMC}(\mathbf{V}), \mathbf{S}) + \lambda_{\rm Reg}\|\mathbf{V}[...,1:3]\|_1, \vspace{-0.2cm}
     \label{EQ:OPT}
\end{equation}
where $\texttt{DMC}(\cdot)$ refers to the differentiable DMC process for extracting surfaces from TSDF-Def tensors, $\mathcal{E}_{\rm Reg}(\cdot,~\cdot)$ measures 
the differences between the rendered depth and silhouette images of two mesh models through the differentiable rasterization \cite{DIFFRAST}, $\|\cdot\|_1$ is the $\ell_1$ norm of a tensor, and $\lambda_{\rm Reg}>0$ is a hyper-parameter to balance the two terms.

Algorithm \ref{OPT} summarizes the whole optimization process. More details can be found in Sec. \textcolor{cvprblue}{1.2} %\ref{APP:LOSS} 
of the \textit{Supplemental Material}. \vspace{-0.3cm}

\begin{algorithm}[h]
\small
  \caption{ \small  Optimization of TSDF-Def Tensor} \label{OPT}
  \KwIn{ 3D mesh model $\mathbf{S}$; the maximum number of iterations \texttt{maxIter}. }
  \KwOut{ The optimal TSDF-Def tensor $\mathbf{V}\in\mathbb{R}^{\scriptscriptstyle K\times K\times K\times 4}$.}
  Place uniformly distributed grids in the cube of $\mathbf{S}$, denoted as $\mathbf{G}\in\mathbb{R}^{\scriptscriptstyle K\times K\times K\times 3}$;\\
  Initialize $\mathbf{V}[...,0]$ as the ground truth TSDF of $\mathbf{S}$ at the location of $\mathbf{G}$,  the deformation $\mathbf{V}[...,1:3]$=0, and the current iteration $\texttt{Iter}=0$; \\
  \While {$\mathtt{{Iter}}<\mathtt{maxIter}$}{
    Recover shape from $\mathbf{V}$ according to DMC, $\texttt{DMC}(\mathbf{V})$\;
    Calculate the reconstruction error, $\mathcal{E}_{\text{Rec}}(\texttt{DMC}(\mathbf{V}),\mathbf{S})$\;
    Optimize $\mathbf{V}$ using ADAM optimizer  based on the reconstruction error\; 
    \texttt{Iter}:=\texttt{Iter}+1\;
  }
  \Return $\mathbf{V}$\;
\end{algorithm}

\subsection{Compact Neural Representation} \label{SEC:CNR}

Observing the similarity of local geometric structures within a typical 3D shape and across different shapes, i.e., redundancy, we further propose a \textit{quantization-aware} neural representation process to summarize the similarity within $\mathcal{V}$, leading to more compact representations with redundancy removed.

\noindent\textbf{
Network Architecture.}
We construct an auto-decoder network architecture to regress these 4D TSDF-Def tensors. Specifically, it comprises a head layer, which increases the channel of its input, and $L$ cascaded upsampling modules, which progressively upscale the feature tensor. We also utilize the PixelShuffle technique \cite{PIXELSHUFFLE} between the convolution and activation layers to achieve upscaling. We refer reviewers to Sec. \textcolor{cvprblue}{2.1} of \textit{Supplemental Material} for more details. For TSDF-Def tensor $\mathbf{V}_i$, the corresponding input to the auto-decoder is the embedded feature, denoted as  $\mathbf{F}_i\in\mathbb{R}^{\scriptscriptstyle K'\times K'\times K'\times C}$, where $K'$ is the resolution satisfying $K'\ll K$ and $C$ is the number of channels.
Moreover, we integrate differentiable quantization to the embedded features and network parameters in the process, which can efficiently reduce the quantization error. 
In all, the compact neural representation process can be written as\vspace{-0.15cm}  
\begin{equation}
    \widehat{\mathbf{V}}_i=\mathcal{D}_{\mathcal{Q}(\bm{{\Theta}})}(\mathcal{Q}(\mathbf{F}_{i})), \vspace{-0.15cm} 
\end{equation}
where $\mathcal{Q}(\cdot)$ stands for the differentiable quantization operator, and $\widehat{\mathbf{V}}_i$ is the regressed TSDF-Def tensor.

\vspace{0.5em}
\noindent\textbf{Loss Function.}
We employ a joint loss function comprising Mean Absolute Error (MAE) and Structural Similarity Index (SSIM) to simultaneously optimize the embedded features $\{\mathbf{F}_i\}$ and the network parameters $\bm{{\Theta}}$. In computing the MAE between the predicted and ground truth TSDF-Def tensors, we concentrate more on the grids near the surface. These surface grids crucially determine the surfaces through their TSDFs and deformations; hence we assign them higher weights during optimization than the grids farther away from the surface. 
The overall loss function for the $i$-th model is written as\vspace{-0.1cm}
\begin{equation}
\begin{aligned}
    \mathcal{L}(\widehat{\mathbf{V}}_i,\mathbf{V}_{i})=&\|\widehat{\mathbf{V}}_i-\mathbf{V}_{i}\|_1 + \lambda_1 \|\mathbf{M}_i\odot(\widehat{\mathbf{V}}_i-\mathbf{V}_{i})\|_1  \\ &+ \lambda_2(1- \texttt{SSIM}(\widehat{\mathbf{V}}_i,\mathbf{V}_{i})), \label{LOSS}
\end{aligned}
\end{equation}
where $\mathbf{M}_i=\mathbbm{1}(|\mathbf{V}_{i}[...,0])|<\tau)$ is the mask, indicating whether a grid is near the surface, i.e., its TSDF is less than the threshold $\tau$, while $\lambda_1$ and $\lambda_2$ are the weights to balance 
each term 
of the loss function.
\iffalse\begin{equation}
    \{\mathcal{F},\bm{\Theta}\}=\mathop{\arg\min}_{\mathcal{F},\bm{\Theta}}\frac{1}{T}\sum_{\mathbf{F}_t\in\mathcal{F}}\mathcal{L}\left(\mathcal{D}_{\bm{\Theta}}(\mathbf{F}_t),\mathbf{V}_t\right),
\end{equation}\fi

\vspace{0.5em}
\noindent\textbf{Entropy Coding.} 
After obtaining the quantized features $\{\widetilde{\mathbf{F}}_i=\mathcal{Q}(\mathbf{F}_i)\}$ and quantized network parameters $\widetilde{\bm{\Theta}}=\mathcal{Q}(\bm{\Theta})$, we adopt the Huffman Codec \cite{HUFFMAN} to further compress them 
into a bitstream. More advanced entropy coding methods can be employed to further improve compression performance.

\begin{figure*}[h]
    \centering \small
    \subfloat[AMA]{\includegraphics[height=0.22\textwidth]{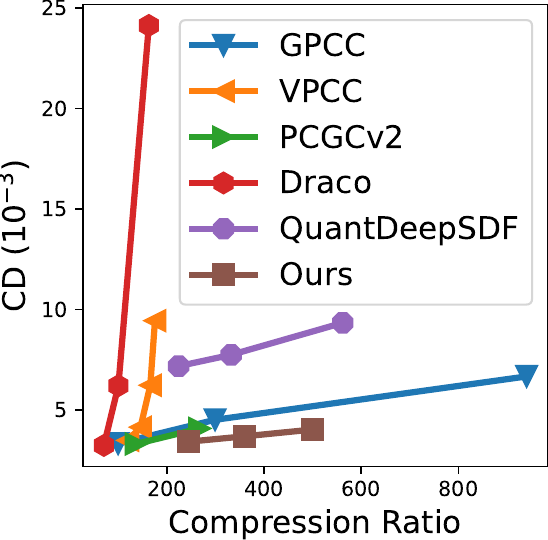}
    \includegraphics[height=0.22\textwidth]{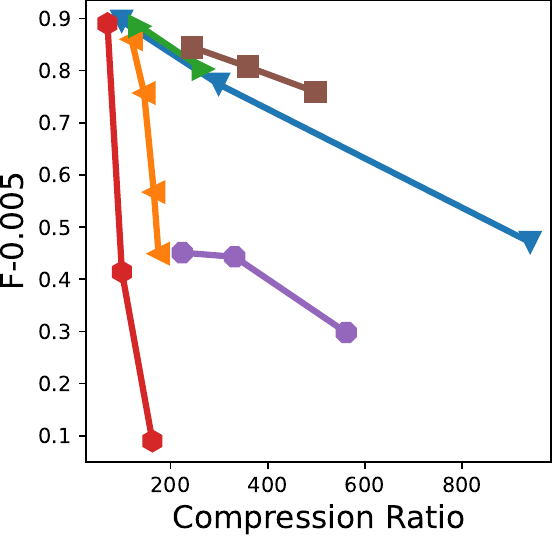}\vspace{-0.2cm}} \quad
    \subfloat[DT4D]{\includegraphics[height=0.22\textwidth]{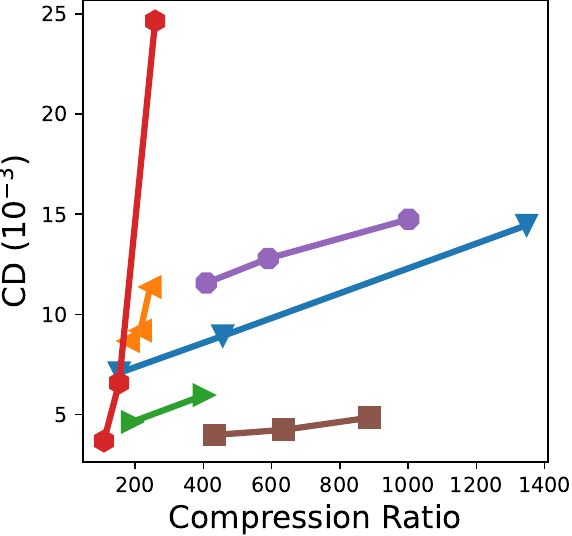}
    \includegraphics[height=0.22\textwidth]{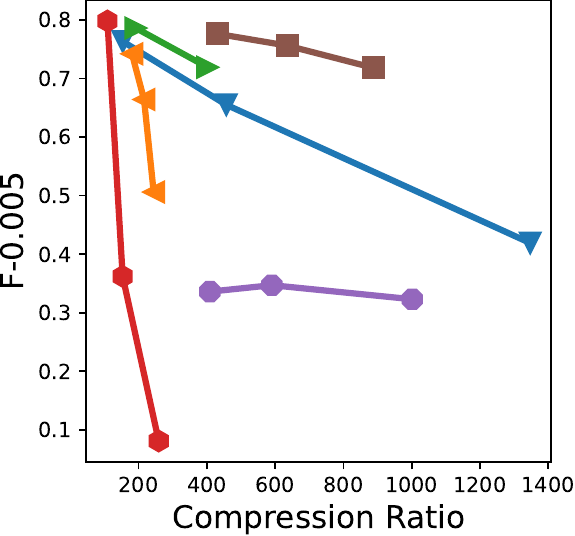}\vspace{-0.2cm}} \\
    \subfloat[Thingi10K]{\includegraphics[height=0.22\textwidth]{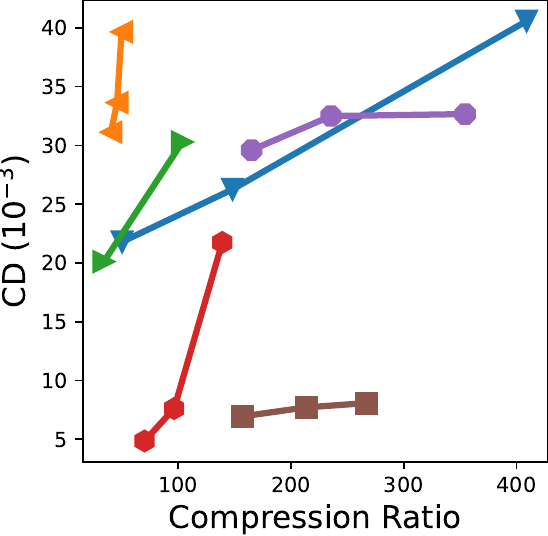}
    \includegraphics[height=0.22\textwidth]{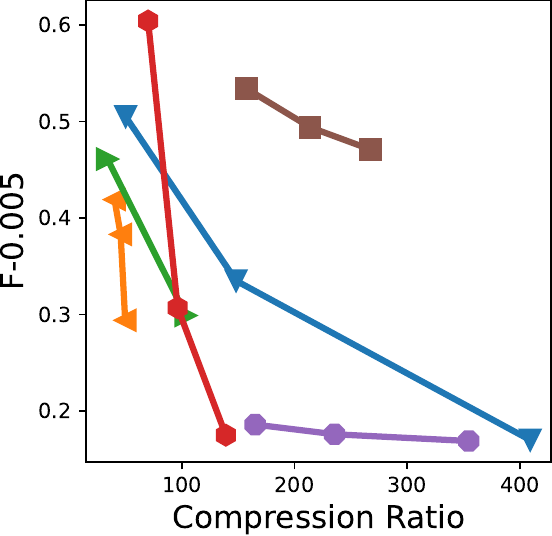}\vspace{-0.2cm}} \quad
    \subfloat[Mixed]{\includegraphics[height=0.22\textwidth]{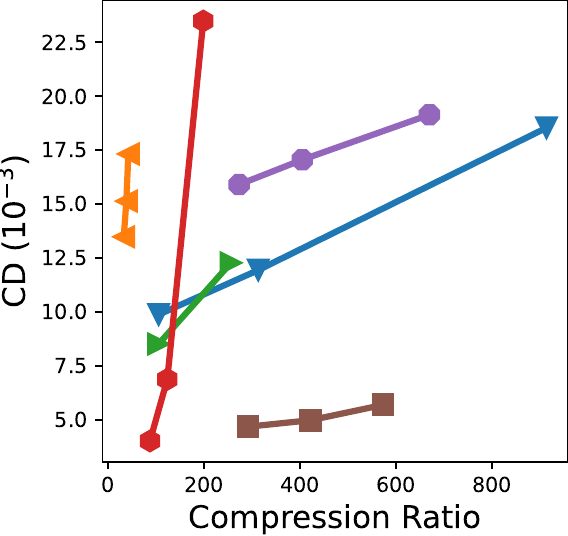} 
    \includegraphics[height=0.22\textwidth]{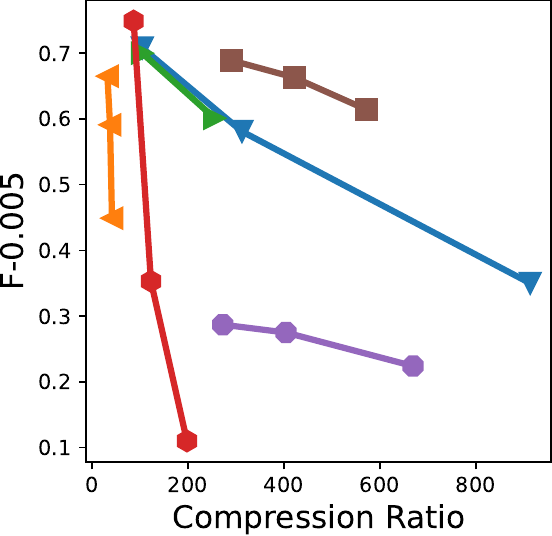}\vspace{-0.2cm}}
    \vspace{-0.2cm}
    \caption{Quantitative comparisons of different methods on four 3D geometry sets. All figures use a common legend. 
    }
    \label{RD}
\end{figure*}

\subsection{Decompression}
To obtain the 3D mesh models from the bitstream, we first decompress the bitstream to derive 
the embedded features, 
$\{\widetilde{\mathbf{F}}_i\}$ and the decoder parameter, $\widetilde{\bm{\Theta}}$. %\JHDEL{from the bitstream}. 
Then, for each $\widetilde{\mathbf{F}}_i$, we feed it to  the decoder $\mathcal{D}_{\widetilde{\bm{\Theta}}}(\cdot)$ to generate its corresponding TSDF-Def tensor, i.e., 
    $\widehat{\mathbf{V}}_i=\mathcal{D}_{\widetilde{\bm{\Theta}}}(\widetilde{\mathbf{F}}_i).$ 
Finally, we utilize DMC to recover each shape from $\widehat{\mathbf{V}}_i$, $\widehat{\mathbf{S}}_i=\texttt{DMC}(\widehat{\mathbf{V}}_i)$, forming the set of decompressed geometry data, $\widehat{\mathcal{S}}=\{\widehat{\mathbf{S}}_i\}_{i=1}^{N}$.

\begin{figure*}[h] \small
\centering
{
\begin{tikzpicture}[]
\node[] (a) at (0,9) {\includegraphics[width=0.13\textwidth]{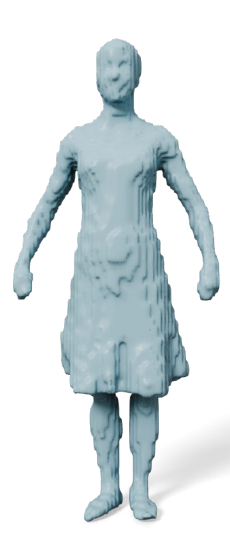}};
\node[] (a) at (0,5.3) {\includegraphics[width=0.13\textwidth]{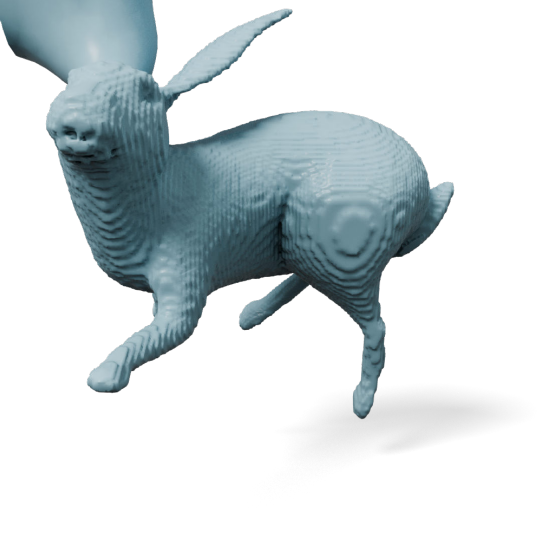} };
\node[] (a) at (0,3.5) {\includegraphics[width=0.13\textwidth]{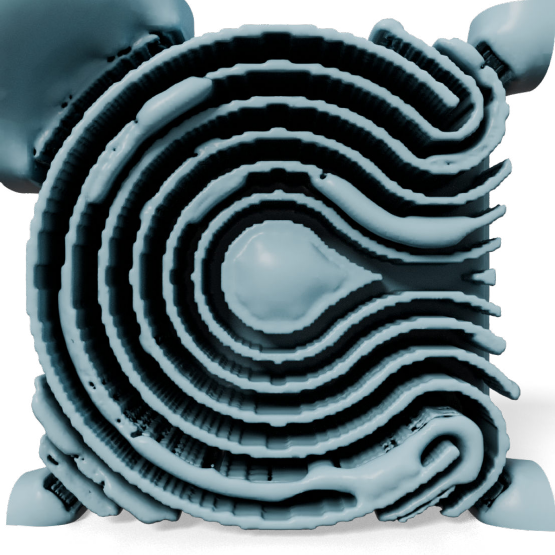} };
\node[] (a) at (0,1.2) {\includegraphics[width=0.13\textwidth]{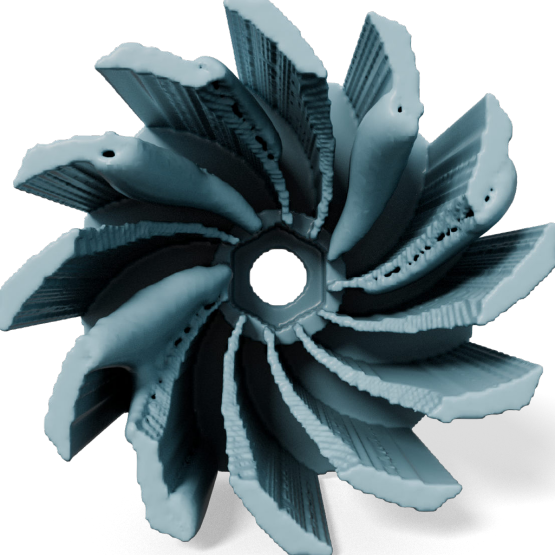} };
\node[] (a) at (0,-0.2) {\small (a) GPCC};

\node[] (a) at (2.5,9) {\includegraphics[width=0.13\textwidth]{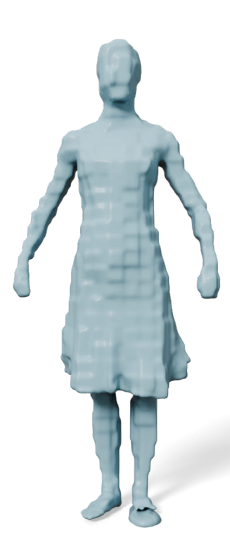}};
\node[] (a) at (2.5,5.3) {\includegraphics[width=0.13\textwidth]{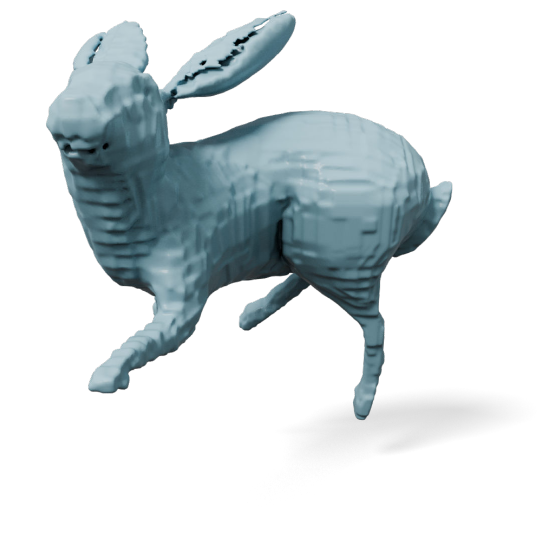} };
\node[] (a) at (2.5,3.5) {\includegraphics[width=0.13\textwidth]{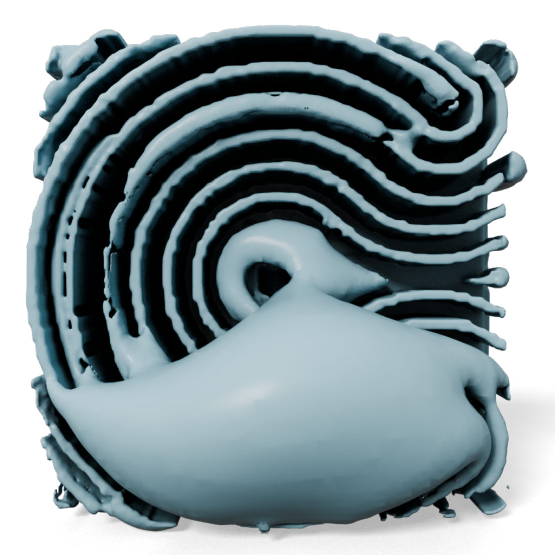} };
\node[] (a) at (2.5,1.2) {\includegraphics[width=0.13\textwidth]{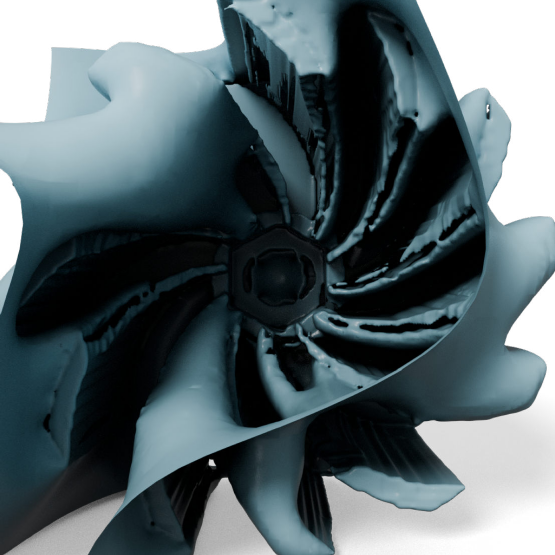} };
\node[] (a) at (2.5,-0.2) {\small (b) VPCC};

\node[] (a) at (5,9) {\includegraphics[width=0.13\textwidth]{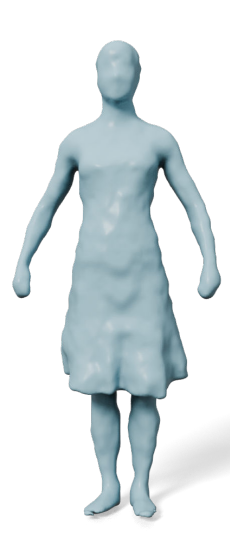}};
\node[] (a) at (5,5.3) {\includegraphics[width=0.13\textwidth]{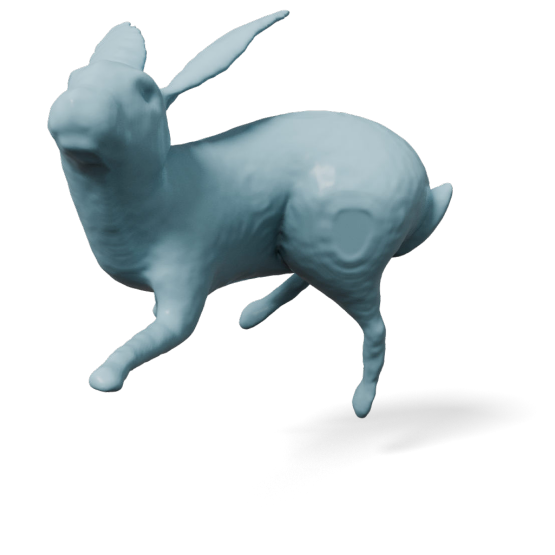} };
\node[] (a) at (5,3.5) {\includegraphics[width=0.13\textwidth]{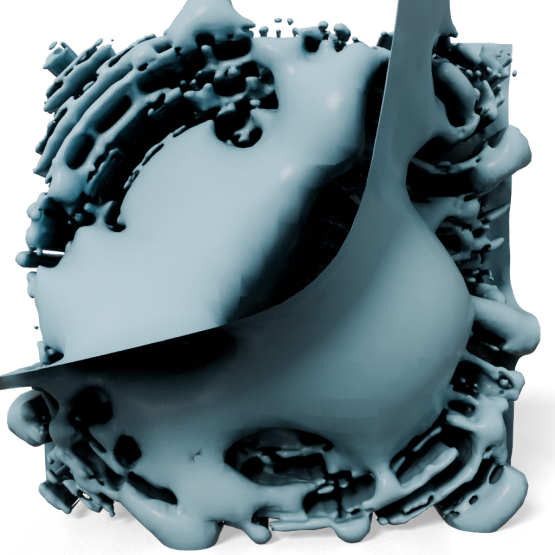} };
\node[] (a) at (5,1.2) {\includegraphics[width=0.13\textwidth]{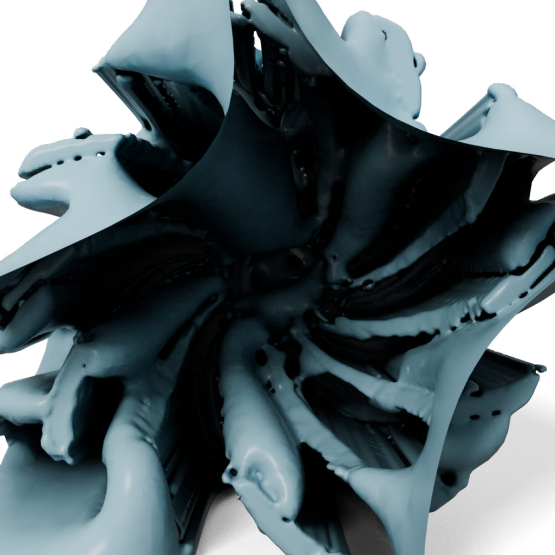} };
\node[] (a) at (5,-0.2) {\small (c) PCGCv2};

\node[] (a) at (7.5,9) {\includegraphics[width=0.13\textwidth]{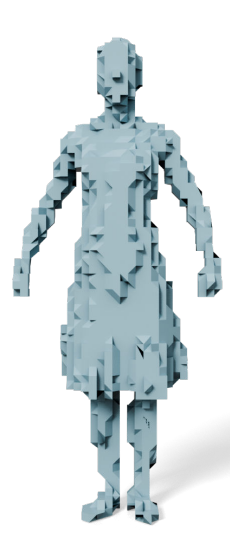}};
\node[] (a) at (7.5,5.3) {\includegraphics[width=0.13\textwidth]{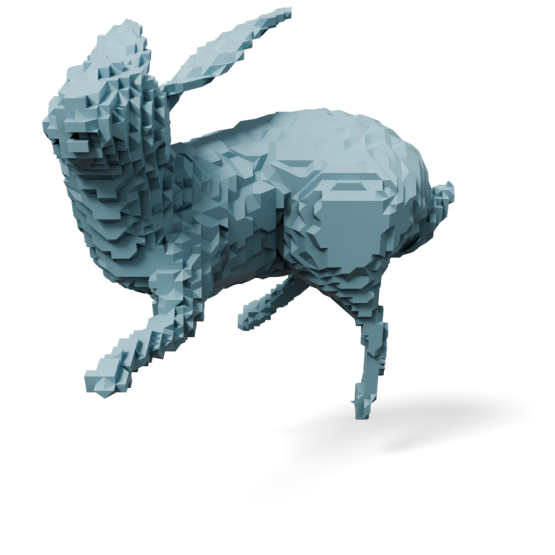} };
\node[] (a) at (7.5,3.5) {\includegraphics[width=0.13\textwidth]{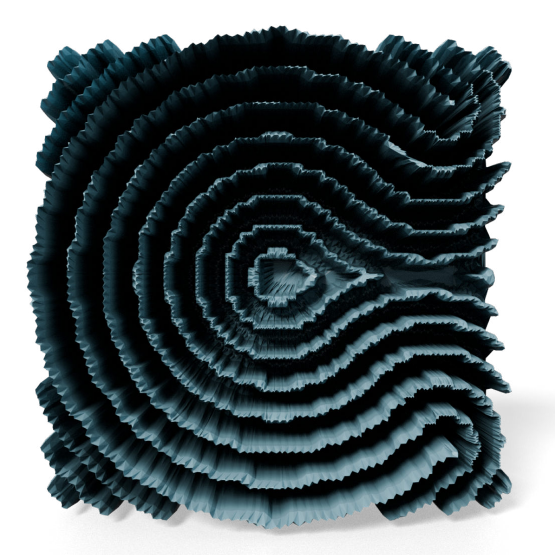} };
\node[] (a) at (7.5,1.2) {\includegraphics[width=0.13\textwidth]{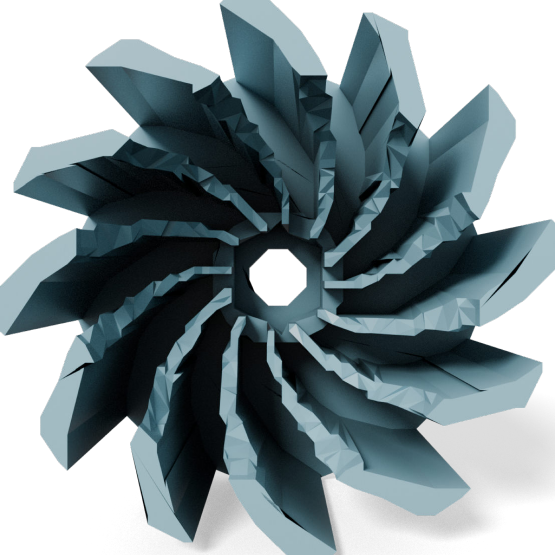} };
\node[] (a) at (7.5,-0.2) {\small (d) Draco};

\node[] (a) at (10,9) {\includegraphics[width=0.13\textwidth]{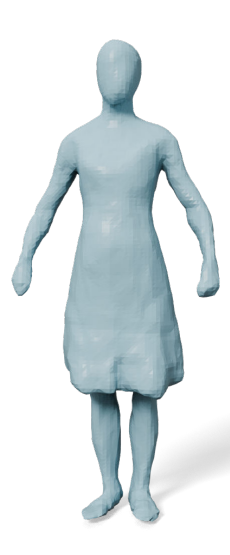}};
\node[] (a) at (10,5.3) {\includegraphics[width=0.13\textwidth]{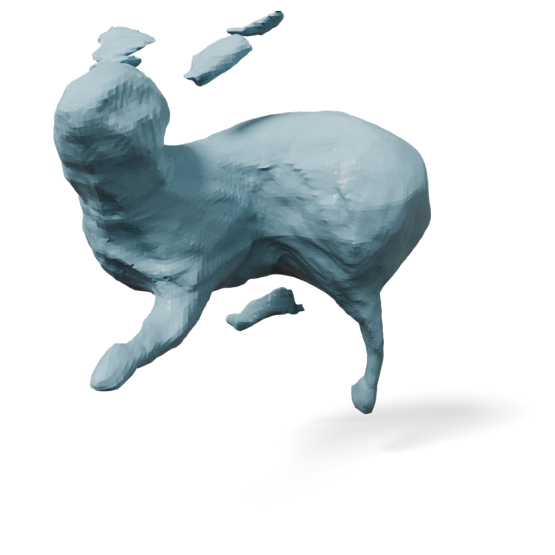} };
\node[] (a) at (10,3.5) {\includegraphics[width=0.13\textwidth]{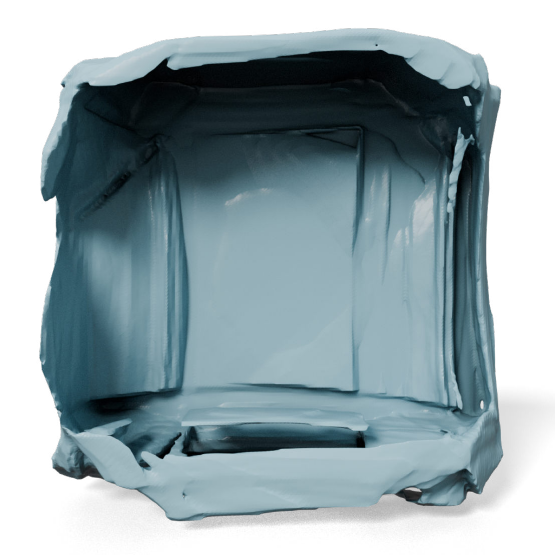} };
\node[] (a) at (10,1.2) {\includegraphics[width=0.13\textwidth]{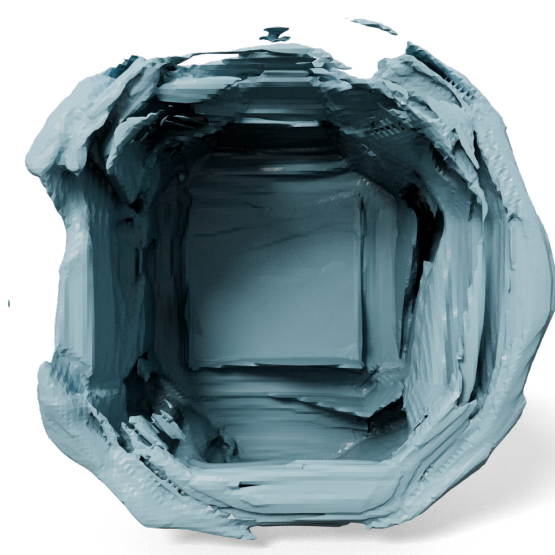} };
\node[] (a) at (10,-0.2) {\small (e) QuantDeepSDF};

\node[] (a) at (12.5,9) {\includegraphics[width=0.13\textwidth]{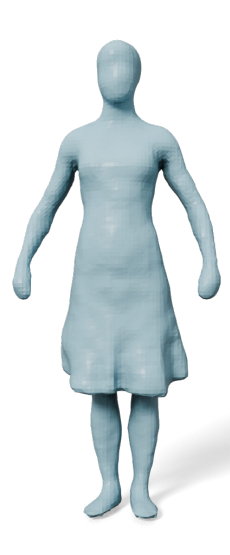}};
\node[] (a) at (12.5,5.3) {\includegraphics[width=0.13\textwidth]{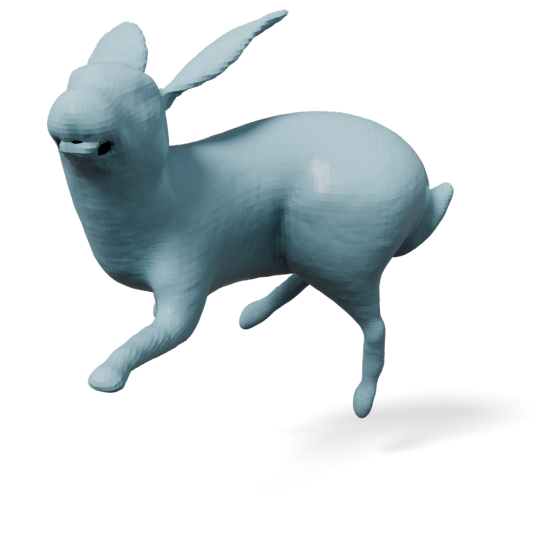} };
\node[] (a) at (12.5,3.5) {\includegraphics[width=0.13\textwidth]{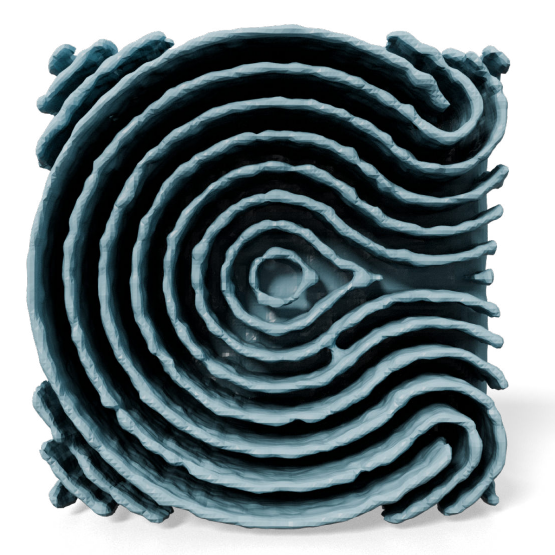} };
\node[] (a) at (12.5,1.2) {\includegraphics[width=0.13\textwidth]{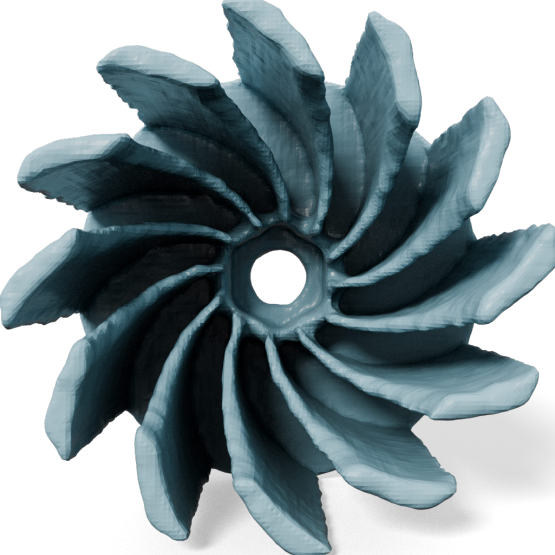} };
\node[] (a) at (12.5,-0.2 ) {\small (f) Ours};

\node[] (a) at (15,9) {\includegraphics[width=0.13\textwidth]{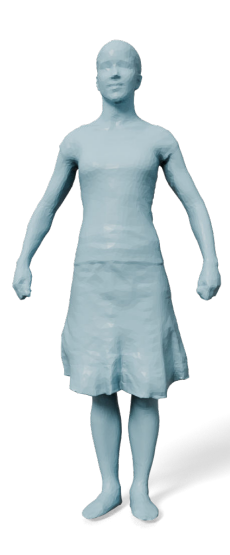}};
\node[] (a) at (15,5.3) {\includegraphics[width=0.13\textwidth]{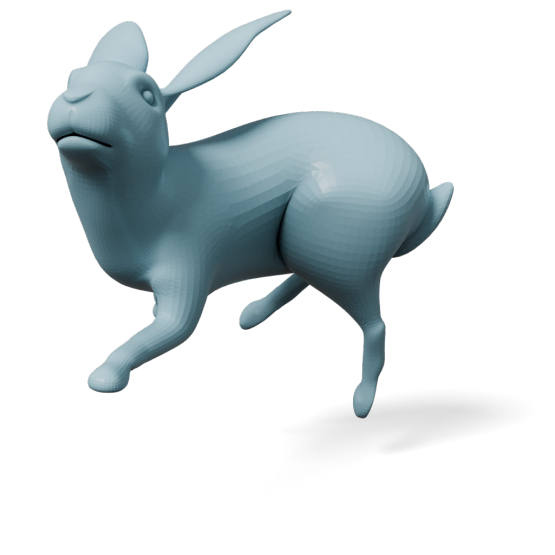} };
\node[] (a) at (15,3.5) {\includegraphics[width=0.13\textwidth]{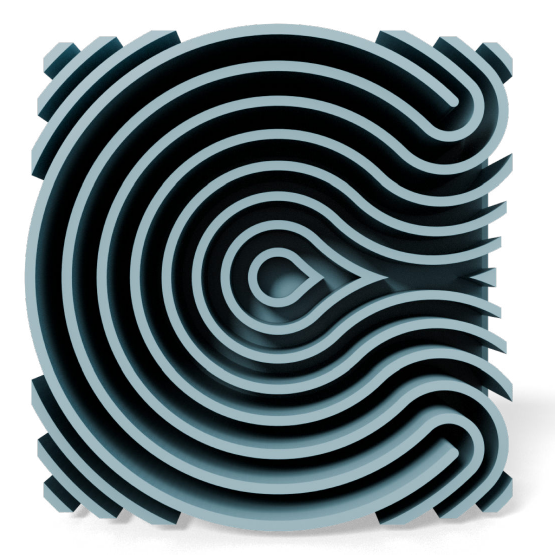} };
\node[] (a) at (15,1.2) {\includegraphics[width=0.13\textwidth]{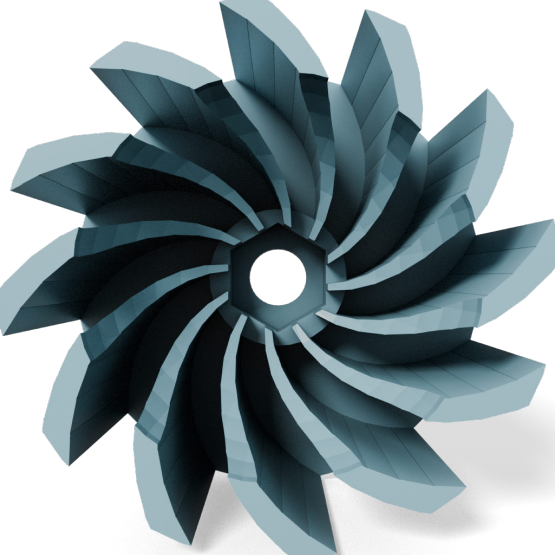} };
\node[] (a) at (15,-0.2) {\small (g) Ori.};

\node[] (a) at (-1.4,9.2) { \rotatebox{90}{\small AMA}};
\node[] (a) at (-1.4,5.3) { \rotatebox{90}{\small DT4D}};
\node[] (a) at (-1.4,3.5) { \rotatebox{90}{\small Thingi10K}};
\node[] (a) at (-1.4,1.2) { \rotatebox{90}{\small Mixed}};

\node[] (a) at (1,0.1) {\scriptsize 312.83};
\node[] (a) at (3.5,0.1) {\scriptsize 41.95};
\node[] (a) at (6,0.1) {\scriptsize 256.01};
\node[] (a) at (8.5,0.1) {\scriptsize 123.20};
\node[] (a) at (11,0.1) {\scriptsize 272.91};
\node[] (a) at (13.5,0.1) {\scriptsize 307.79};
%\node[] (a) at (12.8,0.5) {\small x.xx};

\node[] (a) at (1,2.4) {\scriptsize 148.50};
\node[] (a) at (3.5,2.4) {\scriptsize 49.81};
\node[] (a) at (6,2.4) {\scriptsize 103.90};
\node[] (a) at (8.5,2.4) {\scriptsize 96.52};
\node[] (a) at (11,2.4) {\scriptsize 165.47};
\node[] (a) at (13.5,2.4) {\scriptsize 166.79};

\node[] (a) at (1,4.8) {\scriptsize 457.39};
\node[] (a) at (3.5,4.8) {\scriptsize 244.38};
\node[] (a) at (6,4.8) {\scriptsize 402.70};
\node[] (a) at (8.5,4.8) {\scriptsize 153.45};
\node[] (a) at (11,4.8) {\scriptsize 409.21};
\node[] (a) at (13.5,4.8) {\scriptsize 455.25};

\node[] (a) at (1,6.6) {\scriptsize 299.13};
\node[] (a) at (3.5,6.6) {\scriptsize 165.75};
\node[] (a) at (6,6.6) {\scriptsize 267.42};
\node[] (a) at (8.5,6.6) {\scriptsize 99.94};
\node[] (a) at (11,6.6) {\scriptsize 224.17};
\node[] (a) at (13.5,6.6) {\scriptsize 362.80};

\end{tikzpicture}
}
\vspace{-0.7cm}
\caption{\small Visual comparisons of different compression methods. All numbers in corners represent the compression ratio. \color{blue}{\faSearch~} Zoom in for details. 
}
\label{FIG:RESULT} \vspace{-0.3cm}
\end{figure*}

\section{Experiment}
\subsection{Experimental Setting}
\textbf{Implementation Details.} In the process of optimizing TSDF-Def tensors, we employed the ADAM optimizer \cite{ADAM} for 500 iterations per shape, using a learning rate of 0.01. The resolution of TSDF-Def tensors was $K=128$, and the weight for L1 regularization of deformation was $\lambda_{\rm Reg}=10$. The resolution and the number of channels of the embedded features were $K'=4$ and  $C=16$, respectively. And the decoder is composed of $L=5$ upsampling modules with an up-scaling factor of 2. During the optimization, we set $\lambda_1=5$ and $\lambda_2=10$, and the embedded features and decoder parameters were optimized by the ADAM optimizer for 400 epochs, with a learning rate of 1e-3. We achieved different compression efficiencies by adjusting 
decoder sizes.
We conducted all experiments on an NVIDIA RTX 3090 GPU with Intel(R) Xeon(R) CPU.

\vspace{0.5em}
\noindent\textbf{Datasets.} 
We evaluated our NeCGS on a range of datasets, including models of humans, animals, and CAD designs. For human models, we randomly selected 500 shapes from the AMA dataset \cite{AMA} (378.41 MB\footnote{The original geometry data is kept as triangle meshes, so the storage size is much less than the voxelized point clouds. All data is stored in OBJ format.}). For animal models, we selected 500 shapes from the DT4D dataset \cite{DT4D} (683.80 MB). For CAD models, we chose 1000 shapes from the Thingi10K dataset \cite{THINGI10K} (335.92 MB). Additionally, we created a more challenging dataset, labeled as Mixed, by randomly selecting 200 models from each dataset, totaling 496.16 MB.
In all experiments, we scaled all models in a cube with a range of $[-1,1]^3$ to ensure they are in the same scale.

\vspace{0.5em}
\noindent\textbf{Methods under Comparison.}
In terms of traditional geometry codecs, we chose the three most impactful geometry coding standards with released codes, 
G-PCC\footnote{\href{https://github.com/MPEGGroup/mpeg-pcc-tmc13}{https://github.com/MPEGGroup/mpeg-pcc-tmc13}} and V-PCC\footnote{\href{https://github.com/MPEGGroup/mpeg-pcc-tmc2}{https://github.com/MPEGGroup/mpeg-pcc-tmc2}} from MPEG (see more details about them in \cite{GPCCVPCC, GPCCVPCC2, GPCCVPCC3}), and Draco \footnote{\href{https://github.com/google/draco}{https://github.com/google/draco}} from Google as the baseline methods. 
Additionally, we compared our approach with state-of-the-art deep learning-based compression methods, specifically PCGCv2 \cite{PCGCV2}. Furthermore, we adapted DeepSDF \cite{DEEPSDF} with quantization to serve as another baseline method, denoted as QuantDeepSDF. More details can be found in Sec. \textcolor{cvprblue}{3.2} of \textit{Supplemental Material}. It is worth noting that while some of the chosen baseline methods were originally designed for point cloud compression, we utilized voxel sampling and SPSR \cite{SPSR} to convert them between the forms of point cloud and surface.

\vspace{0.5em}
\noindent\textbf{Evaluation Metrics.}
Following previous reconstruction methods \cite{OCCNET, CONVOCCNET}, we utilize Chamfer Distance (CD), Normal Consistency (NC), F-Score with the thresholds of 0.005 and 0.01 (F1-0.005 and F1-0.01) as the evaluation metrics. Furthermore, to comprehensively compare the compression efficiency of different methods, we use Rate-Distortion (RD) curves. These curves illustrate the distortions at various compression ratios, with CD and F1-0.005 specifically describing the distortion of the decompressed models.
Our goal is to minimize distortion, indicated by a low CD and a high F1-Score, while maximizing the compression ratio. Therefore, for the RD curve representing CD, optimal compression performance is achieved when the curve is closest to the lower right corner. Similarly, for the RD curve representing the F1-Score, the ideal compression performance is when the curve is nearest to the upper right corner.
Their detailed definition can be found in Sec. \textcolor{cvprblue}{3.1} 
of \textit{Supplemental Material}.

\begin{figure}[h]
\centering
\vspace{-0.1cm}
    \subfloat[Ori.]{\includegraphics[width=0.23\linewidth]{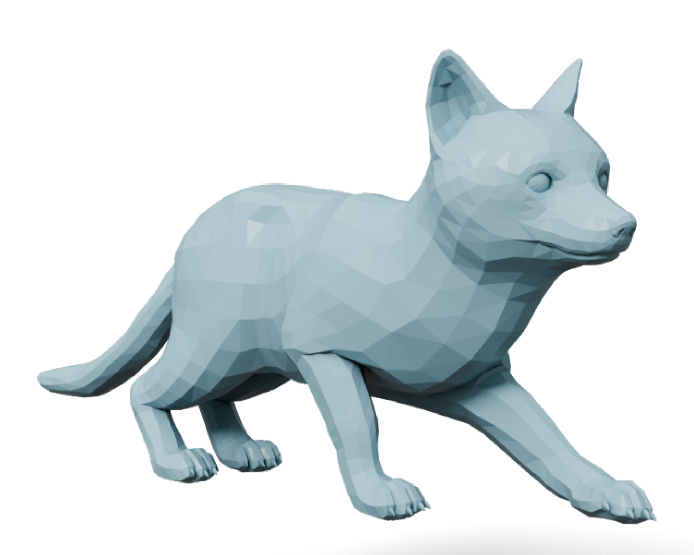}}
    \subfloat[455.25]{\includegraphics[width=0.23\linewidth]{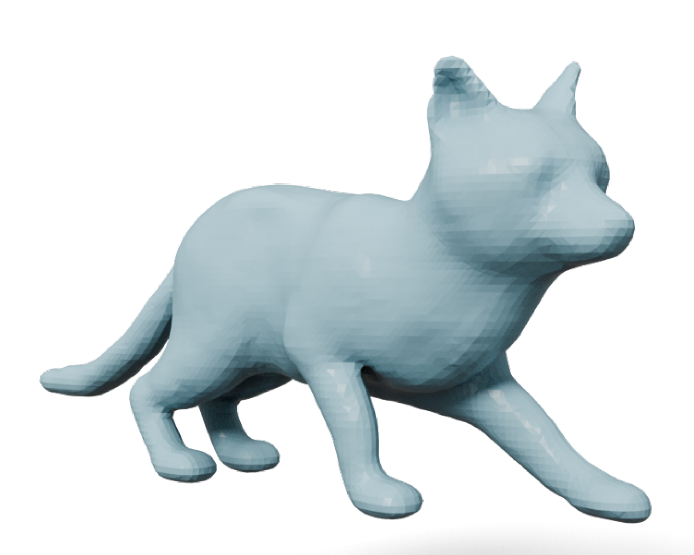}}
    \subfloat[651.85]{\includegraphics[width=0.23\linewidth]{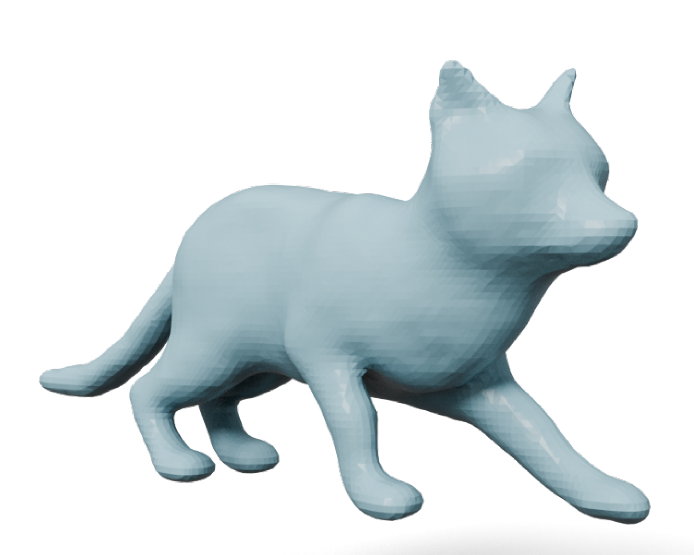}}
    \subfloat[899.73]{\includegraphics[width=0.23\linewidth]{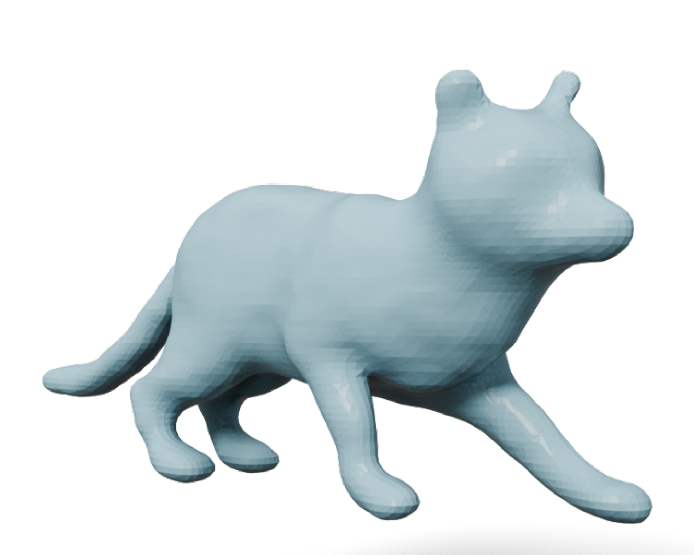}}
    \vspace{-0.3cm}
    \caption{\small Decompressed models under different compression ratios.} \label{RDFIG}
    \vspace{-0.4cm}
\end{figure}

\subsection{Results and Comparisons}
The RD curves of different compression methods under different datasets are shown in Fig. \ref{RD}. 
As the compression ratio increases, the distortion also becomes larger.
It is obvious that our NeCGS can achieve \textbf{much better} compression performance than the baseline methods when the compression ratio is high, even in the challenging Mixed dataset. In particular, our NeCGS achieves a minimum compression ratio of 300, and on the DT4D dataset, the compression ratio even reaches nearly 900, with minimal distortion. Due to the larger model differences within the Thingi10K and Mixed datasets compared to the other two datasets, the compression performance on these two datasets is inferior.

\begin{figure}[t] 
\centering
{
\subfloat[Original]{
\includegraphics[width=0.25\linewidth]{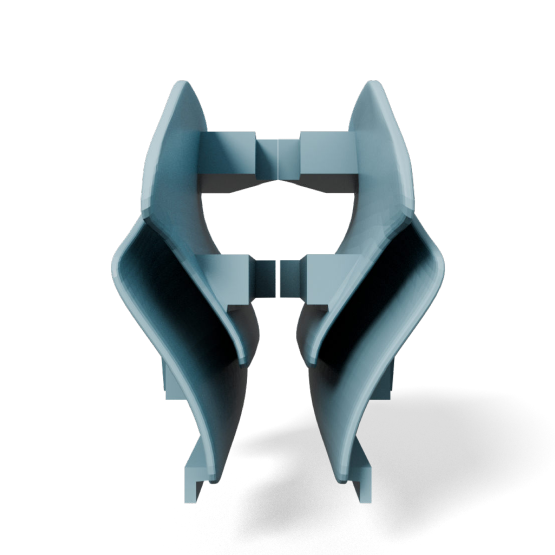}
\includegraphics[width=0.25\linewidth]{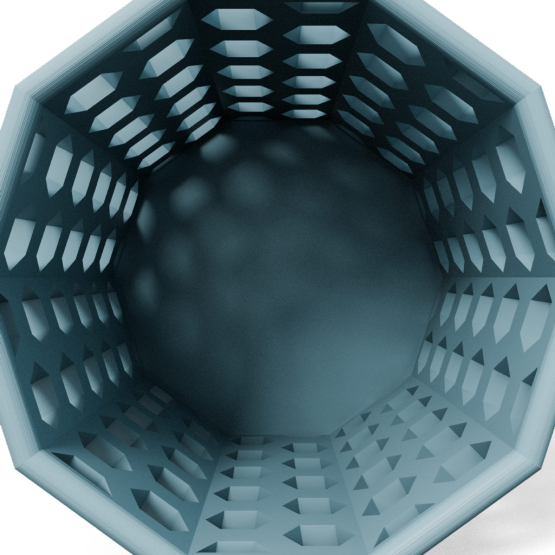} %\quad
\includegraphics[width=0.25\linewidth]{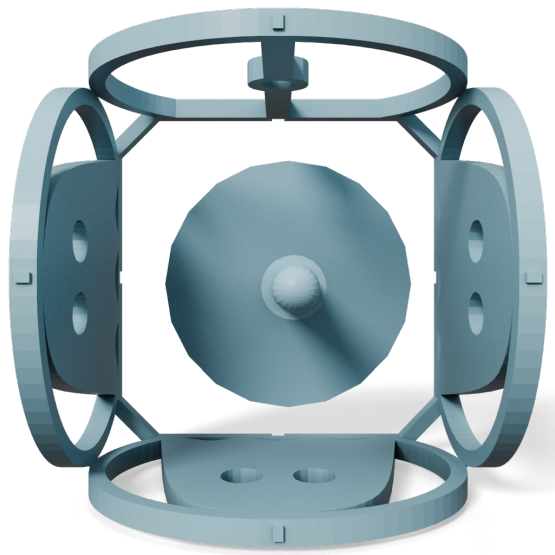}%\quad
\includegraphics[width=0.25\linewidth]{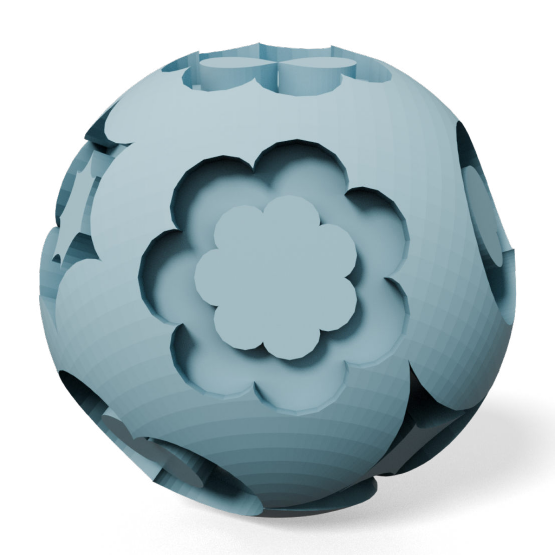}
} \\
\subfloat[Decompressed]{
\includegraphics[width=0.25\linewidth]{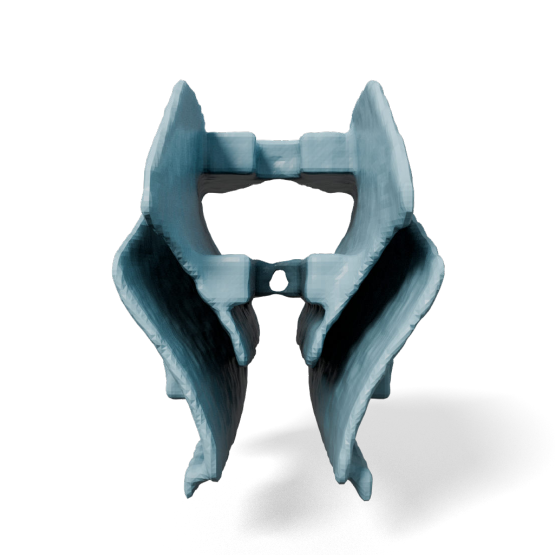}
\includegraphics[width=0.25\linewidth]{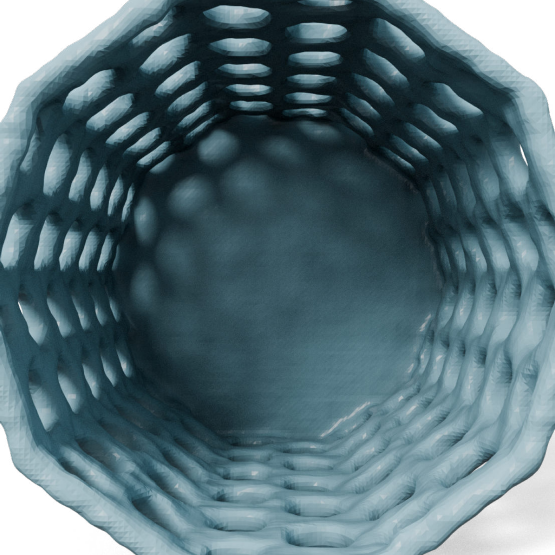}
\includegraphics[width=0.25\linewidth]{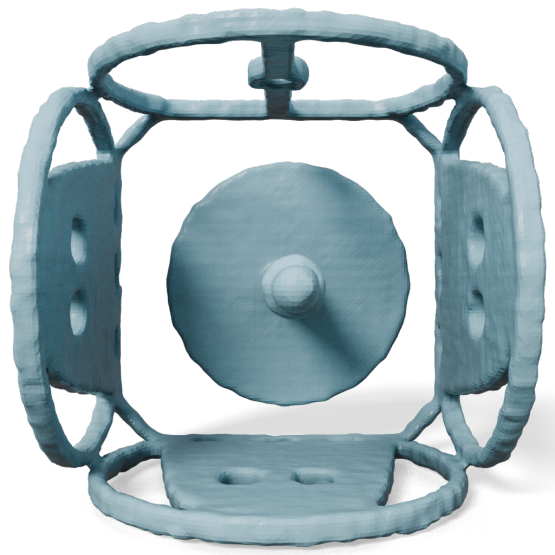}
\includegraphics[width=0.25\linewidth]{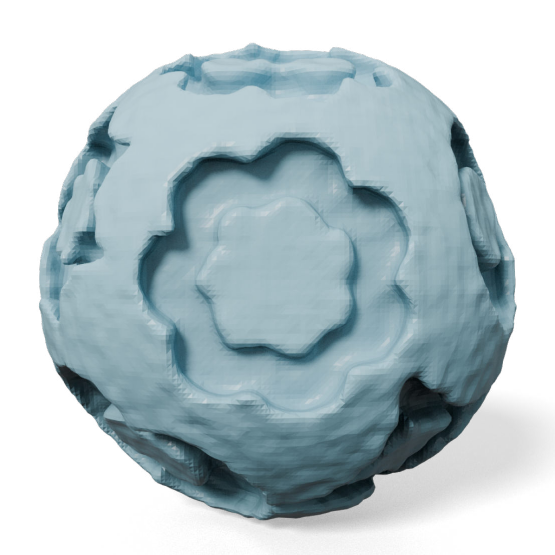}
}
}
\vspace{-0.3cm}
\caption{Visualization of complex shapes from the Thingi10K dataset. Here the compression ratio is 162.  
} \label{VIS:COMPLEX} \vspace{-0.3cm}
\end{figure}

The visual results of different compression methods are shown in Fig. \ref{FIG:RESULT}. Compared to other methods, models compressed using our approach occupy a larger compression ratio and retain more details after decompression. 
Fig. \ref{RDFIG} illustrates the decompressed models under different compression ratios. Even when the compression ratio reaches nearly 900, our method can still retain the details of the models. Fig. \ref{VIS:COMPLEX} shows the visual results on complex models.

\vspace{0.5em}
\noindent\textbf{Efficiency Comparison}. Table \ref{COMP:TIME} lists the compression time of various compression methods for a set and the decompression time for a 3D model.
Compared to the methods for single shapes, i.e., GPCC, PCGCv2, and Draco, our NeCGS consumes more time during compression. While compared with VPCC, which can compress the entire dataset at once, our NecGS is much faster. As for decompression, our NeCGS takes the \textit{least} time. \textit{It is worth noting} that our NeCGS is designed for the offline compression of geometry sets to save storage space, where decompression time assumes a crucial role as it directly impacts the efficiency of subsequent applications on the set, with compression time playing a secondary role.

\begin{table}[h]
    \centering
     \caption{\small Efficiency analysis of different methods.}
     \vspace{-0.3cm}
    \resizebox{.45\textwidth}{!}{\begin{tabular}{l|ccccc}
    \toprule
        Method & GPCC & VPCC & PCGCV2 & Draco & Ours \\
    \hline
       Comp. Time (h) & 0.62 & 39.34 & 1.76 & 0.06 & 10.01 \\
       Decomp. Time (ms) &  562.56 &   762.87 &  100.32  & 365.18   &    98.95    \\
    \bottomrule
    \end{tabular}
    \label{COMP:TIME}}
\end{table}

\subsection{Ablation Study} \label{SEC:ABLATION:STUDY}
In order to illustrate the efficiency of each design of our NeCGS, we conducted extensive ablation study about them %. \siyuadd{}
on the Mixed dataset.

\begin{figure}[h]
    \centering 
    \includegraphics[width=0.19\linewidth]{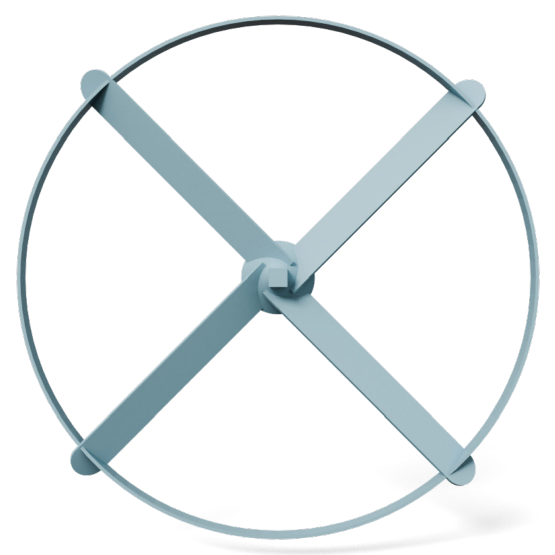}
    \includegraphics[width=0.19\linewidth]{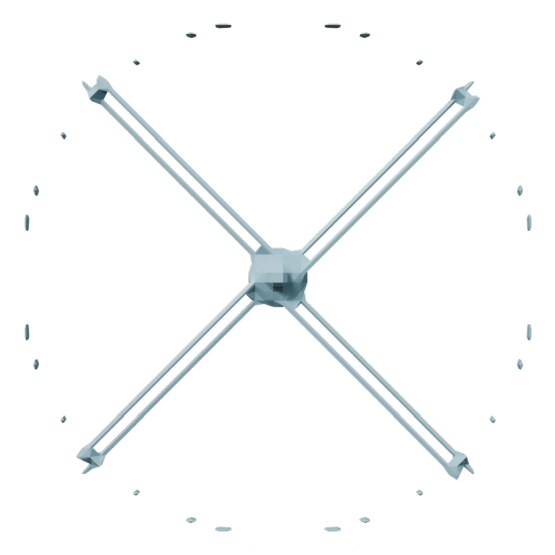}\label{TSDFDEF} 
    \includegraphics[width=0.19\linewidth]{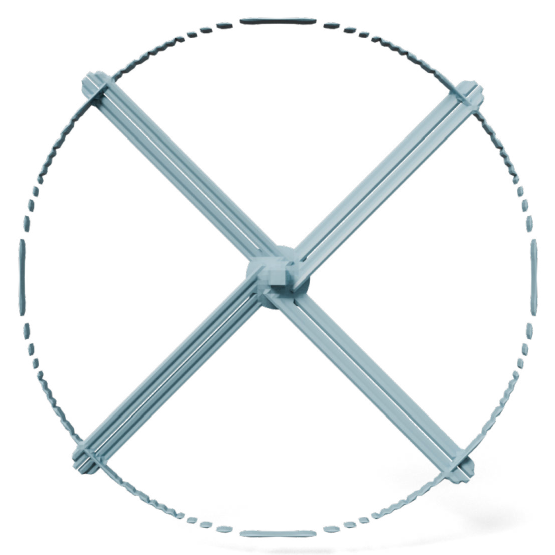}\label{TSDFDEF1}
    \includegraphics[width=0.19\linewidth]{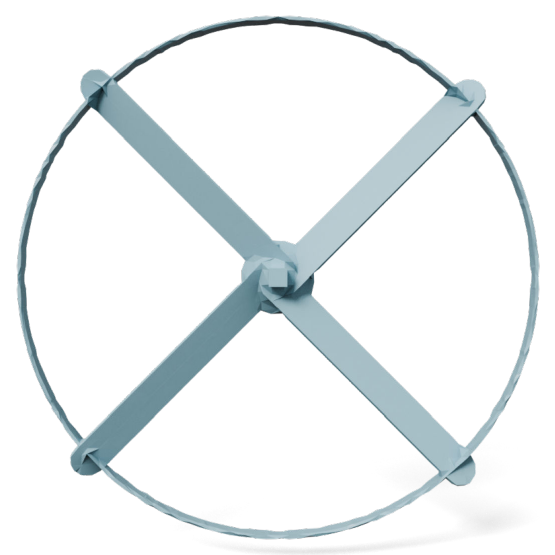} \label{TSDFDEF2}
    \includegraphics[width=0.19\linewidth]{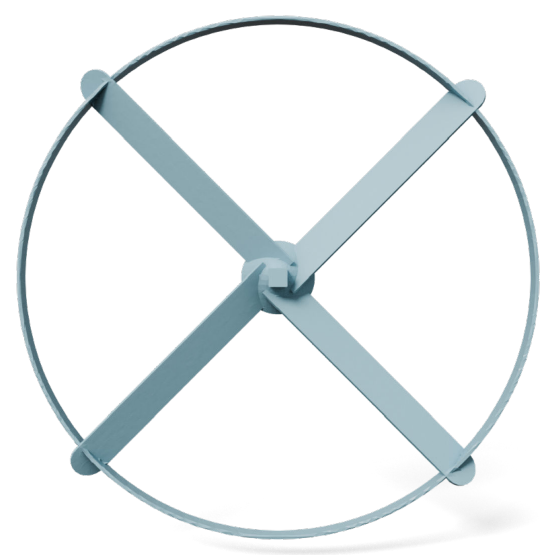}\label{TSDFDEF3}
    \vspace{-0.2cm}
    \caption{\small Models recovered from different regular geometry representations under various tensor resolutions. 
    From \textbf{Left} to \textbf{Right}: Original, TSDF with $K=64$, TSDF with $K=128$, TSDF-Def with $K=64$, and TSDF-Def with $K=128$.
    } \label{ABLATION:DEF}
    \vspace{-0.2cm} 
\end{figure}

\noindent\textbf{Comparison between TSDF and TSDF-Def}.
As shown in Fig. \ref{ABLATION:DEF}, compared with 3D models recovered from TSDF tensors through MC, the models recovered from TSDF-Def tensors through DMC preserve more details of the thin structures, especially when the tensor resolutions are relatively small. Quantitatively, as listed in Table \ref{ABLATION:DEF:TABLE}, our TSDF-Def improves the representation accuracy obviously over TSDF. In addition, the necessity and advantage of the L1 regularization in Eq. \eqref{EQ:OPT} is verified by comparing the results of TSDF-Def w/o L1 Reg. and TSDF-Def in Table \ref{ABLATION:DEF:TABLE}. 

\vspace{-0.2cm}
\begin{table}[h] \small
\caption{\small Quantitative comparisons between TSDF and TSDF-Def on 100 3D shapes with thin and fine structures from the Mixed dataset.%of different RGRs. %\siyuadd{TODO: update} 
} \vspace{-0.3cm}
\centering
\resizebox{.48\textwidth}{!}{
\begin{tabular}{l|c c c c}
\toprule
Representation  & CD ($\times 10^{-3}$) $\downarrow$ & NC $\uparrow$ & F1-0.005 $\uparrow$ & F1-0.01 $\uparrow$ \\
\hline
TSDF  & 6.630  & 0.903 & 0.621 & 0.883 \\
TSDF-Def w/o L1 Reg. & 6.205& 0.910 & 0.649 & 0.907 \\
TSDF-Def   & 5.396 &0.917  & 0.677 & 0.920 \\
\bottomrule
\end{tabular} }
\vspace{-0.3cm}
\label{ABLATION:DEF:TABLE}
\end{table}

\begin{figure}[h]
    \centering\small
    \subfloat[]{
    \includegraphics[height=0.22\linewidth]{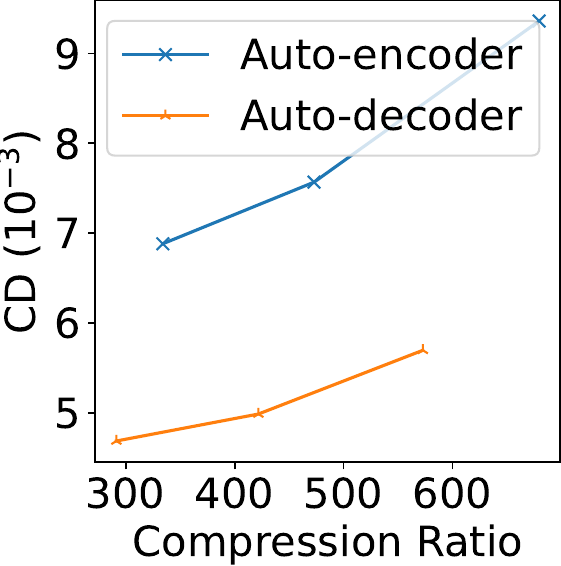} \ 
    \includegraphics[height=0.22\linewidth]{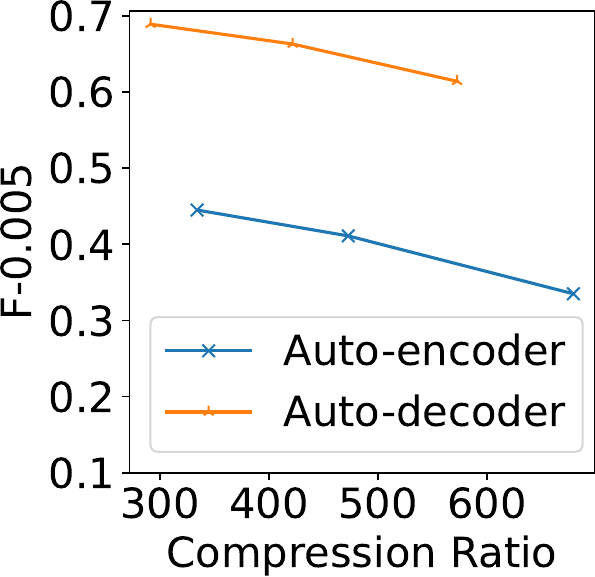} \vspace{-0.2cm} \label{ABLATION:AE:RD} }
    \subfloat[]{
    \includegraphics[height=0.22\linewidth]{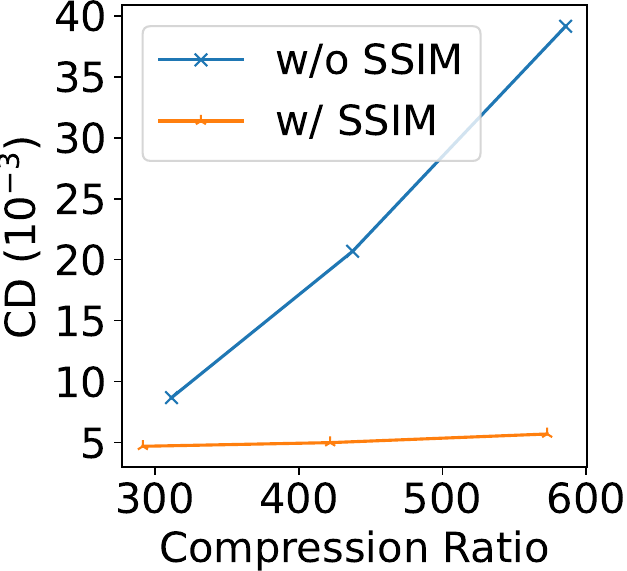} \ 
    \includegraphics[height=0.22\linewidth]{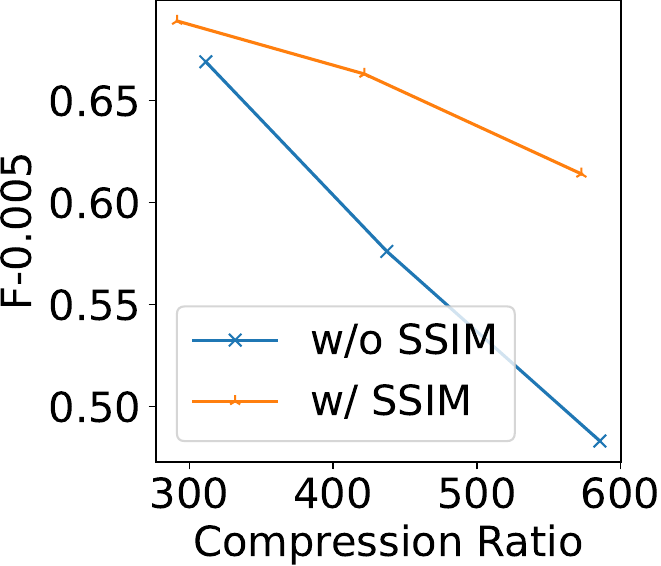} \label{ABLATION:SSIM:RD}\vspace{-0.2cm} }
    \vspace{-0.4cm}
    \caption{\small (a) RD curves of different neural representation structures. (b) RD curves of different regression losses. 
    }
\end{figure}

\noindent\textbf{Neural Representation Structure.} To illustrate the superiority of the auto-decoder framework, we utilized an auto-encoder to regress the TSDF-Def tensor. Technically, we used a ConvNeXt block \cite{CONVNEXT} as the encoder by replacing 2D convolutions with 3D convolutions. Under the auto-encoder framework, we optimize the parameters of the encoder to change the embedded features. The RD curves about these two structures are shown in Fig. \ref{ABLATION:AE:RD}, demonstrating the rationality of our decoder structure. 

\begin{figure}[h]
\centering\small 
\subfloat[\small Original]{\includegraphics[height=0.25\linewidth]{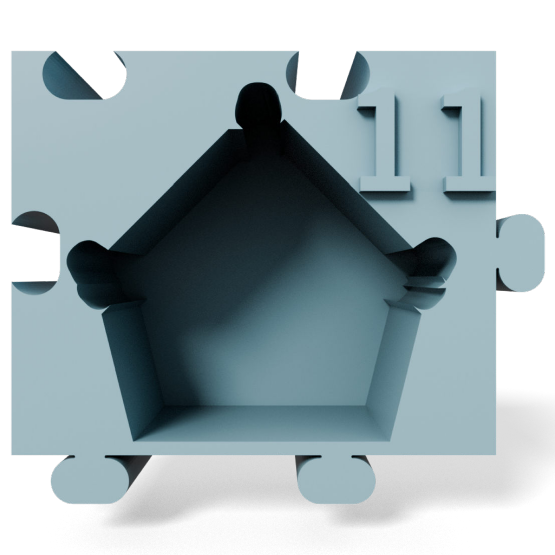}} \quad\quad
\subfloat[\small w/o SSIM]{\includegraphics[height=0.25\linewidth]{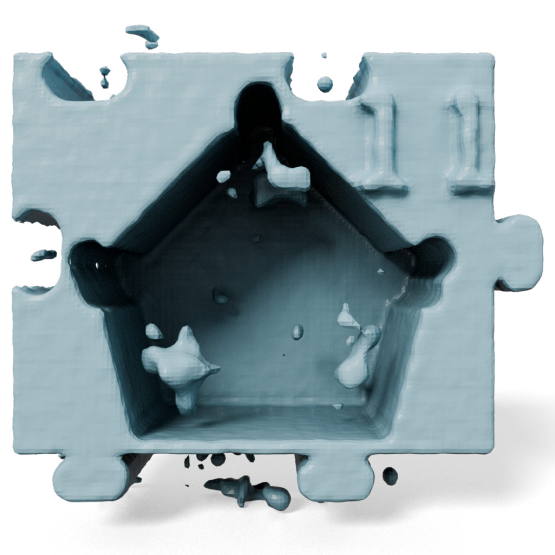}} \quad\quad
\subfloat[\small w/ SSIM]{\includegraphics[height=0.25\linewidth]{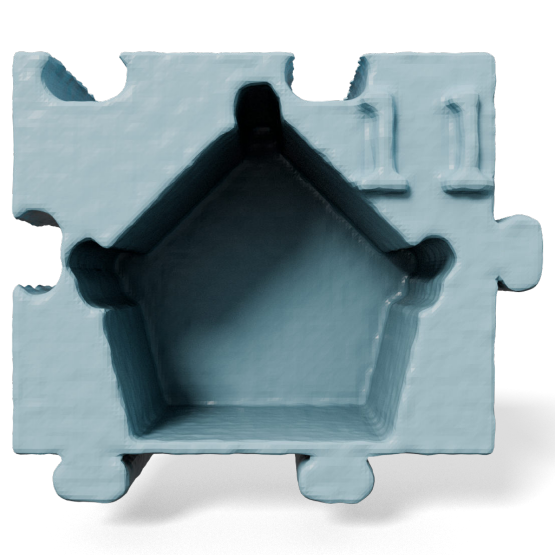}}   
\vspace{-0.3cm}
\caption{\small Visual comparison of regression loss w/ and w/o SSIM item.} \label{ABLATION:SSIM:FIG}
\end{figure}

\noindent\textbf{SSIM Loss.} Compared to MAE, which focuses on one-to-one errors between predicted and ground truth TSDF-Def tensors, the SSIM item in Eq. \ref{LOSS} emphasizes more on the local similarity between tensors, increasing the regression accuracy. To verify this, we removed the SSIM item and kept the others unchanged. Their RD curves are shown in Fig. \ref{ABLATION:SSIM:RD}, and it is obvious that the SSIM item in the regression loss increases the compression performance. The visual comparison is shown in Fig. \ref{ABLATION:SSIM:FIG}, and without SSIM, there are floating parts around the decompressed models.

\begin{figure}
\centering\small \vspace{-0.5cm}
\subfloat[\small Ori.]{\includegraphics[width=0.15\linewidth]{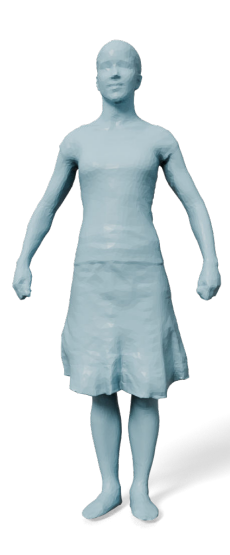}}  \quad\
\subfloat[\small 32]{\includegraphics[width=0.15\linewidth]{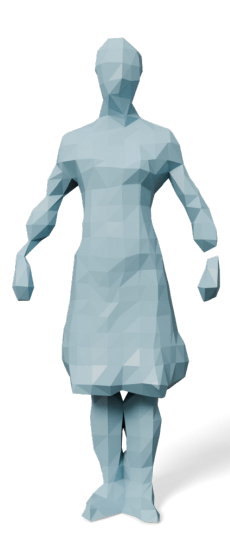}} \quad\
\subfloat[\small 64]{\includegraphics[width=0.15\linewidth]{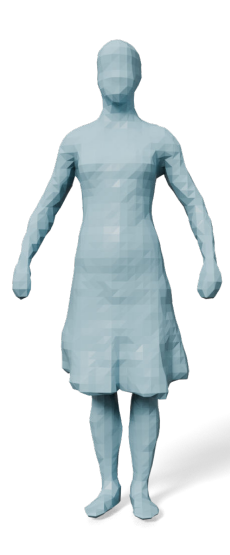}} \quad\
\subfloat[\small 128]{\includegraphics[width=0.15\linewidth]{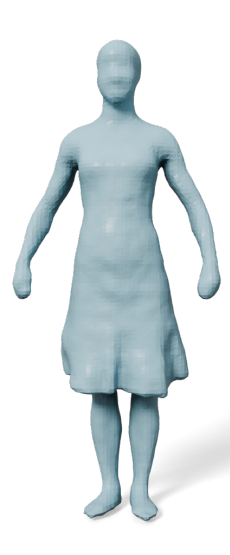}} \quad\
\subfloat[\small 256]{\includegraphics[width=0.15\linewidth]{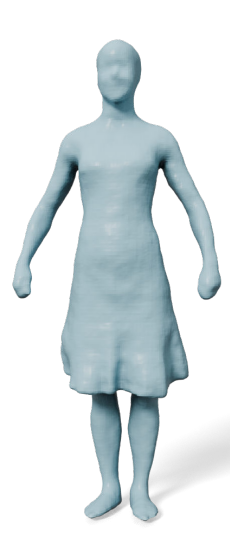}}
\vspace{-0.3cm}
\caption{\small Visual comparison under different resolutions of TSDF-Def tensors.} \label{ABLATION:RES:FIG}
\vspace{-0.4cm}
\end{figure}

\vspace{0.5em}
\noindent\textbf{Resolution of TSDF-Def Tensors.} 
We tested the compression performance at different resolutions of TSDF-Def tensors on the AMA dataset, where the decoder layers are adjusted accordingly. Specifically, we removed the last two layers for a resolution of 32, removed the last layer for a resolution of 64 and added an extra layer for a resolution of 256.
The quantitative and numerical comparisons are shown in Table \ref{ABLATION:RES:TABLE} and Fig. \ref{ABLATION:RES:FIG}, respectively.
Obviously, increasing the resolution can enhance the compression effectiveness, resulting in more detailed structures preserved after decompression.
However, the inference time also increases accordingly due to more layers involved, which directly affects the decompression efficiency.

\begin{table}[h] 
\vspace{-0.3cm}
\caption{\small Quantitative comparisons of different resolutions of TSDF-Def tensors. 
} \vspace{-0.3cm}
\centering \small
\resizebox{0.48\textwidth}{!}{
\begin{tabular}{l|cc|c c c c|c} %
\toprule
Res. & Size (MB)& Com. Ratio& CD ($\times 10^{-3}$) $\downarrow$ & NC $\uparrow$ & F1-0.005 $\uparrow$ & F1-0.01 $\uparrow$    & Infer. Time (ms)\\     
\hline
    32 & 1.037 & 364.89 & 16.982 & 0.872 & 0.215 & 0.542    & 15.30 \\
    64 & 1.408 & 268.75 & 4.271 &  0.927 & 0.721 & 0.966    & 38.97\\     
    128 & 1.493 & 253.45 & 3.436 & 0.952 & 0.842 & 0.991   & 98.95 \\     
    256 & 1.627 & 232.58 & 3.234 & 0.962 & 0.870 & 0.995    & 421.94\\      
\bottomrule
\end{tabular} }
\vspace{-0.4cm}
\label{ABLATION:RES:TABLE}
\end{table}

\begin{figure}[h]
    \centering
    \subfloat[Original]{\includegraphics[width=0.25\linewidth]{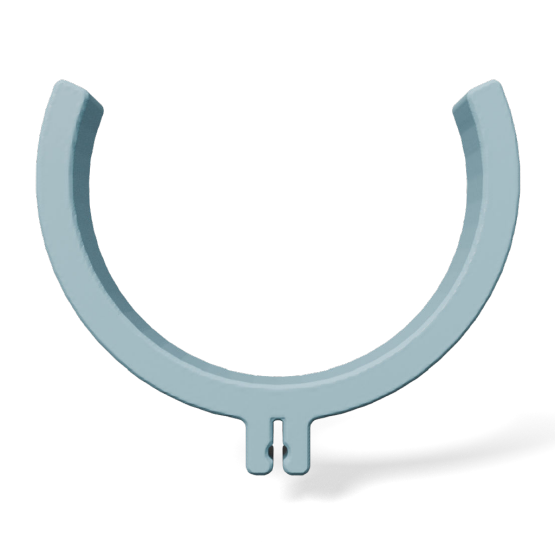}
    \includegraphics[width=0.25\linewidth]{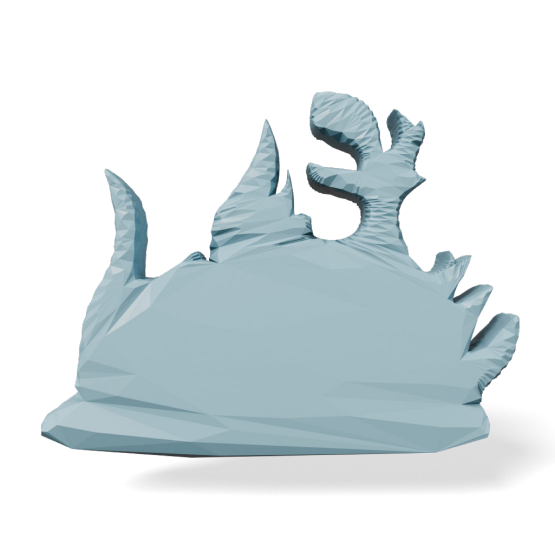}
    \includegraphics[width=0.25\linewidth]{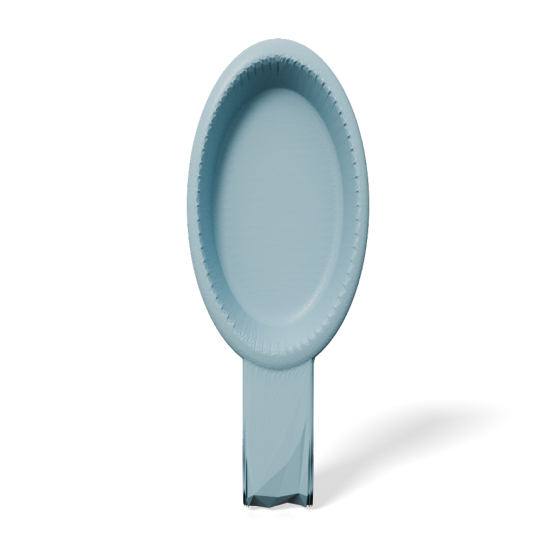}
    \includegraphics[width=0.25\linewidth]{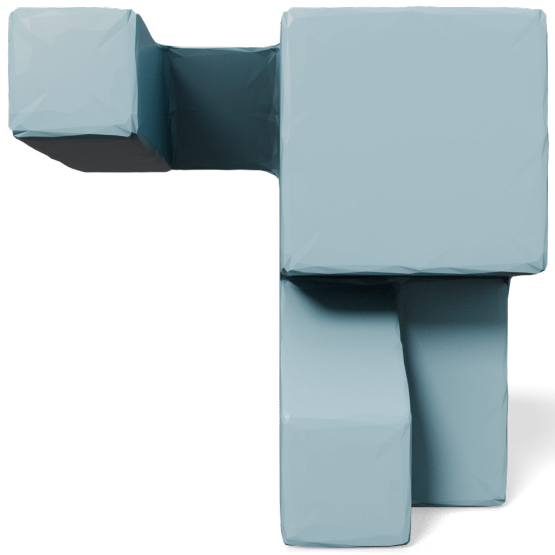}\vspace{-0.2cm}} \\
    \subfloat[Decompressed]{\includegraphics[width=0.25\linewidth]{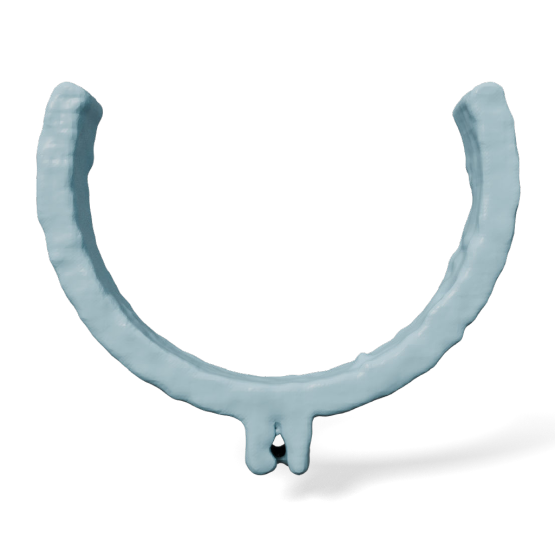}
    \includegraphics[width=0.25\linewidth]{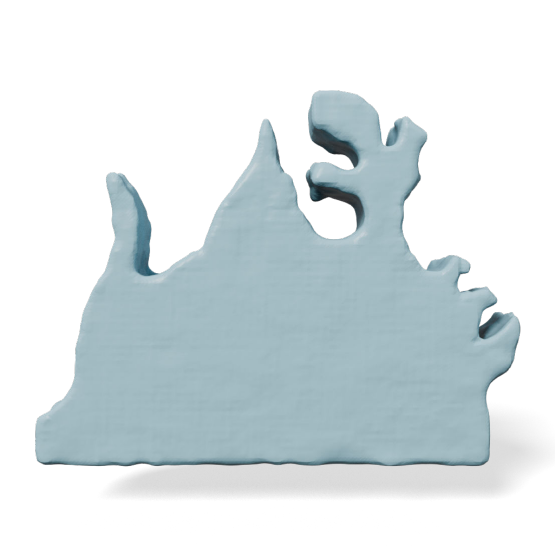}
    \includegraphics[width=0.25\linewidth]{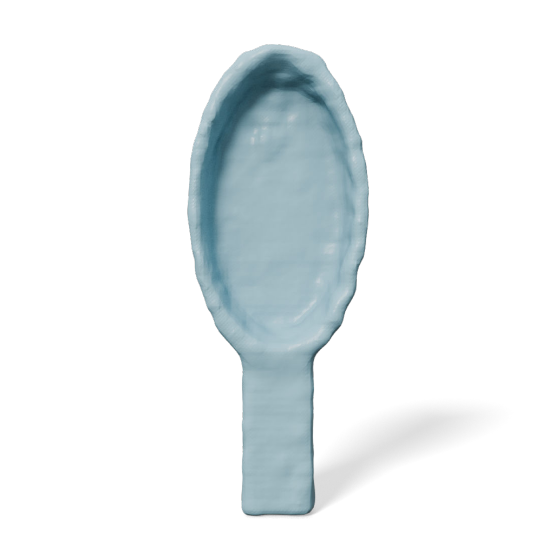}
    \includegraphics[width=0.25\linewidth]{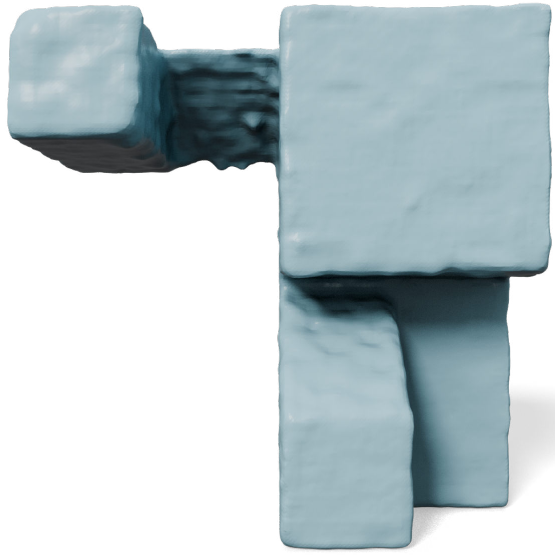}\vspace{-0.2cm}}
    \vspace{-0.3cm}
    \caption{Visual demonstration of the effectiveness of our NeCGS under the dynamic scenario, where new models were added to the Thingi10K dataset.}
    \label{THINGI:EXTRA}
\end{figure}

\noindent\textbf{Effectiveness under Dynamic Scenarios}. Our NeCGS can effectively handle a dynamic scenario, where new 3D models are added to the compressed set, i.e., the trained auto-decoder with existing 3D models contained in the set can be generalized to new 3D models. Specifically, given a new 3D model, we first represent it as a TSDF-Def tensor using Algorithm \ref{OPT}. Then, following the optimization process in Sec. \ref{SEC:CNR}, we optimize solely the corresponding embedded feature while keeping the network parameters of the trained auto-decoder unchanged. The resulting features of the newly added 3D models are entropy-coded into the bitstream. As illustrated in Fig. \ref{THINGI:EXTRA}, obviously, the newly-added 3D models can be well processed, demonstrating our assertion

\section{Conclusion}
We have presented NeCGS, a highly effective neural compression scheme for 3D geometry sets.  NeCGS has achieved remarkable compression performance on various datasets with diverse and detailed shapes, outperforming state-of-the-art compression methods to a large extent. These advantages are attributed to our regular geometry representation and the compression accomplished by a convolution-based auto-decoder. We believe our NeCGS framework will inspire further advancements in the field of geometry compression.

%\input{sec/3_finalcopy}
%\clearpage
{
    \small
    \bibliographystyle{ieeenat_fullname}
    \bibliography{main}
}


\end{document}

%\maketitle
%\input{sec/0_abstract}    
%\input{sec/1_intro_relat}
%\input{sec/2_method_exp}
%\input{sec/3_finalcopy}
%\clearpage
\clearpage
\setcounter{page}{1}
\setcounter{equation}{3}
\setcounter{figure}{12}
\maketitle

In this supplementary material, we provide more details about the regular geometry representation of our NeCGS in Sec. \ref{supp:rgr}, the auto-decoder neural network used in NeCGS in Sec. \ref{supp:decoder}, and additional experimental settings and results in Sec. \ref{supp:exp}.

\section{Details of Regular Geometry Representation} \label{supp:rgr}
\subsection{Tensor Quantization}
Denoted by $\mathbf{x}$ a tensor. We quantize it in a fixed interval, $[a,b]$,  at $(2^N+1)$ levels\footnote{We partition the interval $[a,b]$ into $(2^N+1)$ levels, rather than $2^N$ levels, to ensure the inclusion of the value 0.} by
\begin{equation}
    \mathcal{Q}(\mathbf{x})=\texttt{Round}\left(\frac{\texttt{Clamp}(\mathbf{x},a,b)-a}{s}\right)\times s+a,
\end{equation}
where $s=(b-a)/2^N$. In our experiment, we set $a=-1$ and $b=1$. 

\subsection{Optimization of TSDF-Def Tensor} \label{APP:LOSS}
We set a series of camera pose, $\mathcal{T}=\{\mathbf{T}_i\}_{i=1}^{E}$, around the meshes.
Let $\mathbf{I}^{\scriptscriptstyle\rm{D}}_{1}(\mathbf{T}_i)$ and $\mathbf{I}^{\scriptscriptstyle\rm{D}}_2(\mathbf{T}_i)$ represent the depth images obtained from the reconstructed mesh $\texttt{DMC}(\mathbf{V})$ and the given mesh $\mathbf{S}$ at the pose $\mathbf{T}_i$ respectively. Similarly, let $\mathbf{I}^{\scriptscriptstyle\rm{M}}_{1}(\mathbf{T}_i)$ and $\mathbf{I}^{\scriptscriptstyle\rm{M}}_2(\mathbf{T}_i)$ denote their respective silhouette images at pose $\mathbf{T}_i$. The reconstruction error produced by silhouette and depth images at all pose are 
\begin{equation}
    \mathcal{E}_{\rm M}(\texttt{DMC}(\mathbf{V}),\mathbf{S})=\sum_{\mathcal{T}_i\in\mathcal{T}}\|\mathbf{I}^{\scriptscriptstyle\rm{M}}_{1}(\mathbf{T}_i)-\mathbf{I}^{\scriptscriptstyle\rm{M}}_{2}(\mathbf{T}_i)\|_1
\end{equation}
and 
\begin{equation}
    \mathcal{E}_{\rm D}(\texttt{DMC}(\mathbf{V}),\mathbf{S})=\sum_{\mathcal{T}_i\in\mathcal{T}}\|\left(\mathbf{I}^{\scriptscriptstyle\rm{D}}_{1}(\mathbf{T}_i)-\mathbf{I}^{\scriptscriptstyle\rm{D}}_{2}(\mathbf{T}_i)\right)*\mathbf{I}^{\scriptscriptstyle\rm{M}}_{2}(\mathbf{T}_i)\|_1.
\end{equation}
Then the reconstruction error is defined as 
\begin{equation}
    \mathcal{E}_{\rm Rec}(\texttt{DMC}(\mathbf{V}),\mathbf{S})=\mathcal{E}_{\rm M}(\texttt{DMC}(\mathbf{V}),\mathbf{S})+\lambda_{\rm rec}\mathcal{E}_{\rm D}(\texttt{DMC}(\mathbf{V}),\mathbf{S}),
\end{equation}
where $E=4$ and $\lambda_{\rm rec}=10$ in our experiment.

\section{Auto-decoder-based Neural Compression} \label{supp:decoder}
\subsection{Upsampling Module}
\begin{figure}[h]
    \centering
    \includegraphics[width=0.6\linewidth]{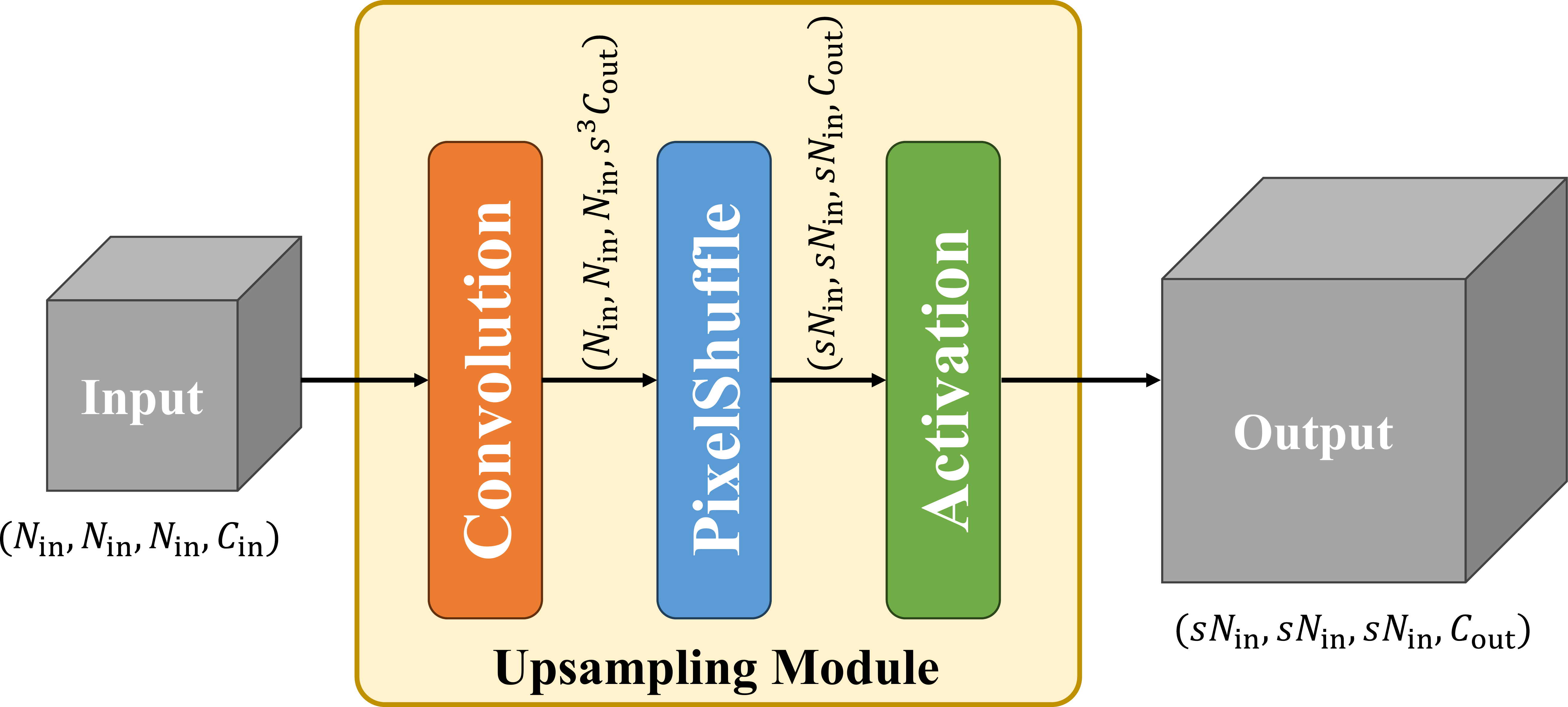}
    \caption{\small Upsampling Module.}
    \label{UP}
\end{figure}

In each upsampling module, we utilize a PixelShuffle layer \cite{PIXELSHUFFLE} between the convolution and activation layers to upscale the input, as shown in Fig. \ref{UP}. The input feature tensor has dimensions $(N_{\rm in}, N_{\rm in}, N_{\rm in}, C_{\rm in})$, with an upsampling scale of $s$ and an output channel count of $C_{\rm out}$.

\section{Experiment} \label{supp:exp}
\subsection{Evaluation Metric} \label{APP:MATRIC}
Let $\mathbf{S}_{\rm Rec}$ and $\mathbf{S}_{\rm GT}$ denote the reconstructed and ground-truth 3D shapes, respectively. We then randomly sample $N_{\rm eval}=10^5$ points on them, obtaining two point clouds, $\mathbf{P}_{\rm Rec}$ and $\mathbf{P}_{\rm GT}$. For each point of $\mathbf{P}_{\rm Rec}$ and $\mathbf{P}_{\rm GT}$, the normal of the triangle face where it is sampled is considered to be its normal vector, and the normal sets of $\mathbf{P}_{\rm Rec}$ and $\mathbf{P}_{\rm GT}$ are denoted as $\mathbf{N}_{\rm Rec}$ and $\mathbf{N}_{\rm GT}$, respectively. Let $\mathtt{NN\_Point}(\mathbf{x},\mathbf{P})$ be the operator that returns the nearest point of $\mathbf{x}$ in the point cloud $\mathbf{P}$.
The CD between them is defined as 
\begin{equation}
\begin{aligned}
    \texttt{CD}(\mathbf{S}_{\rm Rec},\mathbf{S}_{\rm GT})=&\frac{1}{2N_{\rm eval}}\sum_{\mathbf{x}\in\mathbf{P}_{\rm Rec}}\|\mathbf{x}-\mathtt{NN\_Point}(\mathbf{x},\mathbf{P}_{\rm GT})\|_2 \\
    +&\frac{1}{2N_{\rm eval}}\sum_{\mathbf{x}\in\mathbf{P}_{\rm GT}}\|\mathbf{x}-\mathtt{NN\_Point}(\mathbf{x},\mathbf{P}_{\rm Rec})\|_2. \\
\end{aligned}
\end{equation}
Let $\mathtt{NN\_Normal}(\mathbf{x},\mathbf{P})$ be the operator that returns the normal vector of the point $\mathbf{x}$'s nearest point in the point cloud $\mathbf{P}$. 
The NC is defined as 
\begin{equation}
\begin{aligned}
    \texttt{NC}(\mathbf{S}_{\rm Rec},\mathbf{S}_{\rm GT})=&\frac{1}{2N_{\rm eval}}\sum_{\mathbf{x}\in\mathbf{P}_{\rm Rec}}|\mathbf{N}_{\rm Rec}(\mathbf{x})\cdot\mathtt{NN\_Normal}(\mathbf{x},\mathbf{P}_{\rm GT})| \\
    +&\frac{1}{2N_{\rm eval}}\sum_{\mathbf{x}\in\mathbf{P}_{\rm GT}}|\mathbf{N}_{\rm GT}(\mathbf{x})\cdot\mathtt{NN\_Normal}(\mathbf{x},\mathbf{P}_{\rm Rec})|.
\end{aligned}
\end{equation}
F-Score is defined as the harmonic mean between the precision and the recall of points that lie within a certain distance threshold $\epsilon$ between $\mathbf{S}_{\rm Rec}$ and $\mathbf{S}_{\rm GT}$,
\begin{equation}
    \mathtt{F-Score}(\mathbf{S}_{\rm Rec},\mathbf{S}_{\rm GT},\epsilon)=\frac{2\cdot\mathtt{Recall}\cdot\mathtt{Precision}}{\mathtt{Recall}+\mathtt{Precision}},
\end{equation}
where
\begin{equation}
    \begin{aligned}
        \mathtt{Recall}(\mathbf{S}_{\rm Rec},\mathbf{S}_{\rm GT},\epsilon)&=\left|\left\{\mathbf{x}_1\in\mathbf{P}_{\rm Rec}, \rm{s.t.} \min_{\mathbf{x}_2\in\mathbf{P}_{\rm GT}}\|\mathbf{x}_1-\mathbf{x}_2\|_2<\epsilon\right\}\right|, \\
        \mathtt{Precision}(\mathbf{S}_{\rm Rec},\mathbf{S}_{\rm GT},\epsilon)&=\left|\left\{\mathbf{x}_2\in\mathbf{P}_{\rm GT}, \rm{s.t.} \min_{\mathbf{x}_1\in\mathbf{P}_{\rm Rec}}\|\mathbf{x}_1-\mathbf{x}_2\|_2<\epsilon\right\}\right|.
    \end{aligned}
\end{equation}

\begin{figure}[h]
    \centering
    \includegraphics[width=0.6\linewidth]{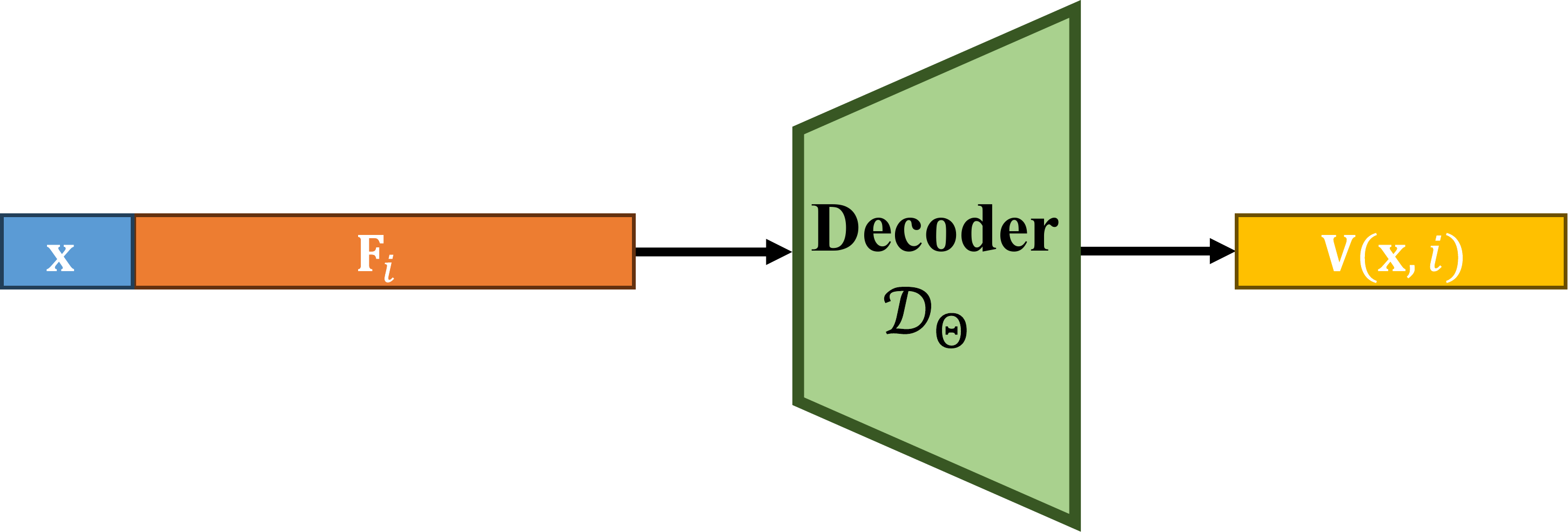}
    \caption{Pipeline of QuantDeepSDF.}
    \label{DEEPSDF}
\end{figure}

\subsection{QuantDeepSDF} \label{APP:DEEPSDF}
Compared to DeepSDF \cite{DEEPSDF}, our QuantDeepSDF incorporates the following two modifications:
\begin{itemize}
\item The decoder parameters are quantized to enhance compression efficiency.
\item To maintain consistency with our NeCGS, the points sampled during training are drawn from TSDF-Def tensors.
\end{itemize}
The pipeline of QuantDeepSDF is shown in Fig. \ref{DEEPSDF}. 
Specifically, the decoder is an MLP, where the input is the concatenated  vector of coordinate $\mathbf{x}\in\mathbb{R}^{3}$ and the $i$-th embedded feature vector $\mathbf{F}_i\in\mathbb{R}^{C}$, and the output is the corresponding TSDF-Def value. In our experiment, the decoder consists of 8 layers, and the compression ratio is controled by changing the width of each layer.

\subsection{Auto-Encoder in Ablation Study}
Different from the auto-encoder used in our framework, where the embed features are directly optimized, auto-encoder utilizes an encoder to produce the embedded features, where the inputs are the TSDF-Def tensors. And the decoder is kept the same as our framework. During the optimization, the parameters of encoder and decoder are optimized. Once optimized, the embedded features produced by the encoder and decoder parameters are compressed into bitstreams.

\subsection{More Visual Results}
\begin{figure*}[h] \small
\centering
{
\begin{tikzpicture}[]
\node[] (a) at (0,8.1) {\includegraphics[width=0.13\textwidth]{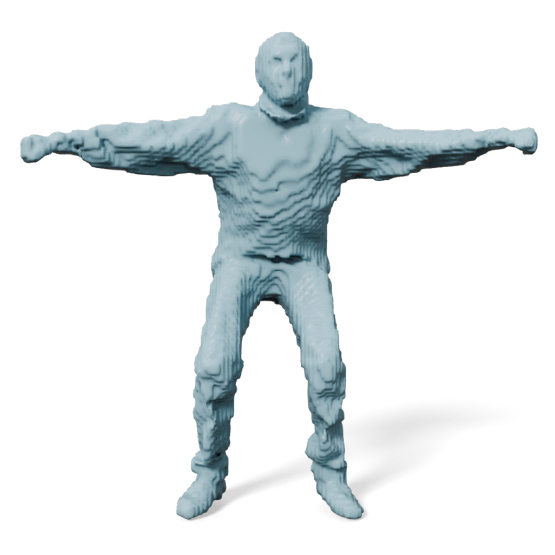}};
\node[] (a) at (0,5.8) {\includegraphics[width=0.13\textwidth]{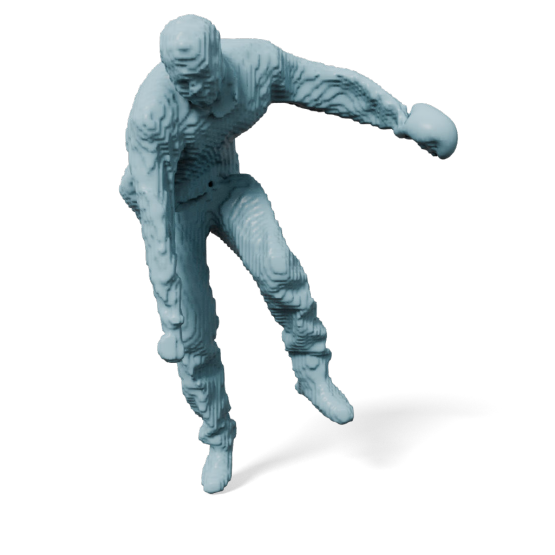} };
\node[] (a) at (0,3.5) {\includegraphics[width=0.13\textwidth]{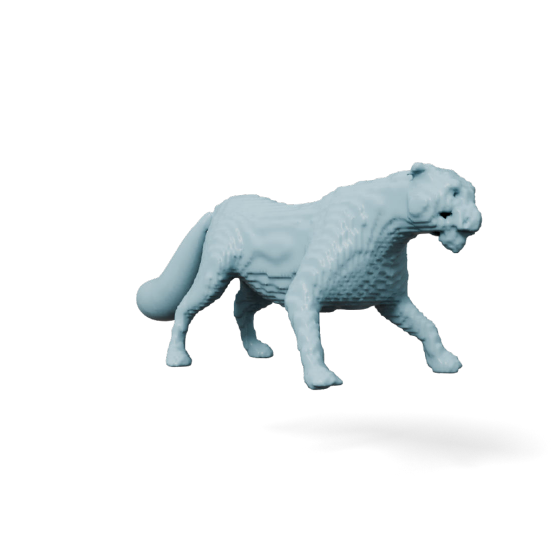} };
\node[] (a) at (0,1.2) {\includegraphics[width=0.13\textwidth]{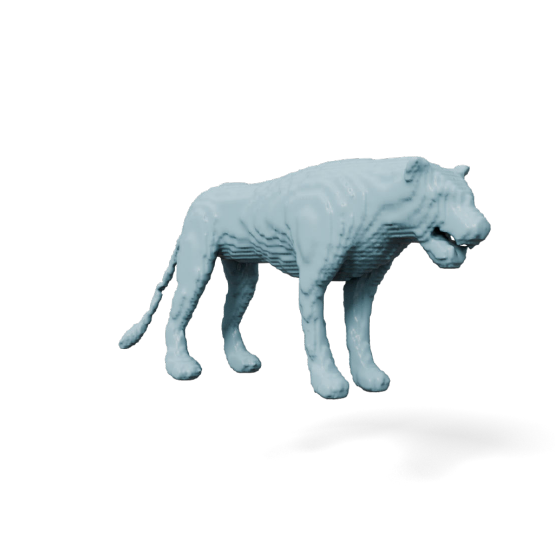} };
\node[] (a) at (0,-1.1) {\includegraphics[width=0.13\textwidth]{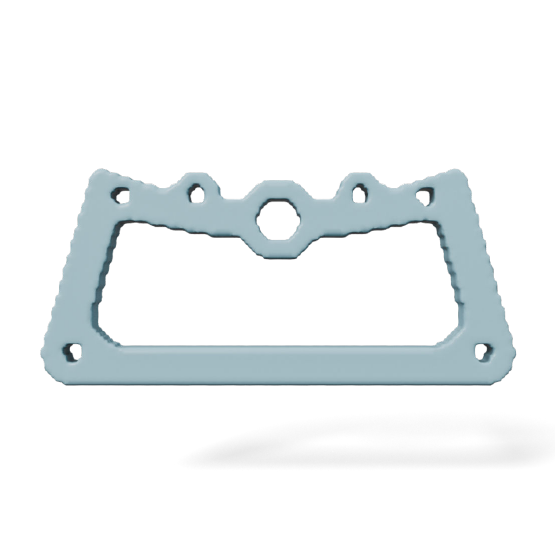} };
\node[] (a) at (0,-3.4) {\includegraphics[width=0.13\textwidth]{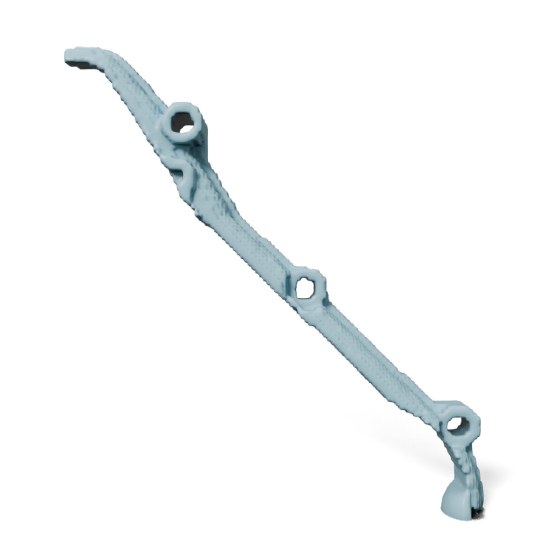} };

\node[] (a) at (0,-4.8) {\small (a) GPCC};

\node[] (a) at (2.5,8.1) {\includegraphics[width=0.13\textwidth]{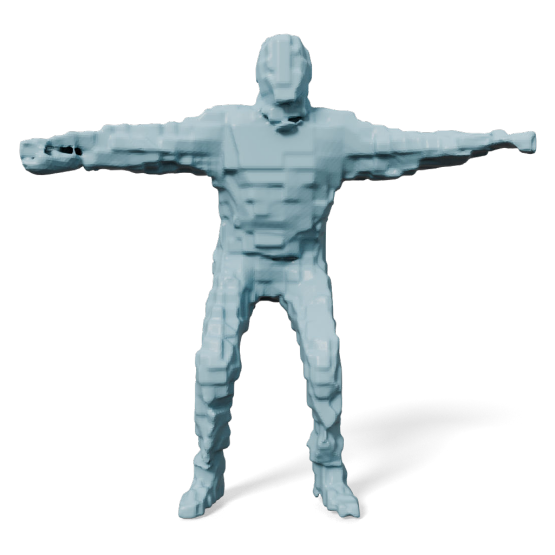}};
\node[] (a) at (2.5,5.8) {\includegraphics[width=0.13\textwidth]{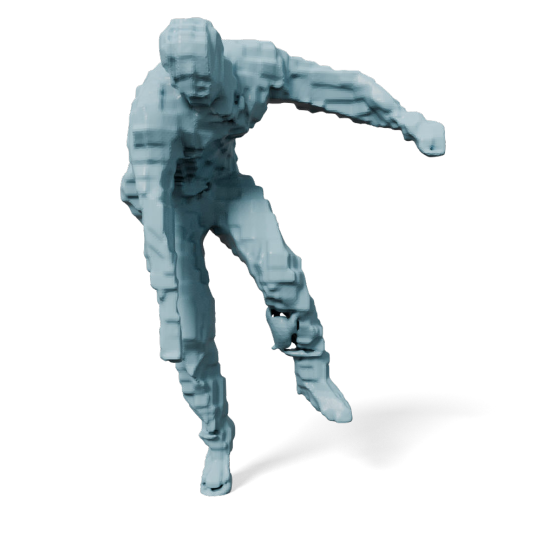} };
\node[] (a) at (2.5,3.5) {\includegraphics[width=0.13\textwidth]{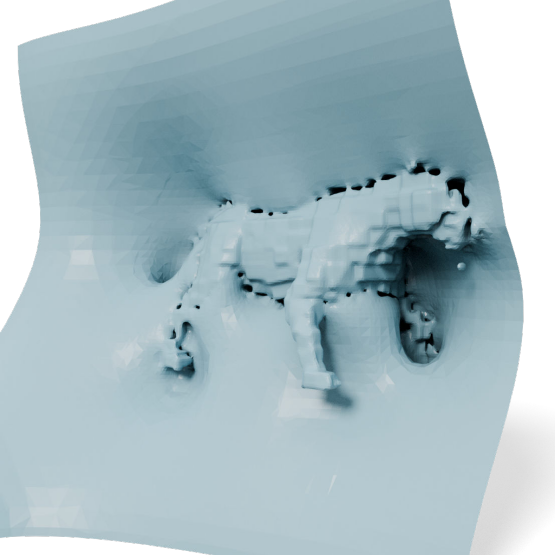} };
\node[] (a) at (2.5,1.2) {\includegraphics[width=0.13\textwidth]{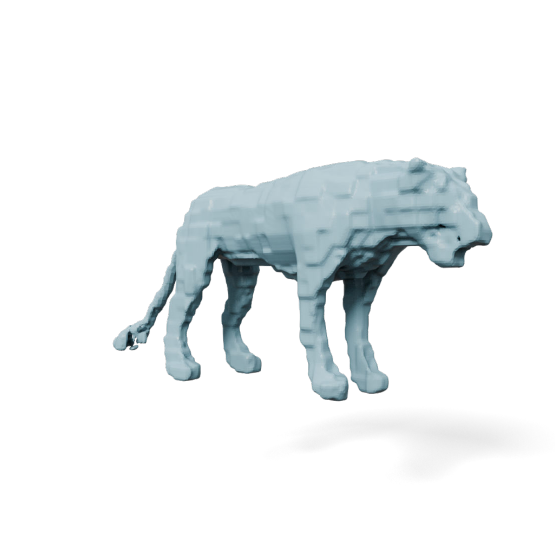} };
\node[] (a) at (2.5,-1.1) {\includegraphics[width=0.13\textwidth]{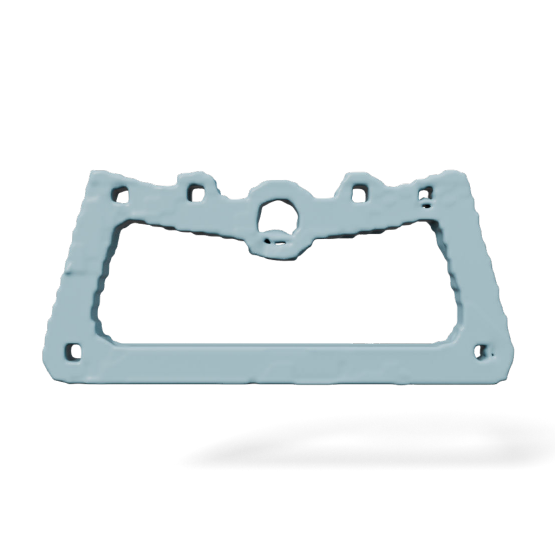} };
\node[] (a) at (2.5,-3.4) {\includegraphics[width=0.13\textwidth]{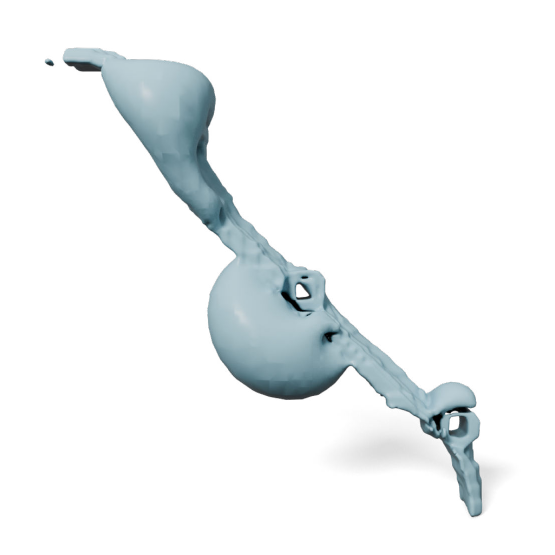} };

\node[] (a) at (2.5,-4.8) {\small (b) VPCC};

\node[] (a) at (5,8.1) {\includegraphics[width=0.13\textwidth]{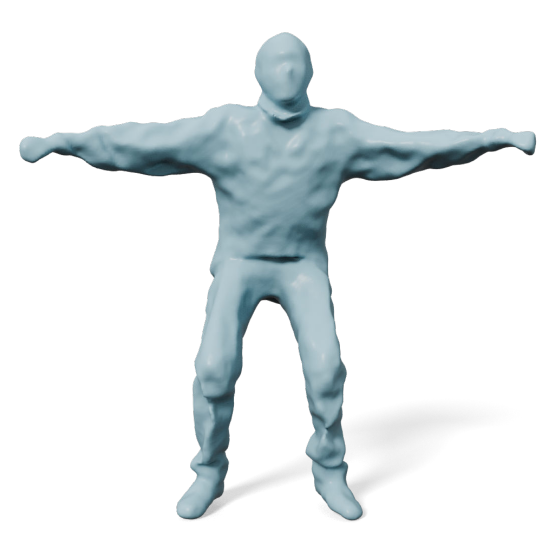}};
\node[] (a) at (5,5.8) {\includegraphics[width=0.13\textwidth]{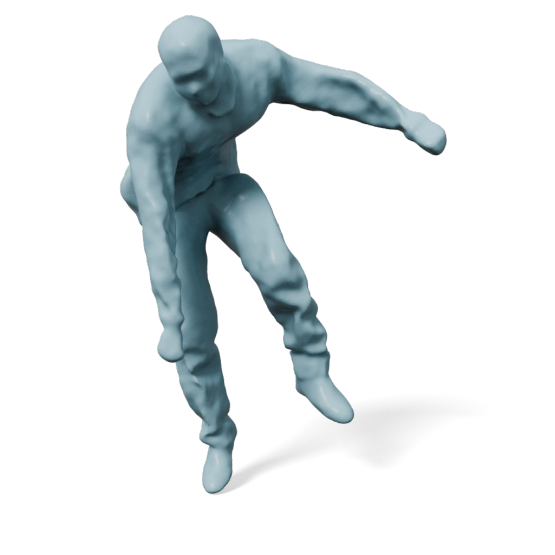} };
\node[] (a) at (5,3.5) {\includegraphics[width=0.13\textwidth]{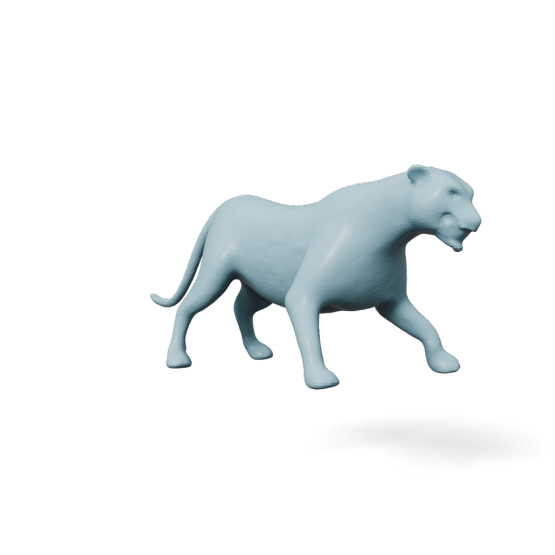} };
\node[] (a) at (5,1.2) {\includegraphics[width=0.13\textwidth]{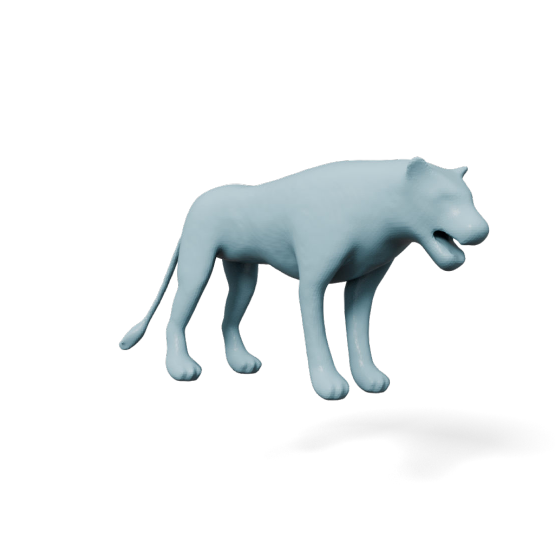} };
\node[] (a) at (5,-1.1) {\includegraphics[width=0.13\textwidth]{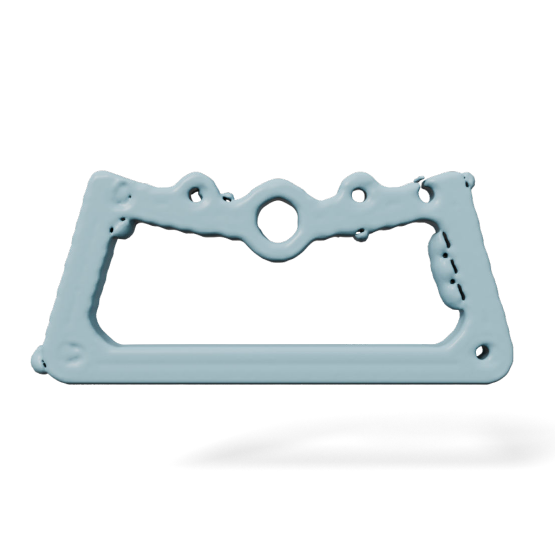} };
\node[] (a) at (5,-3.4) {\includegraphics[width=0.13\textwidth]{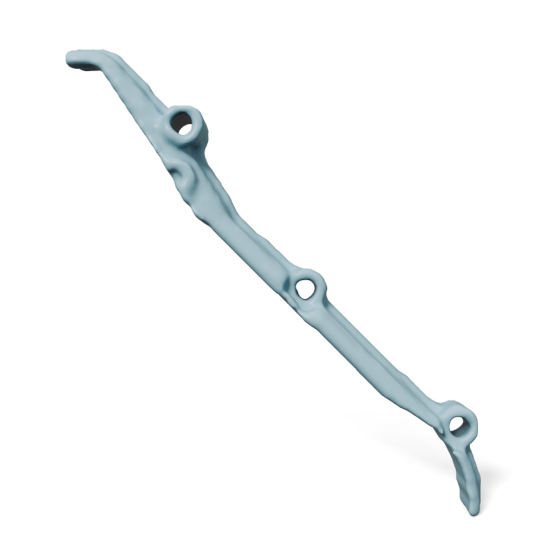} };

\node[] (a) at (5,-4.8) {\small (c) PCGCv2};

\node[] (a) at (7.5,8.1) {\includegraphics[width=0.13\textwidth]{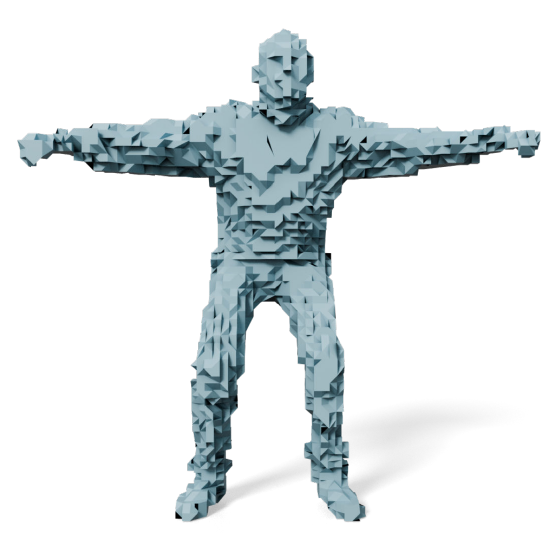}};
\node[] (a) at (7.5,5.8) {\includegraphics[width=0.13\textwidth]{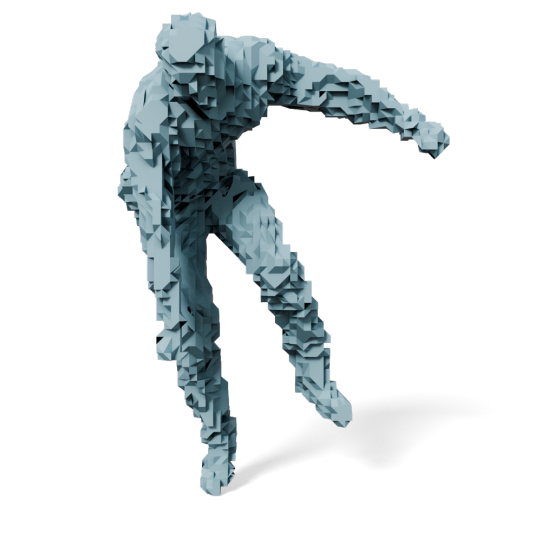} };
\node[] (a) at (7.5,3.5) {\includegraphics[width=0.13\textwidth]{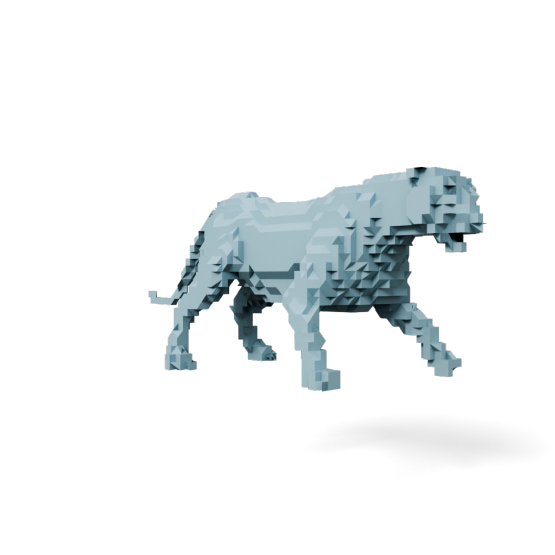} };
\node[] (a) at (7.5,1.2) {\includegraphics[width=0.13\textwidth]{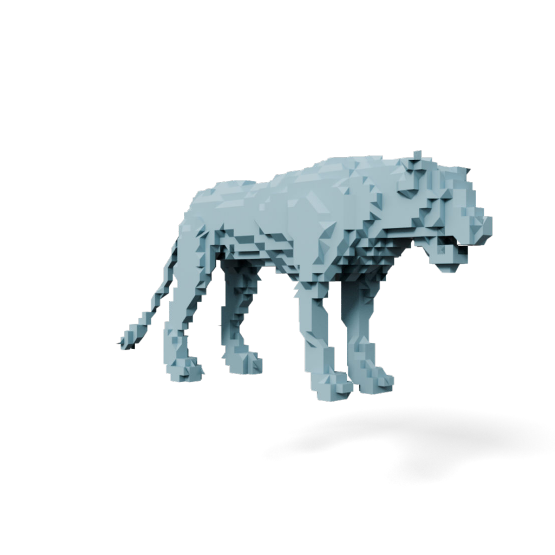} };
\node[] (a) at (7.5,-1.1) {\includegraphics[width=0.13\textwidth]{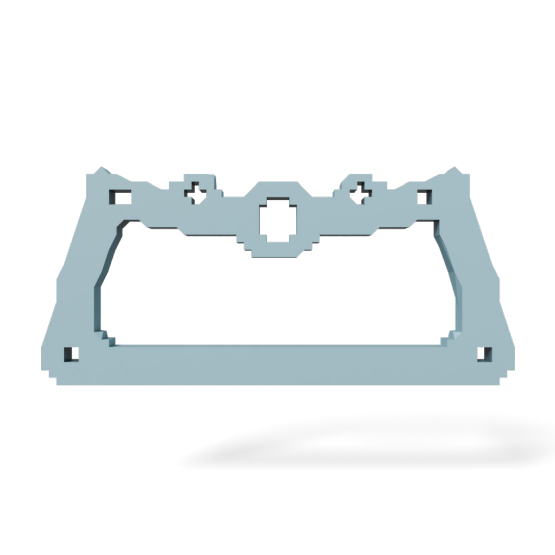} };
\node[] (a) at (7.5,-3.4) {\includegraphics[width=0.13\textwidth]{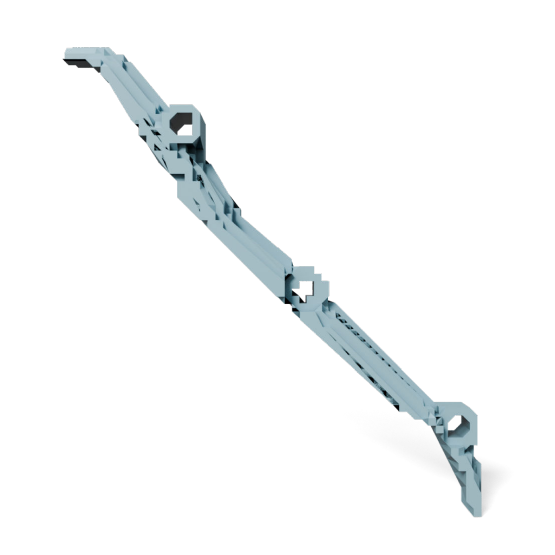} };

\node[] (a) at (7.5,-4.8) {\small (d) Draco};

\node[] (a) at (10,8.1) {\includegraphics[width=0.13\textwidth]{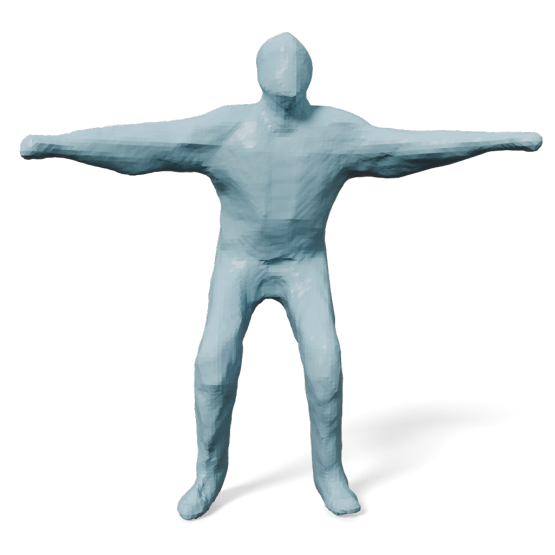}};
\node[] (a) at (10,5.8) {\includegraphics[width=0.13\textwidth]{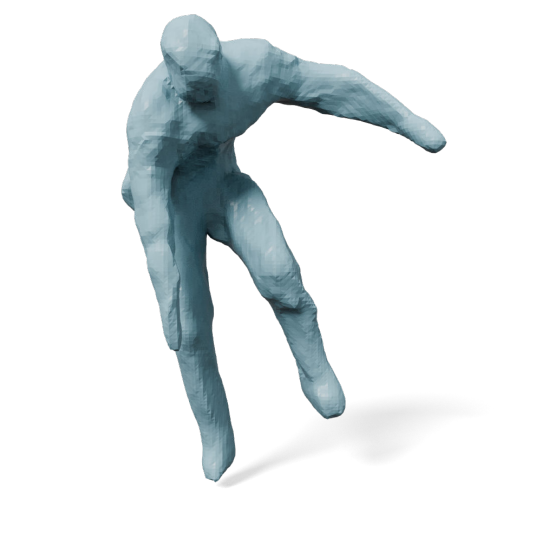} };
\node[] (a) at (10,3.5) {\includegraphics[width=0.13\textwidth]{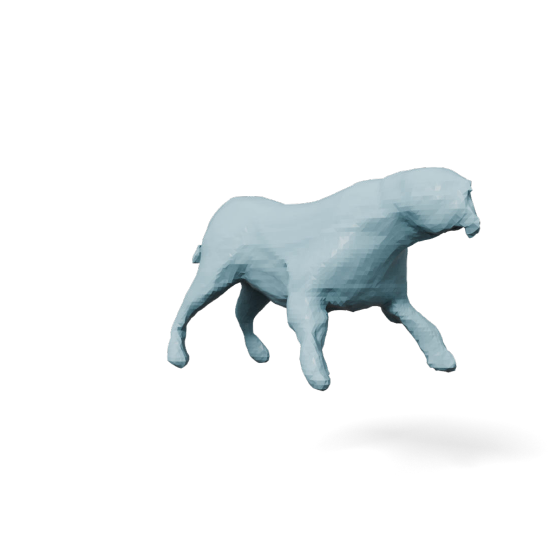} };
\node[] (a) at (10,1.2) {\includegraphics[width=0.13\textwidth]{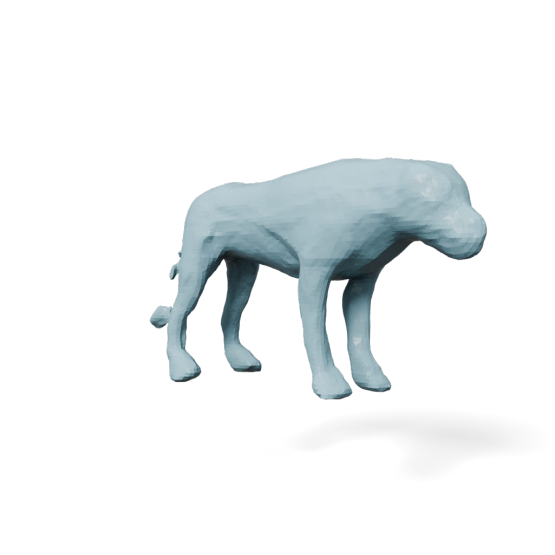} };
\node[] (a) at (10,-1.1) {\includegraphics[width=0.13\textwidth]{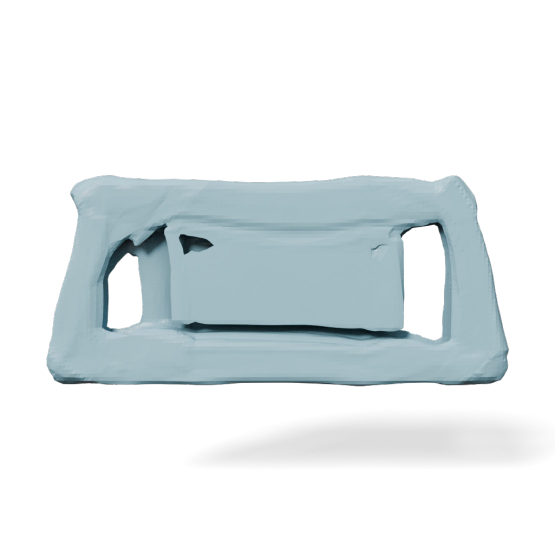} };
\node[] (a) at (10,-3.4) {\includegraphics[width=0.13\textwidth]{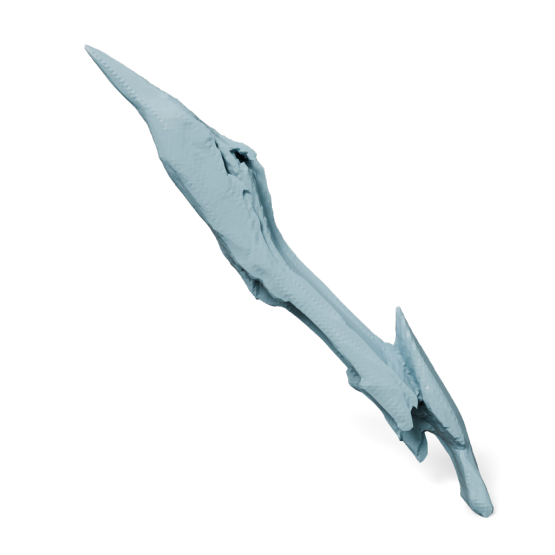} };

\node[] (a) at (10,-4.8) {\small (e) QuantDeepSDF};

\node[] (a) at (12.5,8.1) {\includegraphics[width=0.13\textwidth]{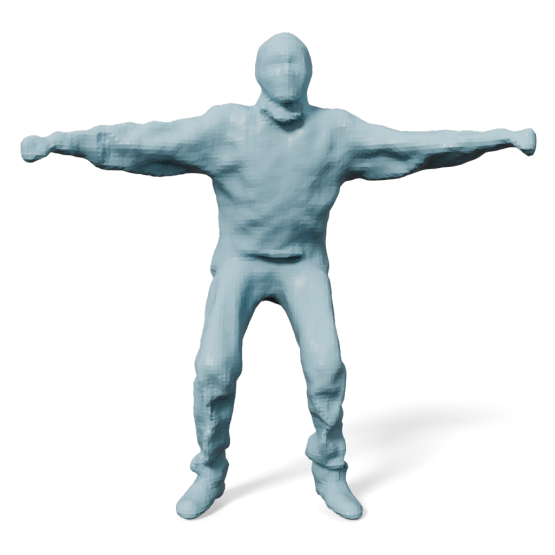}};
\node[] (a) at (12.5,5.8) {\includegraphics[width=0.13\textwidth]{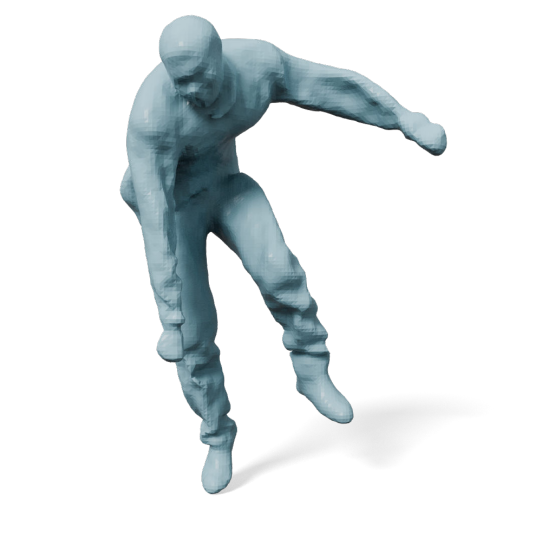} };
\node[] (a) at (12.5,3.5) {\includegraphics[width=0.13\textwidth]{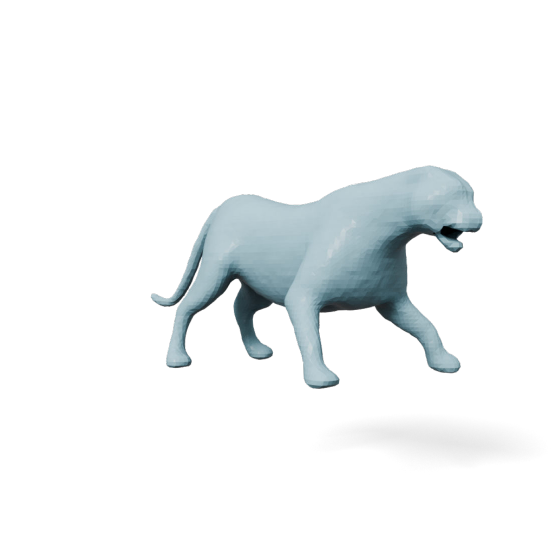} };
\node[] (a) at (12.5,1.2) {\includegraphics[width=0.13\textwidth]{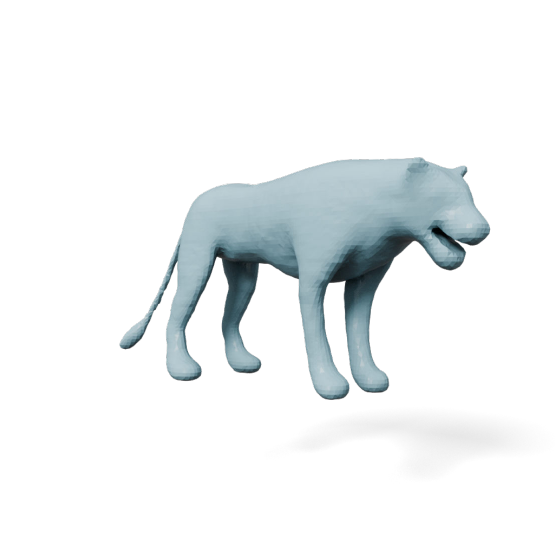} };
\node[] (a) at (12.5,-1.1) {\includegraphics[width=0.13\textwidth]{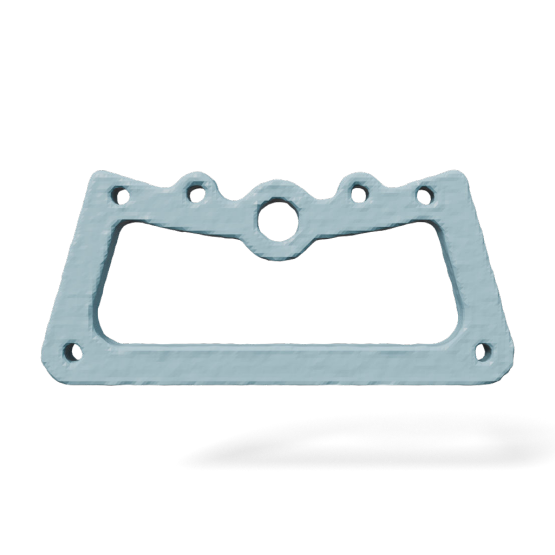} };
\node[] (a) at (12.5,-3.4) {\includegraphics[width=0.13\textwidth]{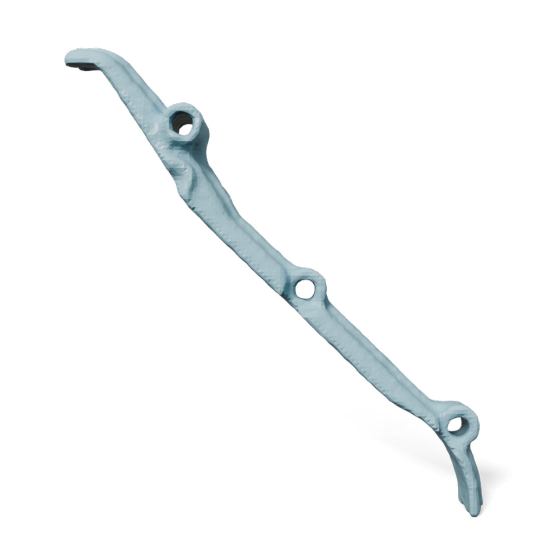} };

\node[] (a) at (12.5,-4.8) {\small (f) Ours};

\node[] (a) at (15,8.1) {\includegraphics[width=0.13\textwidth]{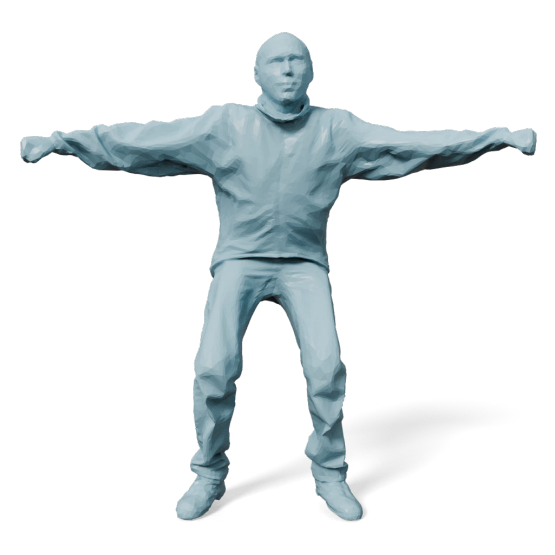}};
\node[] (a) at (15,5.8) {\includegraphics[width=0.13\textwidth]{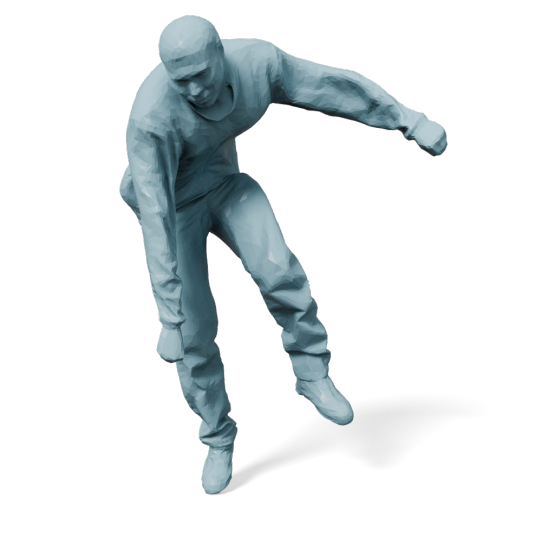} };
\node[] (a) at (15,3.5) {\includegraphics[width=0.13\textwidth]{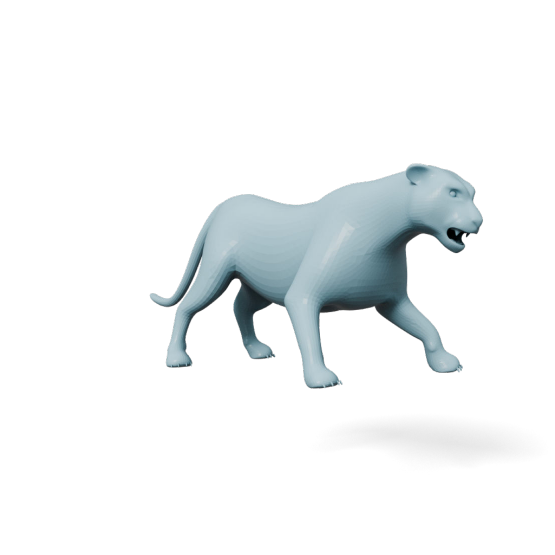} };
\node[] (a) at (15,1.2) {\includegraphics[width=0.13\textwidth]{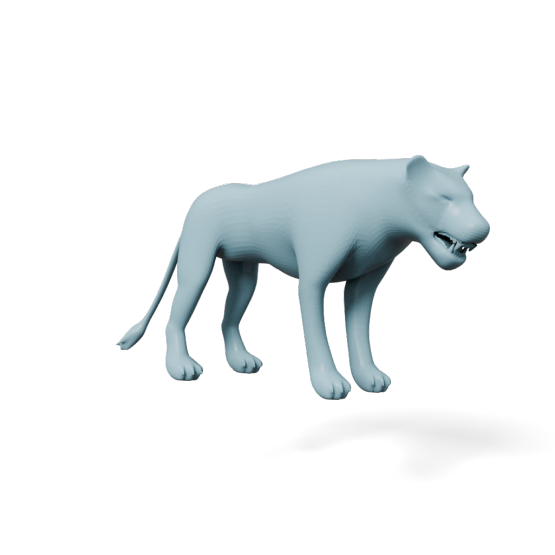} };
\node[] (a) at (15,-1.1) {\includegraphics[width=0.13\textwidth]{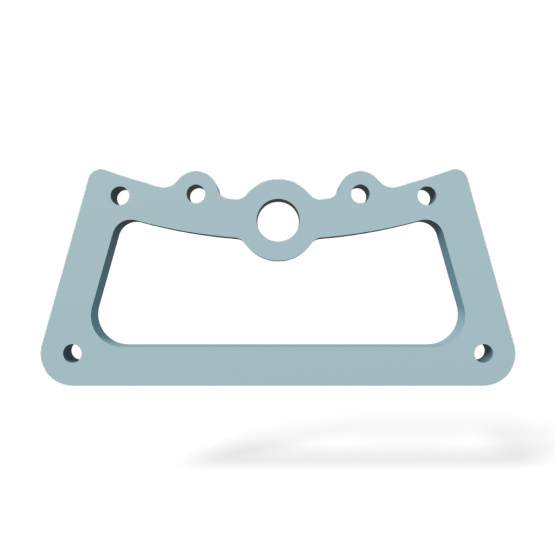} };
\node[] (a) at (15,-3.4) {\includegraphics[width=0.13\textwidth]{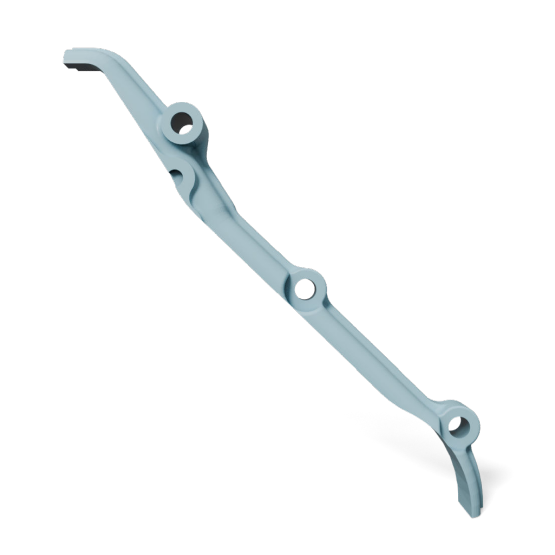} };

\node[] (a) at (15,-4.8) {\small (g) Original};

\node[] (a) at (1,-2) {\scriptsize 148.76};
\node[] (a) at (3.5,-2) {\scriptsize 49.81};
\node[] (a) at (6,-2) {\scriptsize 107.86};
\node[] (a) at (8.5,-2) {\scriptsize 95.64};
\node[] (a) at (11,-2) {\scriptsize 165.47};
\node[] (a) at (13.5,-2) {\scriptsize 166.79};

\node[] (a) at (1,-4.3) {\scriptsize 148.64};
\node[] (a) at (3.5,-4.3) {\scriptsize 49.81};
\node[] (a) at (6,-4.3) {\scriptsize 106.78};
\node[] (a) at (8.5,-4.3) {\scriptsize 98.64};
\node[] (a) at (11,-4.3) {\scriptsize 165.47};
\node[] (a) at (13.5,-4.3) {\scriptsize 166.79};

\node[] (a) at (1,2.6) {\scriptsize 457.12};
\node[] (a) at (3.5,2.6) {\scriptsize 244.38};
\node[] (a) at (6,2.6) {\scriptsize 402.32};
\node[] (a) at (8.5,2.6) {\scriptsize 158.67};
\node[] (a) at (11,2.6) {\scriptsize 409.21};
\node[] (a) at (13.5,2.6) {\scriptsize 455.25};

\node[] (a) at (1,0.3) {\scriptsize 457.64};
\node[] (a) at (3.5,0.3) {\scriptsize 244.38};
\node[] (a) at (6,0.3) {\scriptsize 404.25};
\node[] (a) at (8.5,0.3) {\scriptsize 157.67};
\node[] (a) at (11,0.3) {\scriptsize 409.21};
\node[] (a) at (13.5,0.3) {\scriptsize 455.25};

\node[] (a) at (1,7.2) {\scriptsize 291.18};
\node[] (a) at (3.5,7.2) {\scriptsize 165.75};
\node[] (a) at (6,7.2) {\scriptsize 267.89};
\node[] (a) at (8.5,7.2) {\scriptsize 96.78};
\node[] (a) at (11,7.2) {\scriptsize 224.17};
\node[] (a) at (13.5,7.2) {\scriptsize 362.80};

\node[] (a) at (1,4.9) {\scriptsize 290.19};
\node[] (a) at (3.5,4.9) {\scriptsize 165.75};
\node[] (a) at (6,4.9) {\scriptsize 264.16};
\node[] (a) at (8.5,4.9) {\scriptsize 98.16};
\node[] (a) at (11,4.9) {\scriptsize 224.17};
\node[] (a) at (13.5,4.9) {\scriptsize 362.80};

\end{tikzpicture}
}
\vspace{-0.7cm}
\caption{\small Visual comparisons of different compression methods. All numbers in corners represent the compression ratio. 
\color{blue}{\faSearch~} Zoom in for details. 
}
\label{FIG:MORE:RESULT:VIS} \vspace{-0.3cm}
\end{figure*}

\begin{figure*}
    \centering
    \includegraphics[width=0.19\linewidth]{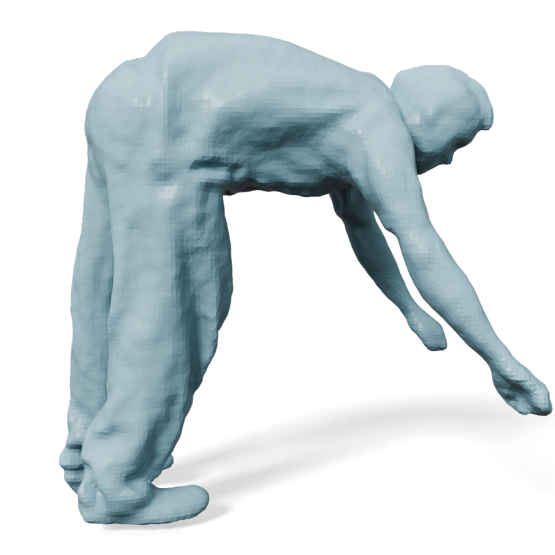}
    \includegraphics[width=0.19\linewidth]{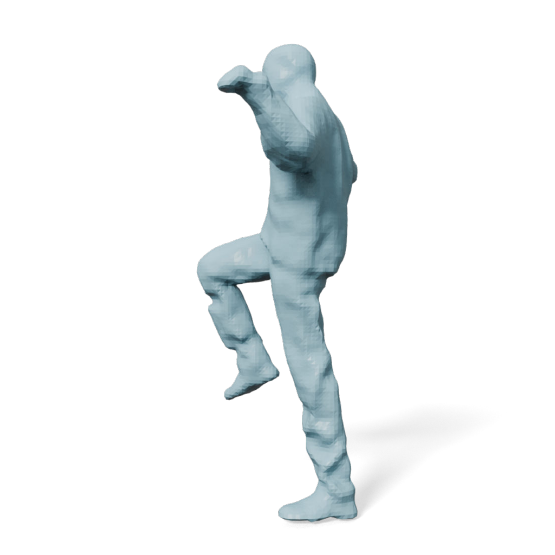}
    \includegraphics[width=0.19\linewidth]{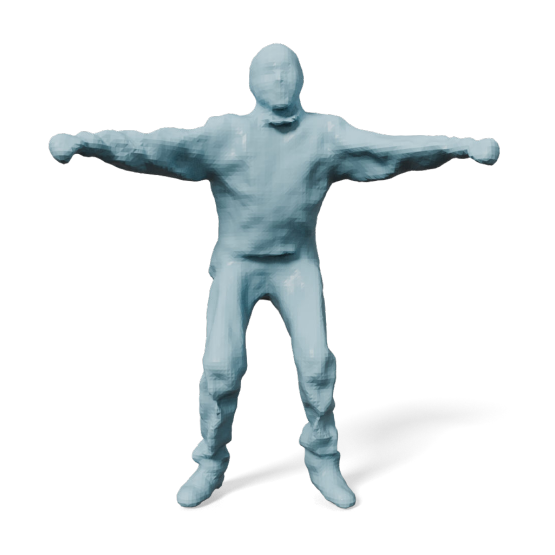}
    \includegraphics[width=0.19\linewidth]{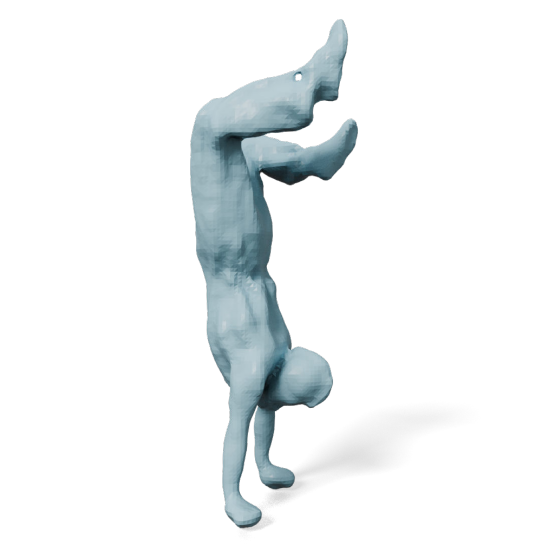}
    \includegraphics[width=0.19\linewidth]{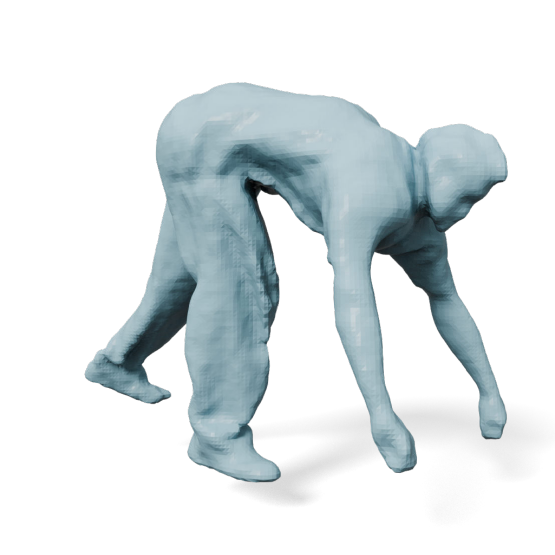} \\
    \includegraphics[width=0.19\linewidth]{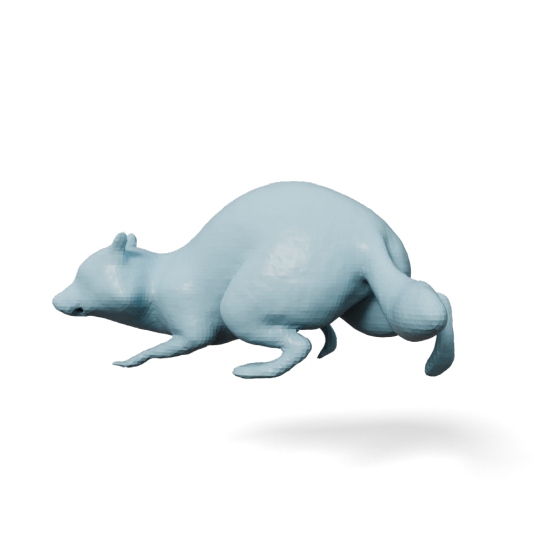}
    \includegraphics[width=0.19\linewidth]{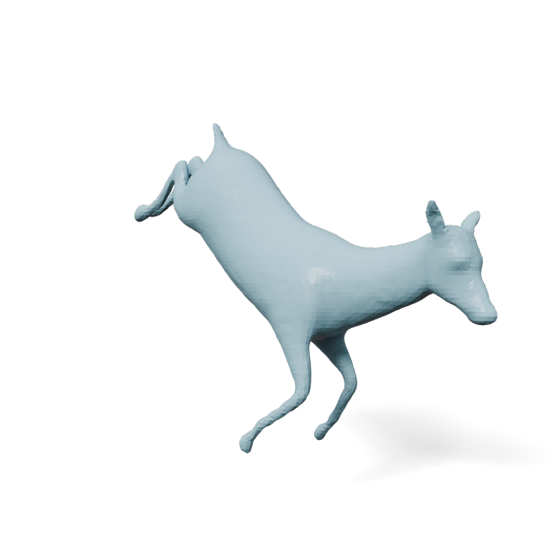}
    \includegraphics[width=0.19\linewidth]{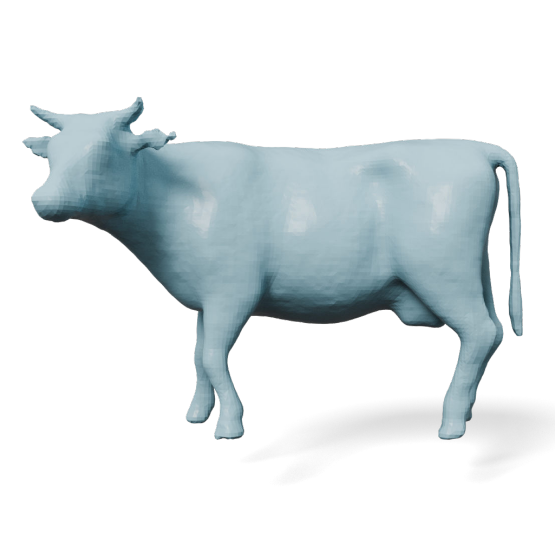}
    \includegraphics[width=0.19\linewidth]{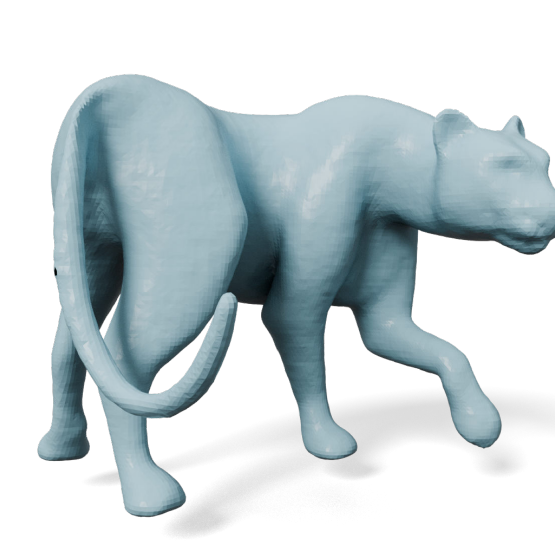}
    \includegraphics[width=0.19\linewidth]{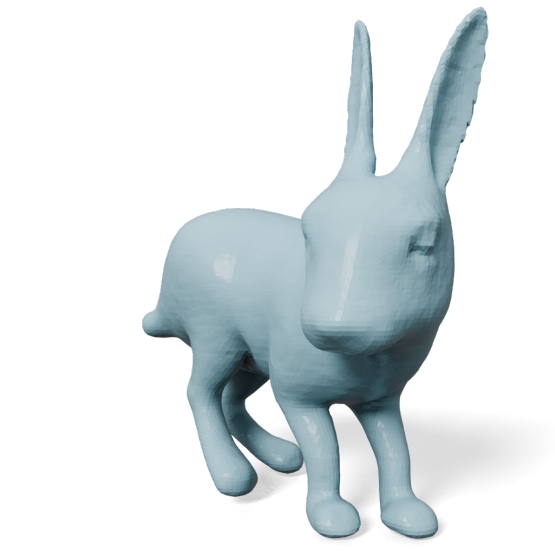} \\
    \includegraphics[width=0.19\linewidth]{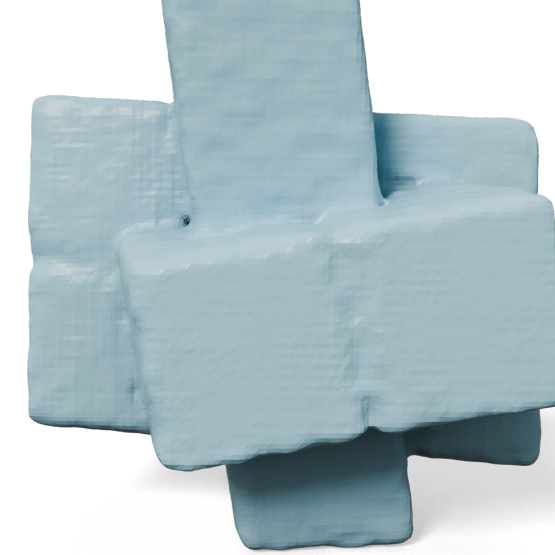}
    \includegraphics[width=0.19\linewidth]{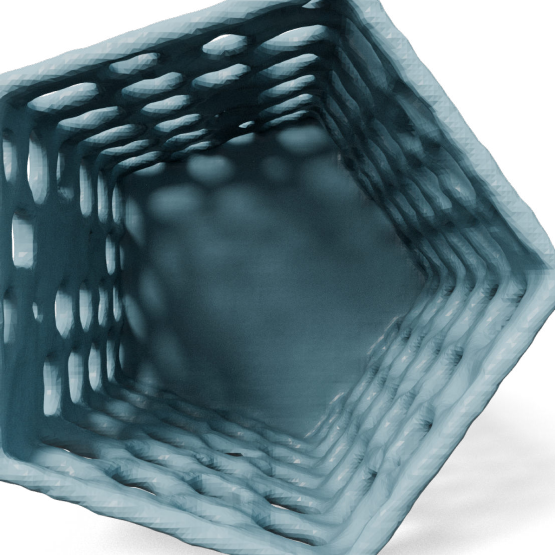}
    \includegraphics[width=0.19\linewidth]{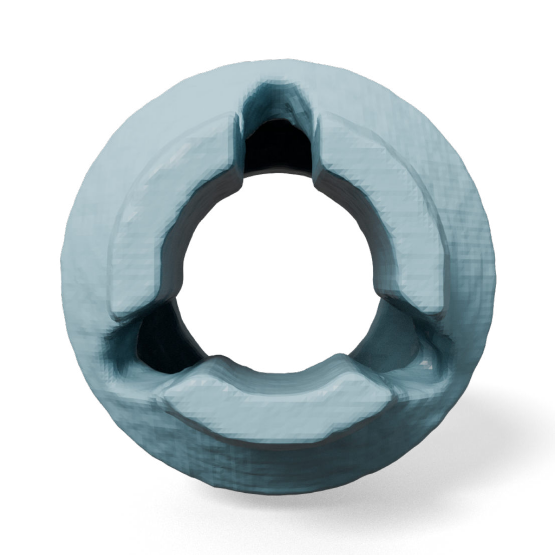}
    \includegraphics[width=0.19\linewidth]{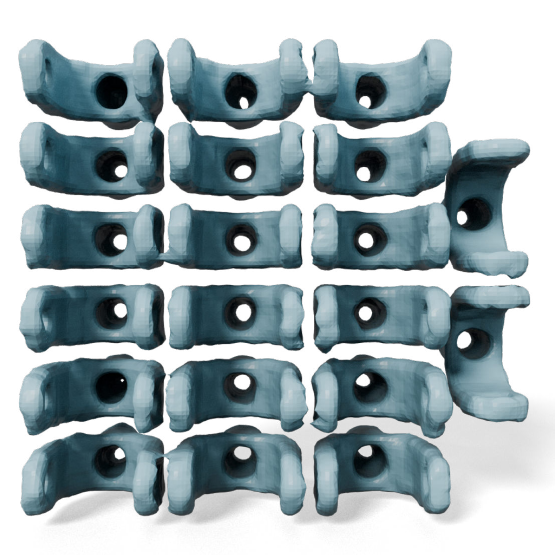}
    \includegraphics[width=0.19\linewidth]{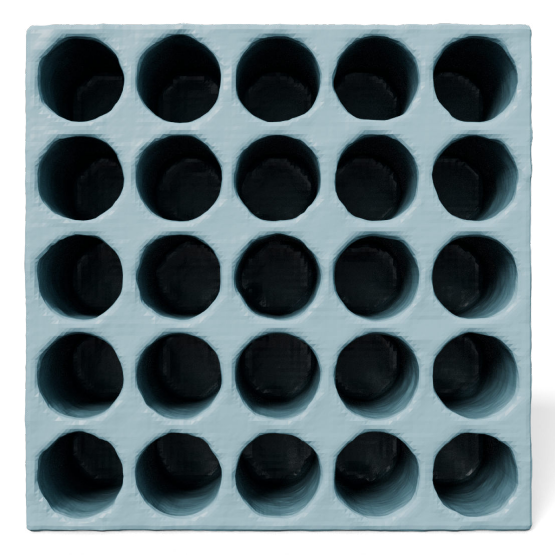} 
    \caption{Visualization of the decompressed models with detailed structures. \color{blue}{\faSearch~} Zoom in for details.}
    \label{FIG:DECOMPRESSED}
\end{figure*}

\begin{figure}
    \centering \small
    \subfloat[Original]{\includegraphics[width=0.45\linewidth]{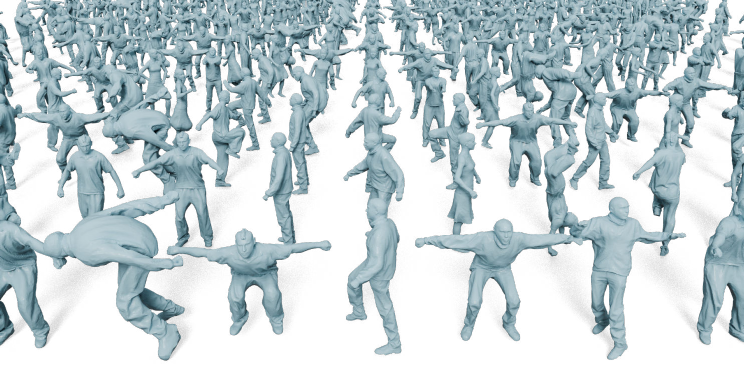}\vspace{-0.3cm}} \quad
    \subfloat[253.45]{\includegraphics[width=0.45\linewidth]{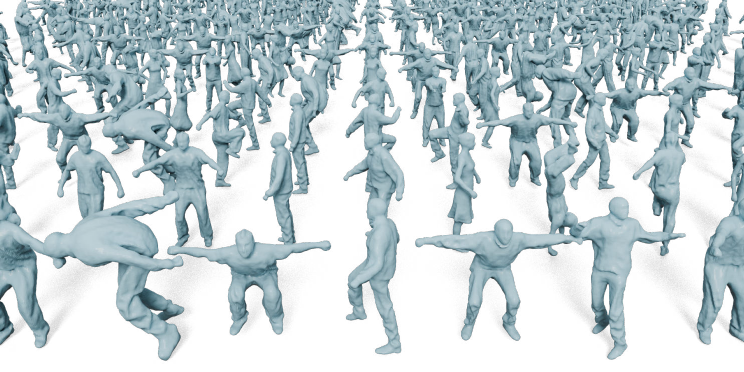}\vspace{-0.3cm}}\\
    \subfloat[362.80]{\includegraphics[width=0.45\linewidth]{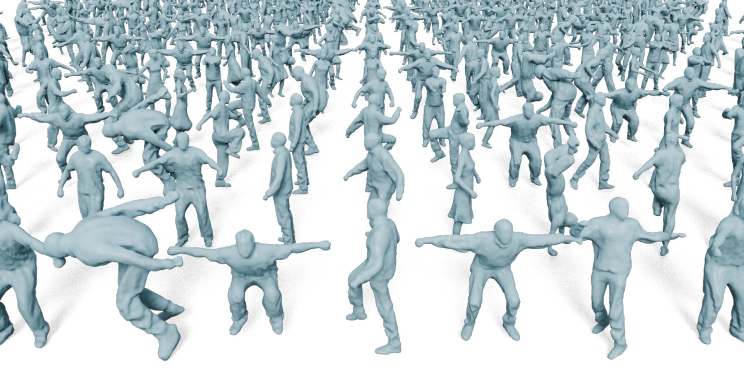}\vspace{-0.3cm}}\quad
    \subfloat[500.54]{\includegraphics[width=0.45\linewidth]{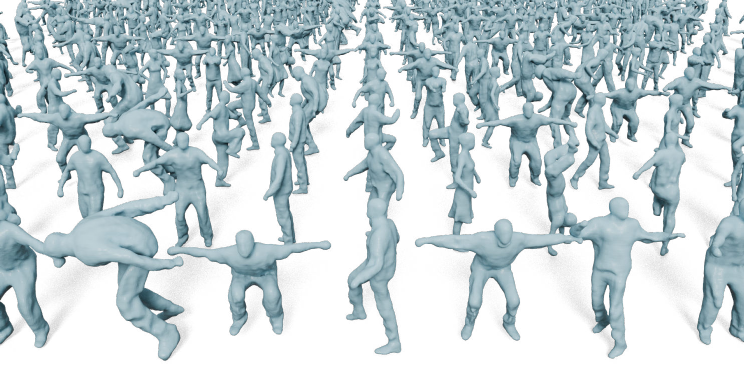}\vspace{-0.3cm}}
    \caption{\small Visual results of the decompressed models from the AMA dataset  under different compression ratios.}
    \label{MORE:VIS:AMA}
\end{figure}

\begin{figure}
    \centering \small
    \subfloat[Original]{\includegraphics[width=0.45\linewidth]{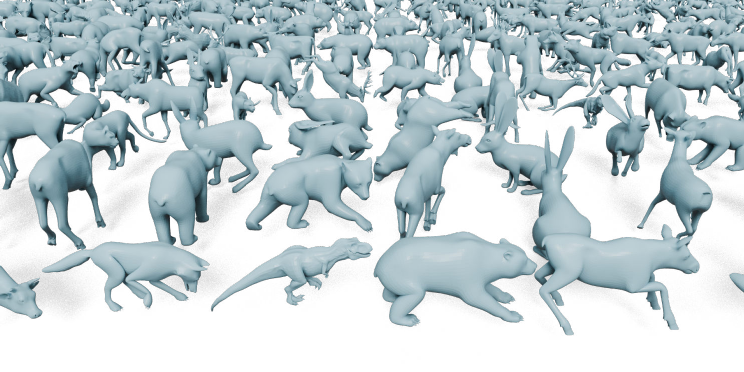}\vspace{-0.3cm}}\quad
    \subfloat[455.26 ]{\includegraphics[width=0.45\linewidth]{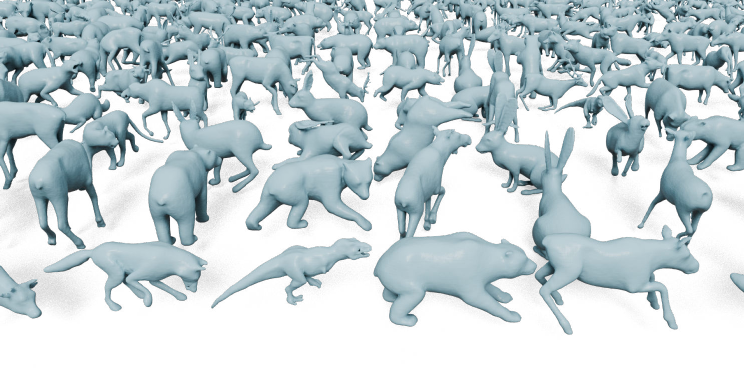}\vspace{-0.3cm}} \\
    \subfloat[651.85]{\includegraphics[width=0.45\linewidth]{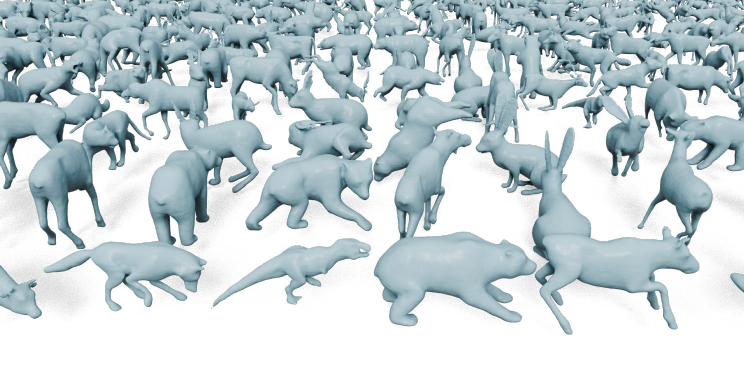}\vspace{-0.3cm}}\quad
    \subfloat[899.73]{\includegraphics[width=0.45\linewidth]{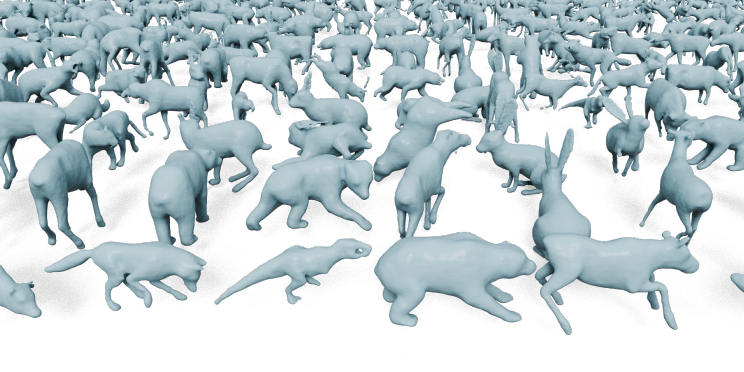}\vspace{-0.3cm}}
    \caption{\small Visual results of the decompressed models from the DT4D dataset  under different compression ratios.}
    \label{MORE:VIS:DT4D}
\end{figure}

\begin{figure}
    \centering \small
    \subfloat[Original]{\includegraphics[width=0.45\linewidth]{figures/TEASER/rec_mesh_thingi_gt.pdf}\vspace{-0.3cm}}\quad
    \subfloat[166.79]{\includegraphics[width=0.45\linewidth]{figures/TEASER/rec_mesh_thingi_pred.pdf}\vspace{-0.3cm}}\\
    \subfloat[219.84]{\includegraphics[width=0.45\linewidth]{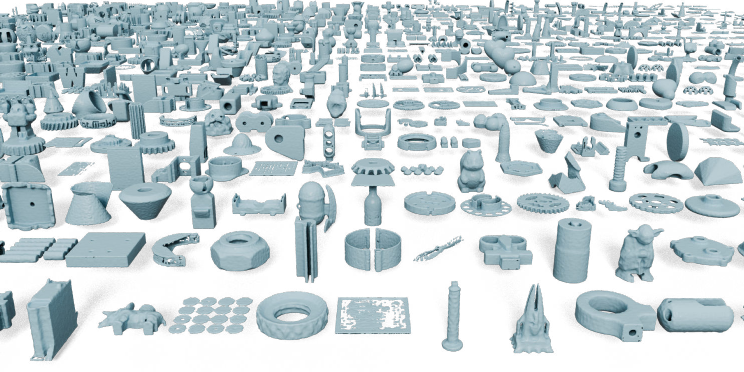}\vspace{-0.3cm}}\quad
    \subfloat[273.32]{\includegraphics[width=0.45\linewidth]{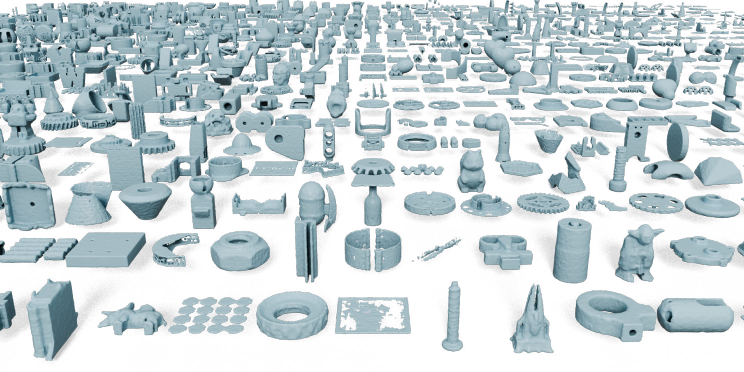}\vspace{-0.3cm}}
    \caption{\small Visual results of the decompressed models from the Thingi10K dataset  under different compression ratios.}
    \label{MORE:VIS:THINGI10K}
\end{figure}

Fig. \ref{FIG:MORE:RESULT:VIS} shows more visual comparison between different methods on the AMA dataset, DT4D dataset, and Thingi10K dataset. Obviously, the decompressed models through our method have less distortion than baseline methods. % the decompressed models from the AMA dataset, DT4D dataset, and Thingi10K dataset. 
Fig. \ref{FIG:DECOMPRESSED} showcases the decompressed models with detailed structures obtained using our method, demonstrating that the detailed structures are well-preserved after decompression.
%demonstrates the decompressed models with detailed structures through our method, where the detailed structures are preserved after decompressed.
Figs. \ref{MORE:VIS:AMA}, \ref{MORE:VIS:DT4D} and \ref{MORE:VIS:THINGI10K} depicts the visual results of the decompressed models from these datasets 
under various compression ratios. With the compression ratio increasing, the decompressed models still preserve the detailed structures, without large distortion.

\newpage
{
    \small
    \bibliographystyle{ieeenat_fullname}
   \bibliography{main}
}